\documentclass[9pt,twocolumn,twoside]{osajnl}

\journal{ol} 

\setboolean{shortarticle}{true} 

\title{Design and simplified calibration of a Mueller imaging polarimeter for material classification}

\author[*]{Yvain Qu{\'e}au}
\author[ ]{Florian Leporcq}
\author[ ]{Ayman Alfalou}

\affil[ ]{L@bISEN Yncrea-Ouest, VISION Team, ISEN Brest, 20 rue Cuirass{\'e} Bretagne CS 42807 29228, Brest Cedex, France}
\affil[*]{yvain.queau@isen-ouest.yncrea.fr}

\ociscodes{(110.5405) Polarimetric imaging; (120.5410) Polarimetry}

\doi{\url{https://doi.org/10.1364/OL.43.004941}}

\begin{abstract}
This study is concerned with the design of a Mueller imaging polarimeter for the visualization of spatially-varying Mueller matrix fields. A simplified calibration procedure is advocated, where all the optical elements are calibrated simultaneously rather than independently as in the state-of-the-art. This is shown to significantly reduce the bias inherent to sequential calibration methods. In addition, this procedure requires no reference sample, it allows calibration both in transmission or in reflection modes, and it relies on ready-to-use cameras. Put together, these novelties should help non-specialists in optics designing and calibrating a Mueller imaging polarimeter for applications such as material classification.
\end{abstract}

\setboolean{displaycopyright}{true}

\usepackage{xspace}
\usepackage{color}
\usepackage{amsmath}
\usepackage{amssymb}
\usepackage{import}
\usepackage{mathtools}
\usepackage{enumitem}

\makeatletter
\DeclareRobustCommand\onedot{\futurelet\@let@token\@onedot}
\def\@onedot{\ifx\@let@token.\else.\null\fi\xspace}

\def\eg{\emph{e.g}\onedot} 
\def\ie{\emph{i.e}\onedot}

\def\wrt{w.r.t\onedot} 

\makeatother

\def\presuper#1#2%
  {\mathop{}%
   \mathopen{\vphantom{#2}}^{#1}%
   \kern-\scriptspace%
   #2}
\def\presub#1#2%
  {\mathop{}%
   \mathopen{\vphantom{#2}}_{#1}%
   \kern-\scriptspace%
   #2}   

\begin{document}

\maketitle


The polarization properties of a medium can be measured through Mueller polarimetry~\cite{Chipman2009}, and have proven to be of fundamental importance in many applications such as biomedical diagnosis~\cite{Qi2017} or material classification~\cite{Vaughn2012}. Although the design and calibration of a Mueller polarimeter has long been investigated in the Optics community, most of existing works focus on the accurate measure of a single Mueller matrix, using \eg a HgCdTe photodetector~\cite{Goldstein1992} or a photodiode~\cite{Zallat2012}. On the other hand, practitioners need two-dimensional visualizations of the Mueller matrix field, in order to identify spatially-varying properties. Yet, non-specialists in Optics would probably favor a solution based on commercial cameras, which requires no reference sample~\cite{Compain1999} or post-processing to remove calibration errors~\cite{Hauge1978,Goldstein1990,Chenault1992,Bhattacharyya2017}, and which can handle both transmission and reflection modes~\cite{Carmagnola2014}. Calibration procedures that meet such requirements do already exist~\cite{Collins1999,Zallat2012} but they consider a dedicated calibration procedure for each optical element. One objective of this study is to show that one should rather simultaneously calibrate all the optical elements, for the sake of both simplicity and accuracy. Overall, this results in an easy-to-implement calibration procedure which simultaneously meets all the aforementioned requirements, and should hopefuly help non-specialists in Optics in the design of a Mueller imaging polarimeter.

\begin{figure}[!ht]
\centering
\def\svgwidth{.4\linewidth}
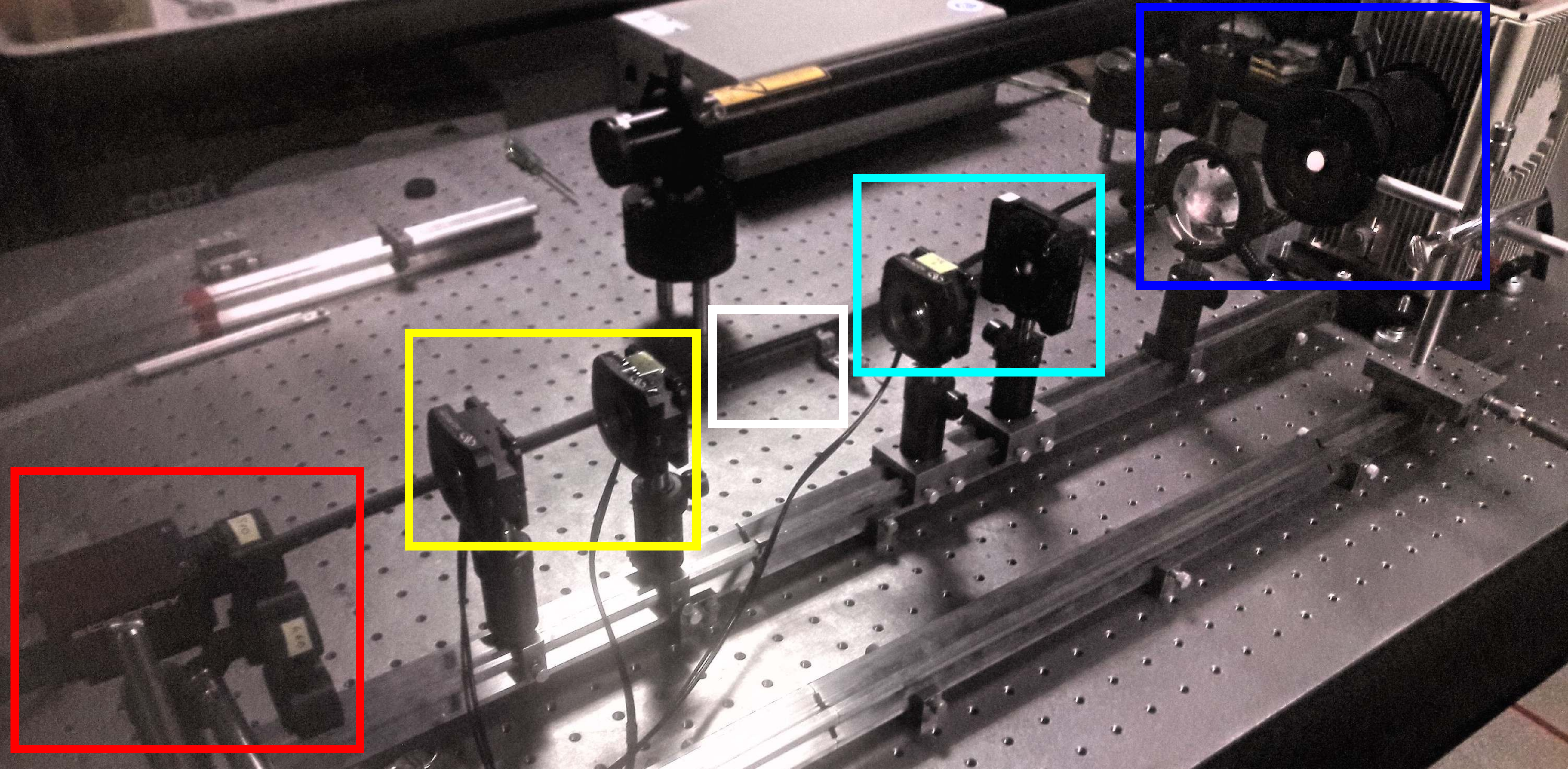
\caption{Polarimeter used in our experiments, comprising a Kohler illumination (blue), a PSG (cyan) consisting of a polarizer $P^G$ and a retarder $R^G$, a medium $M$ to be analyzed (white), a PSA (yellow) consisting of a retarder $R^A$ and a polarizer $P^A$, and a CCD camera with interference filters (red).}
\label{fig:1}
\end{figure}


We consider a dual-rotating Mueller polarimeter composed of the following elements, from source to detector (see Figure~\ref{fig:1}): 
\begin{itemize}[noitemsep]
\item A Kohler illumination system emitting a parallel and uniform white lighting;
\item A polarization state generator (PSG) comprising:
\setlist{nolistsep}
\begin{itemize}[noitemsep]
\item a linear polarizer $P^G$ with angle $\theta^G$; 
\item a retarder $R^G$ with controllable fast axis (azimuth) $\alpha^G$; 
\end{itemize} 
\item A medium $M$ to be analyzed;
\item A polarization state analyzer (PSA) comprising:
\setlist{nolistsep}
\begin{itemize}[noitemsep]
\item a retarder $R^A$ with controllable azimuth $\alpha^A$; 
\item a linear polarizer $P^A$ with angle $\theta^A$; 
\end{itemize} 
\item A CCD camera equipped with interference filters\footnote{In our experiments, we used \texttt{Newport 10LP-VIS-B} polarizers and \texttt{Newport 10RP64-532} zero-order waveplates, with a \texttt{Stingray F-033B} graylevel camera, \texttt{Newport 10BPF10} band-pass filters and \texttt{Newport AG-PR100P} piezo rotation stages to control the angles from $0^\circ$ to $340^\circ$ with a resolution of $0.001^\circ$.}.
\end{itemize}

In dual-rotating Mueller polarimetry~\cite{Collins1999,Sanz2011,Carmagnola2014,Smith2002}, both polarizers are kept fixed while several images are acquired under varying azimuthal angles of the retarders. The optical properties of the medium, represented by its Mueller matrix $M$, can be obtained by solving a system of equations having the following form: 
\begin{equation}
\begin{bmatrix}
I &
Q &
U &
V
\end{bmatrix}^\top
\propto 
\underbrace{
P^A \, R^A}_{A} 
\, M \, 
\underbrace{
R^G \, P^G \begin{bmatrix}
1 &
0 & 
0 &
0
\end{bmatrix}^\top}_G,
\label{eq:Stokes}
\end{equation}
where $\left[1,0,0,0\right]^\top$  is the Stokes vector of the (unpolarized) light entering the PSG, the lhs is the Stokes vector of the (polarized) light entering the detector, with $I$ the intensity measured by the camera, and the Mueller matrices in the rhs are given the same name as the optical element they represent.

Estimating the Mueller matrix $M$ in \eqref{eq:Stokes} from a set of intensity measurements $I$ requires knowledge of the incident Stokes vector $G$ and of the first row of the PSA matrix $A$. That is to say, the matrices $P^A$, $R^A$, $R^G$ and $P^G$ need being calibrated. However, as shown in
\begin{equation}
{\tiny
R(\alpha;\alpha_0,\delta_0) \propto \begin{bmatrix}
1 & 0 & 0 & 0 \\
0 & \cos \delta_0 \sin^2 2(\alpha-\alpha_0)  & (1 - \cos \delta_0) \cos 2(\alpha-\alpha_0)  & \sin \delta_0 \sin 2(\alpha-\alpha_0) \\
& + \cos^2 2(\alpha-\alpha_0) & \times \sin 2(\alpha-\alpha_0) & \\
0 & (1 - \cos \delta_0) \cos 2(\alpha-\alpha_0)  & \cos \delta_0 \cos^2 2(\alpha-\alpha_0)  & - \sin \delta_0 \cos 2(\alpha-\alpha_0) \\
& \times \sin 2(\alpha-\alpha_0) & + \sin^2 2(\alpha-\alpha_0) & \\
0 & - \sin \delta_0 \sin 2(\alpha-\alpha_0) & \sin \delta_0 \cos 2(\alpha-\alpha_0) & \cos \delta_0
\end{bmatrix}
} 
\label{eq:Muller_retarder}
\end{equation}
and 
\begin{equation}
{\scriptsize
P(\theta;\theta_0) \propto \begin{bmatrix}
1 & \cos 2(\theta-\theta_0) & \sin 2(\theta-\theta_0) & 0 \\
\cos 2(\theta-\theta_0) & \cos^2 2(\theta-\theta_0) & \cos 2(\theta-\theta_0) \sin 2(\theta-\theta_0) & 0 \\
\sin 2(\theta-\theta_0) & \cos 2(\theta-\theta_0) \sin 2(\theta-\theta_0) & \sin^2 2(\theta-\theta_0) & 0 \\
0 & 0 & 0 & 0
\end{bmatrix},
}
\label{eq:Muller_polarizer} 
\end{equation}
these matrices have closed-form expressions involving the angles $\theta^G$, $\theta^A$, $\alpha^G$ and $\alpha^A$, which are defined \wrt unkown reference angles (indexed with a zero). Moreover, the Mueller matrices of the retarders involve the delays $\delta^{G/A}_0$, which also need to be calibrated as functions of the wavelength.

Inaccurate calibration has long been identified as a source of serious bias in Mueller polarimetry \cite{Goldstein1990}, but as mentioned earlier there is a surprising lack of literature on accurate and simple calibration techniques. The rest of this study describes two such methods based on maximum likelihood estimation, which has recently been shown to overcome the eigenvalue method~\cite{Hu2014}. They can be used to calibrate \textit{all} the polarimeter parameters either in transmission (the medium is then the air and $M$ is the identity matrix) or in reflection (the medium is a mirror and $M$ is a diagonal matrix with elements $\left[1,1,-1,-1\right]^\top$). 

\subsection{Sequential polarimeter calibration}

\begin{figure*}[!ht]
\centering
\begin{tabular}{cccc}
\includegraphics[width=.2 \linewidth]{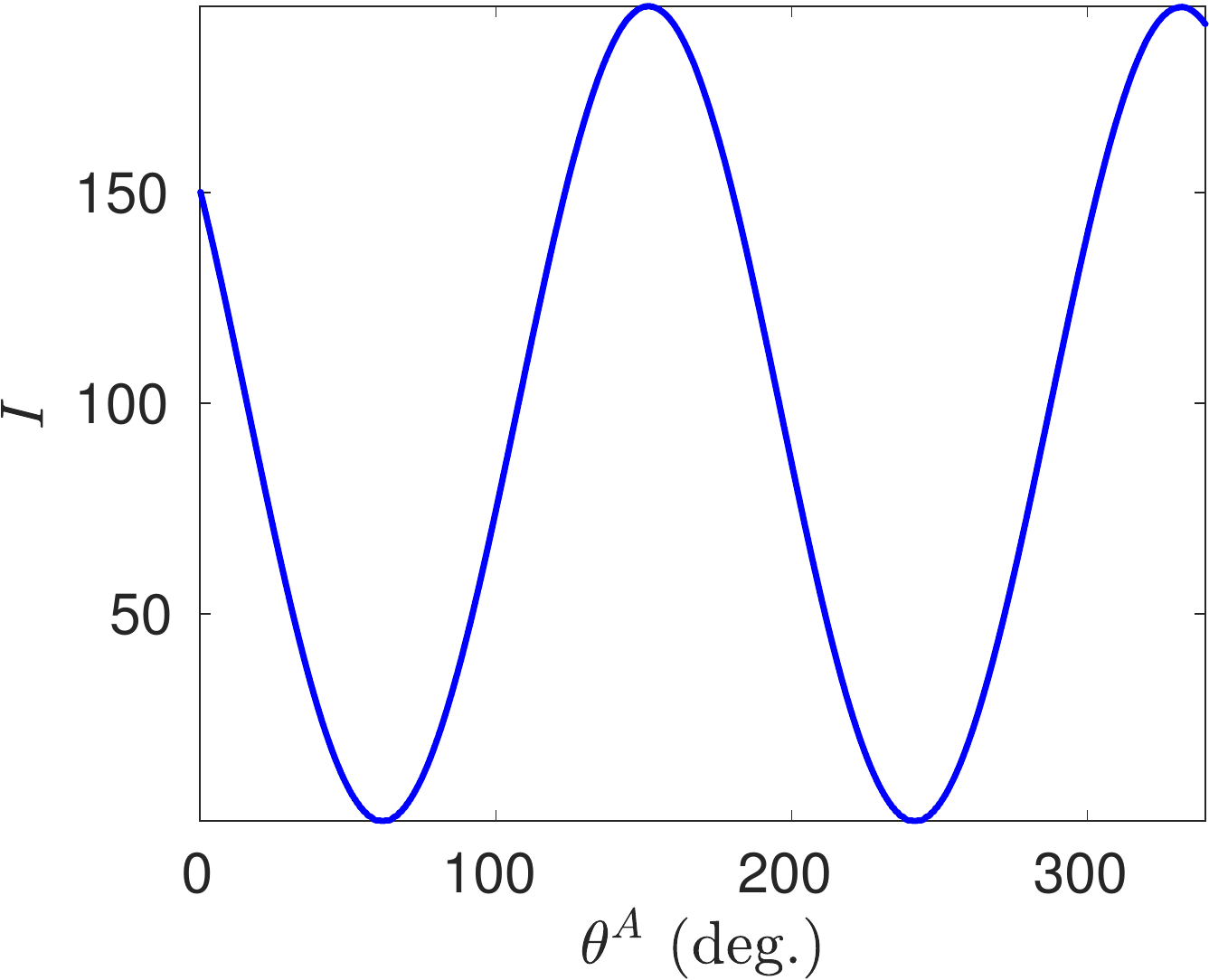} \qquad\quad&\quad\qquad
\includegraphics[width=.2 \linewidth]{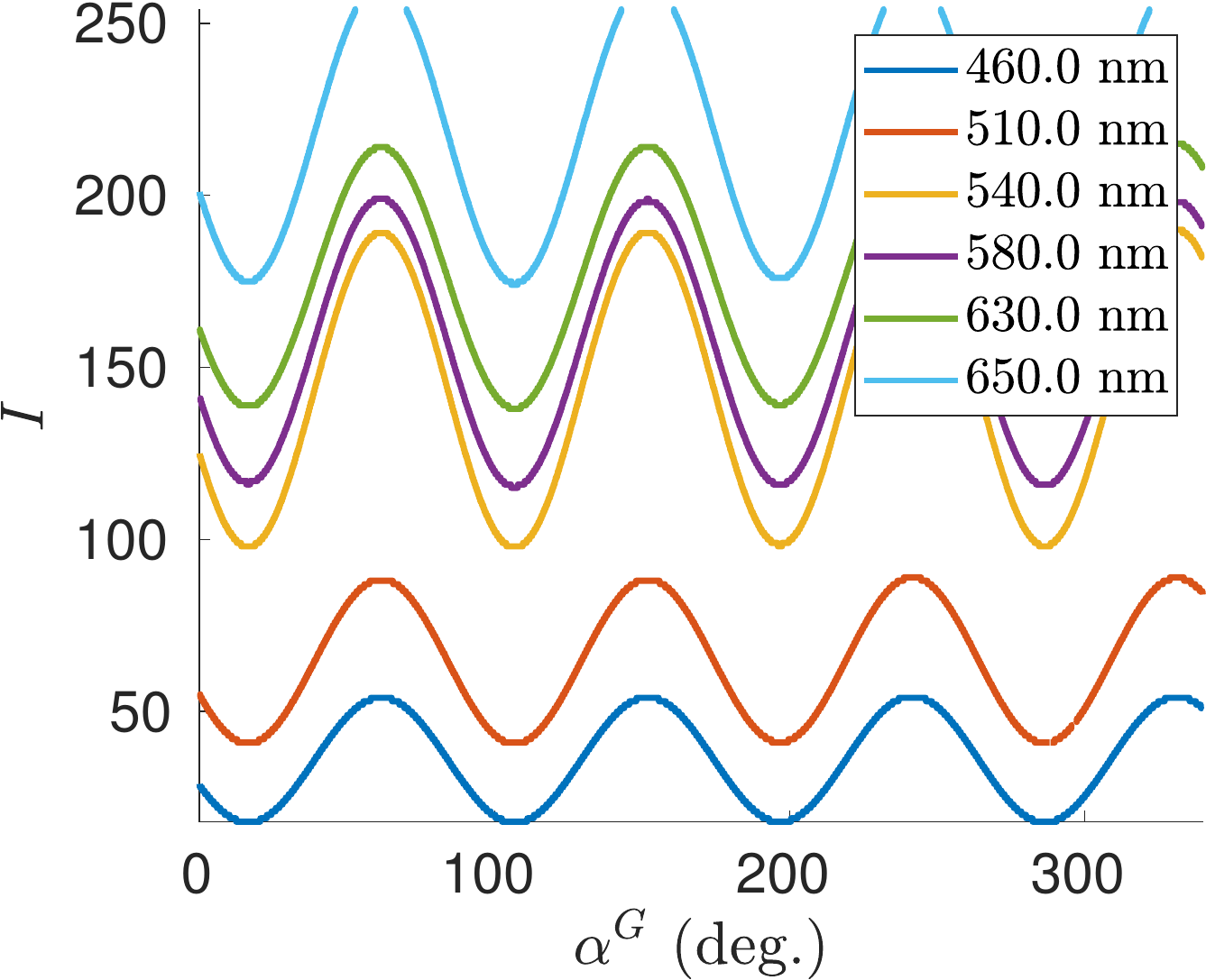}&
\includegraphics[width=.2 \linewidth]{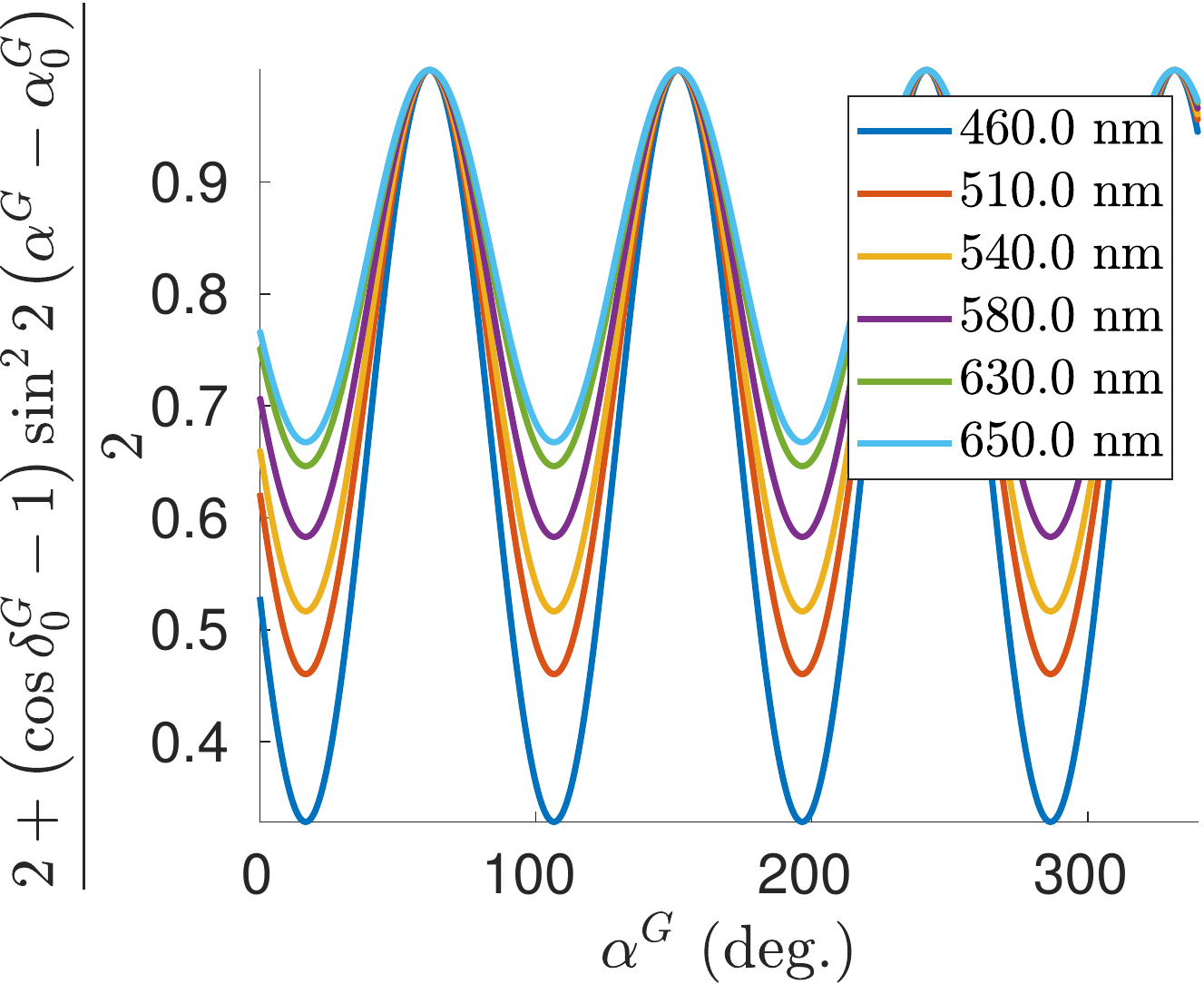} & 
\includegraphics[width=.2 \linewidth]{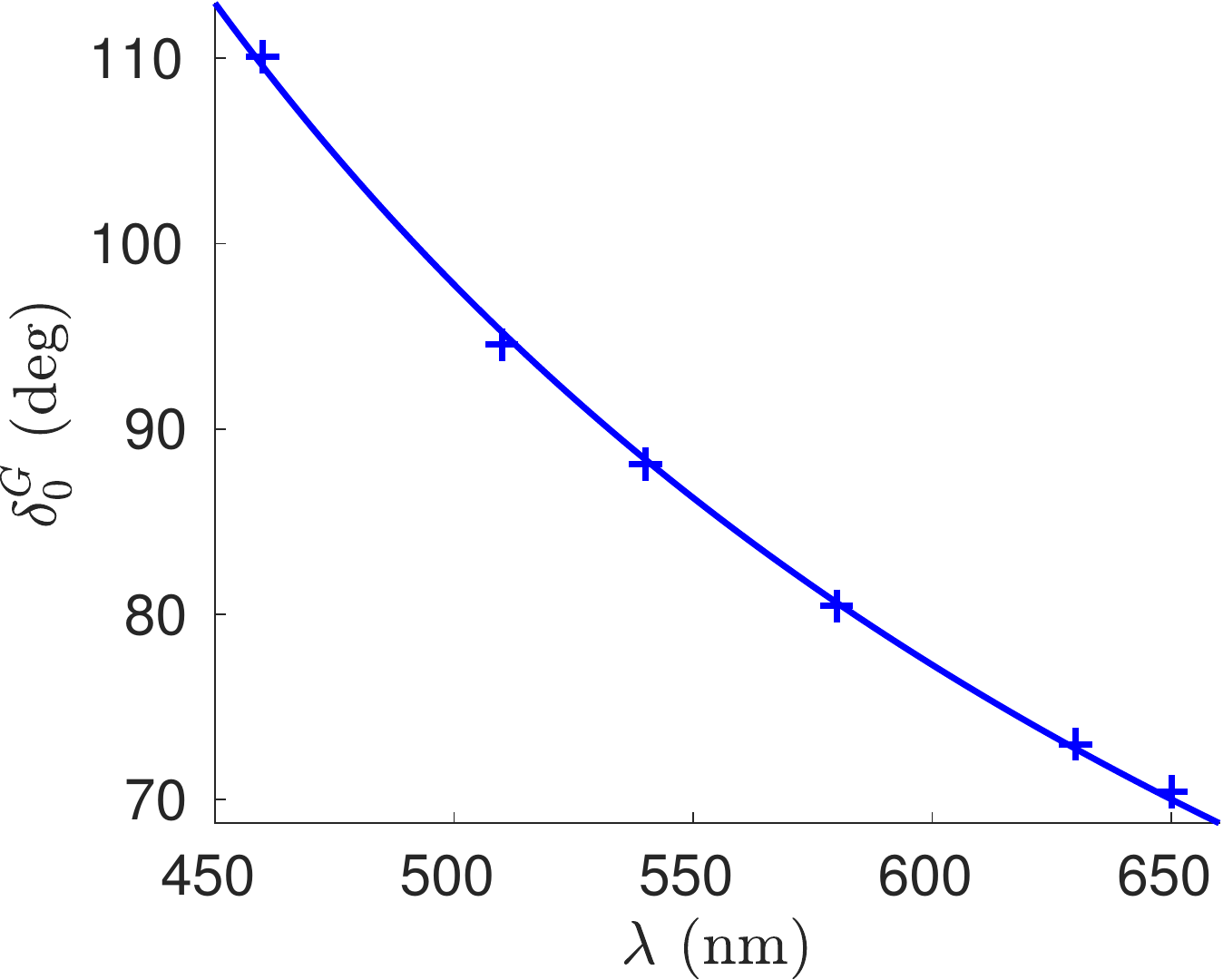} \\
\includegraphics[width=.2 \linewidth]{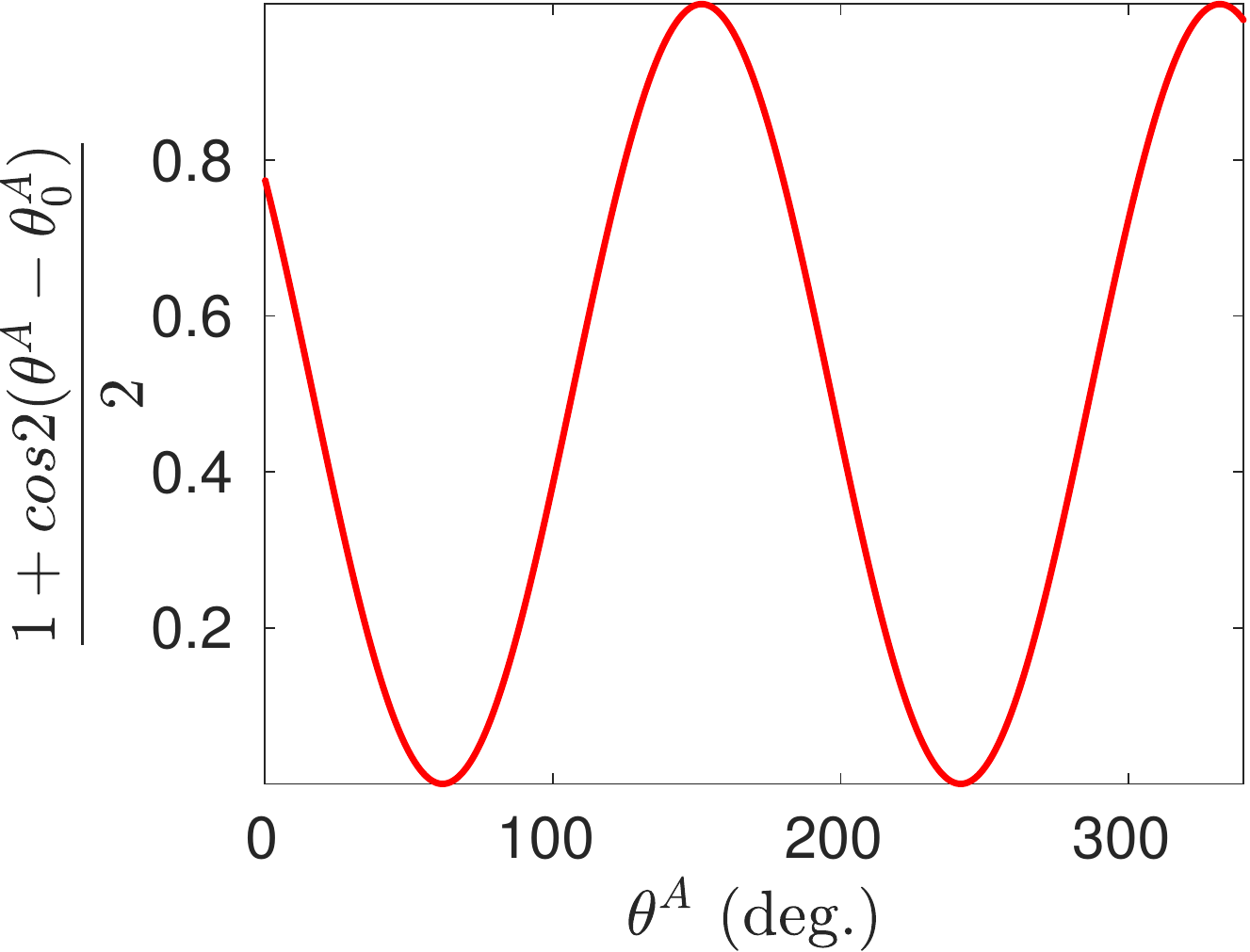}\qquad\quad&\quad \qquad
\includegraphics[width=.2 \linewidth]{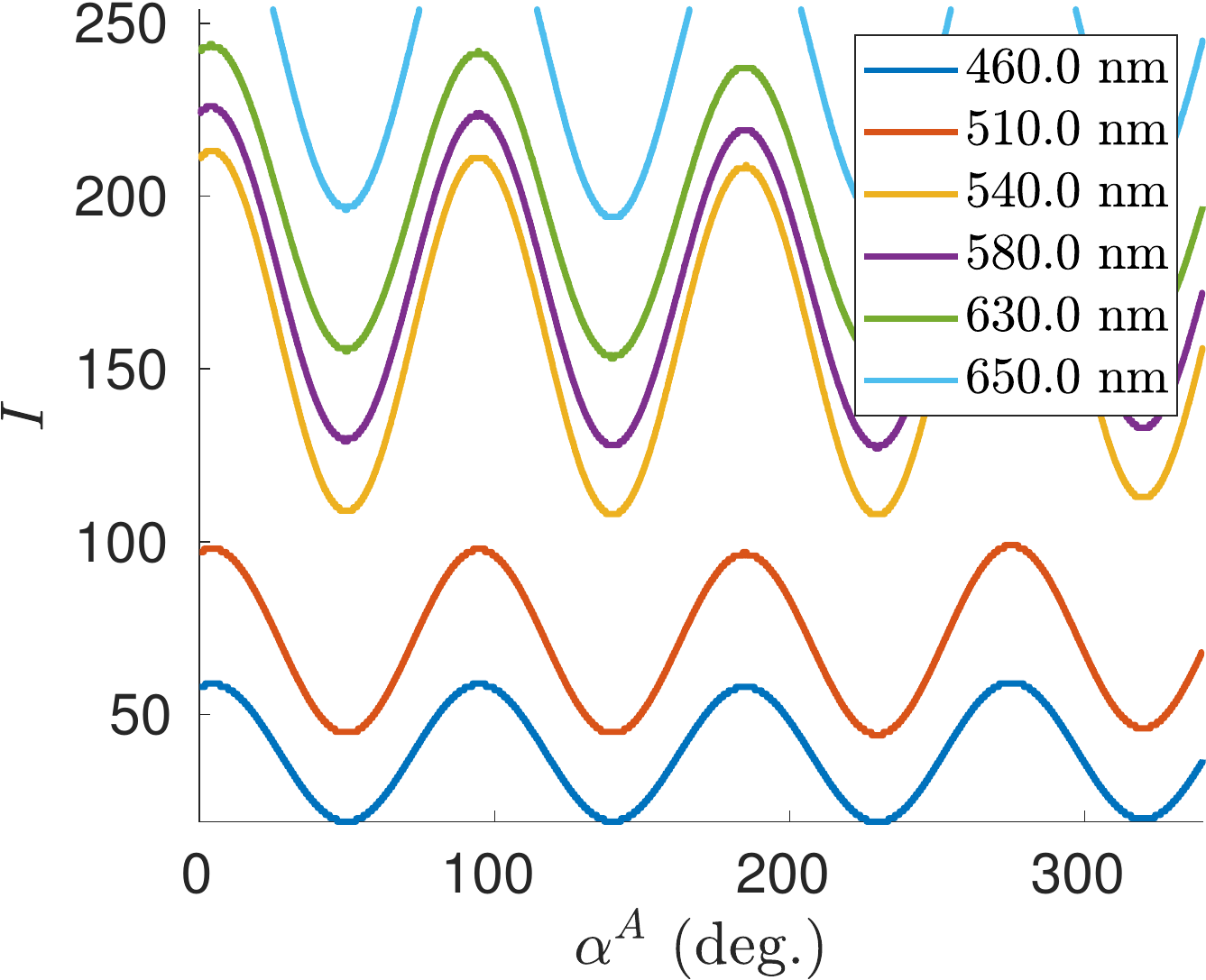}& 
\includegraphics[width=.2 \linewidth]{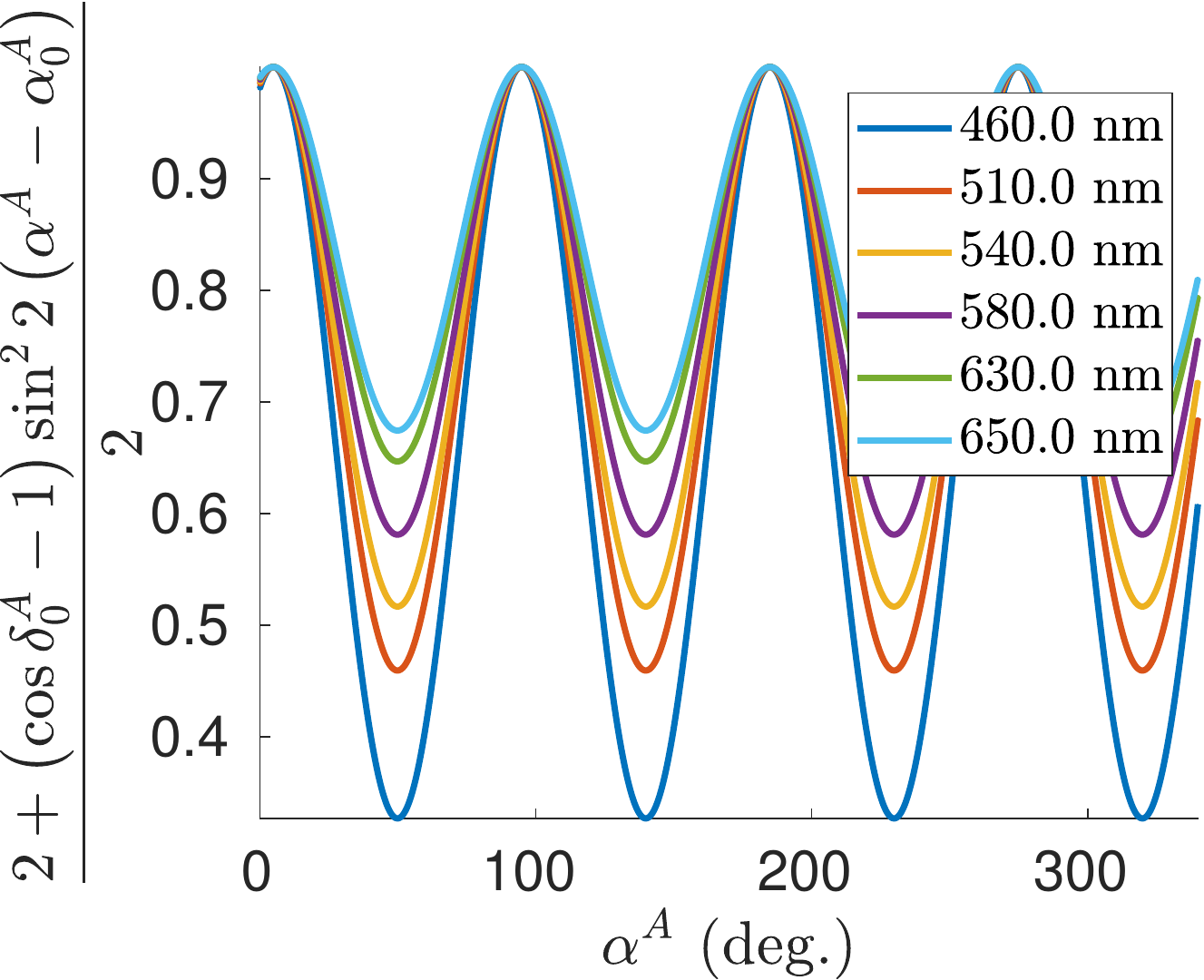}& 
\includegraphics[width=.2 \linewidth]{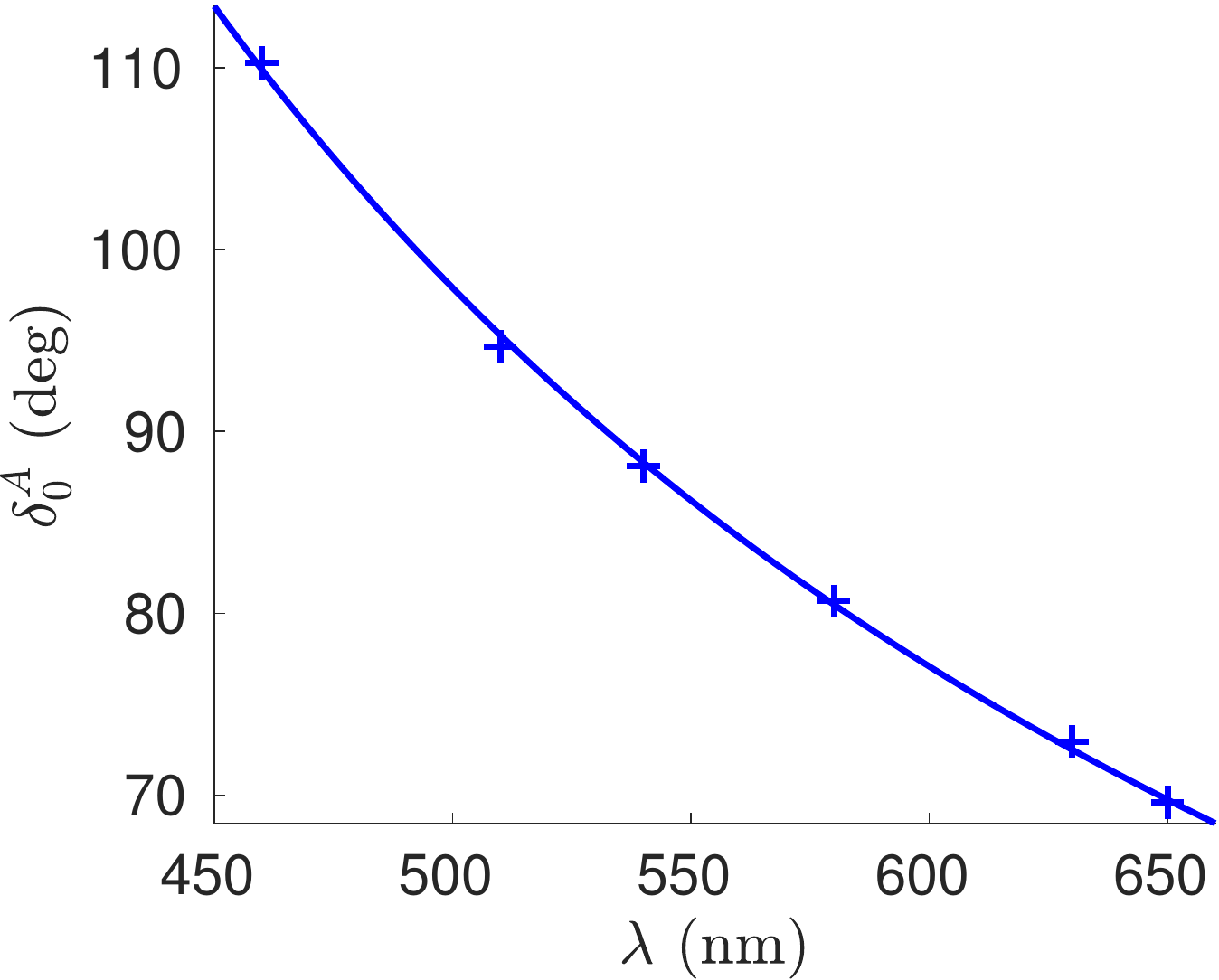} 
\end{tabular}
\caption{Sequential calibration results. First column illustrates the PSA polarizer calibration: (top) RAW measurements, using $n = 3400$\protect\footnotemark orientation values taken every $0.1^\circ$ between $0^\circ$ and $340^\circ$), and (bottom) simulated normalized intensities, using the estimate $\theta^A_0 = 151.71^\circ$. Columns two to four illustrate the calibration of the PSG (top) and PSA (bottom) retarders. Second column: RAW intensities, for $m = 6$ wavelengths and $n = 3400$ azimuths values taken every $0.1^\circ$ between $0^\circ$ and $340^\circ$. Third column: simulated normalized intensities, using the estimated angles $\alpha^G_0 = 61.58^\circ$ and $\alpha^A_0 = 94.86^\circ$ and delays $\delta^G_0$ and $\delta^A_0$. Fourth column: estimated delay values (crosses) and fitted delay function (solid line). Note the choice $n=3400$ overdetermines the estimation much more than necessary, in order for sequential calibration to consitute a reasonable reference for comparison.}
\label{fig:polarizer}
\end{figure*}

We first describe a sequential calibration procedure where the optical elements are added to the setup and calibrated one after the other, as advocated \eg, in \cite{Zallat2012,Collins1999}. 
To calibrate the PSA polarizer, the PSG polarizer is present but both retarders are removed ($R^A = R^G = I_4$). Our goal is to calibrate the orientation $\theta^A_0$ of the PSA polarizer \wrt that $\theta^G_0$ of the PSG polarizer\footnote{In the rest of this study, the origin of axes is that of the PSG linear polarizer ($\theta^G_0 = 0$), and this polarizer is kept fixed during all the experiments ($\theta^G = \theta^G_0$). \eqref{eq:Muller_polarizer} is thus a matrix with ones in the $2\times 2$ top-left block and zeros elsewhere.}. For this purpose, we take a series of $n$ measurements $I_1 \dots I_n$ under varying angle $\theta^A_1 \dots \theta^A_n$. Let $a$ be the proportionality coefficient in~\eqref{eq:Stokes}, and assume this relationship is satisfied up to a homoskedastic, zero-mean Gaussian noise. Expanding the first row of \eqref{eq:Stokes}, replacing the PSA and PSG polarizer matrices by their expressions, and assuming additive, zero-mean and homoskedastic Gaussian noise, the maximum likelihood estimate for the couple $(\theta^A_0,a)$\footnote{The proportionality constant $a$ is seen here as a hidden parameter to estimate, instead of being arbitrarily taken as the maximum intensity, which might induce errors due to quantization.} is the solution of the following nonlinear least-squares optimization problem, which we solve using Levenberg-Marquardt's algorithm~\cite{Marquardt1963}: 
\begin{equation}
\underset{\theta^A_0,a}{\min~} \sum_{j=1}^n \left( a \dfrac{1 + \cos 2(\theta^A_j-\theta^A_0)}{2} - I_j \right)^2.
\end{equation}
The left column in Figure~\ref{fig:polarizer} shows an example of results obtained with this approach, while calibrating the polarimeter shown in Figure~\ref{fig:1}.


To calibrate the PSG retarder, the PSA retarder is removed ($R^A = I_4$), and the angles of both polarizers are set to zero ($\theta^G = \theta^G_0$ and $\theta^A = \theta^A_0$). 
The unknowns are the angle $\alpha^G_0$ and the delay $\delta^G_0$ (which is a function of the wavelength). To estimate them, we take $m$ series of shots under different wavelength $\lambda^i,~i \in \{1,\dots,m\}$\footnote{This can be accomplished either by using a multispectral camera, or by placing narrow-band interference filters before a monochromatic CCD sensor.}, and for each series $i$ we record $n$ measurements $\prescript{}{k}{I}^i, k \in \{1,\dots,n\}$ under varying angle $\prescript{}{k}{\alpha}^G,~k \in \{1,\dots,n\}$. Let us assume again homoskedastic, zero-mean Gaussian noise, and denote by $b^i$ the proportionality constant (which is wavelength-dependent, due to the sensor response being wavelength-dependent) and by $\delta^{G,i}_{0}$ the delay for the wavelength $\lambda^i$. By expanding the first row of \eqref{eq:Stokes} along with \eqref{eq:Muller_retarder} and \eqref{eq:Muller_polarizer}, the maximum likelihood estimation for the set of unknown parameters $\left(\alpha^G_0,\left\{(b^i,\delta_0^{G,i})\right\}_{i \in \{1,\dots,m\}}\right)$ is attained by solving the following nonlinear least-squares problem using, \eg, Levenberg-Marquardt's algorithm: 
%
\begin{equation}
\underset{\alpha^G_0,\left\{(b^i,\delta_0^{G,i})\right\}_{i}}{\min~}  \sum_{i=1}^m \sum_{k=1}^n \left( b^i \dfrac{2 + \left( \cos \delta^{G,i}_{0} - 1 \right) \sin^2 2(\prescript{}{k}{\alpha}^G-\alpha^G_0) }{2} - \prescript{}{k}{I}^i \right)^2.
\end{equation}
Then, from the estimated values $\{\delta_0^{G,i}\}_i$ of the delays we can obtain the delay value for any wavelength $\lambda$ according to Cauchy's approximation\footnote{We used Cauchy's approximation since we focus on the visible spectrum, yet a more accurate model such as Sellmeier's could have been employed.}
\begin{equation}
\delta_{0}^G(\lambda) = \dfrac{\kappa^G_1}{\lambda} + \dfrac{\kappa^G_2}{\lambda^3},
\label{eq:delay}
\end{equation}
where $(\kappa^G_1,\kappa^G_2)$ can be obtained by solving in a least-squares manner the system of linear equations formed by the $m$ equations~(\ref{eq:delay}) with the estimated values $\{\delta_0^{G,i}\}_i$ and the chosen wavelengths $\{\lambda^{i}\}_i$.
Columns two to four in Figure~\ref{fig:polarizer} show  examples of results for the calibration of the PSG and PSA retarders of Figure \ref{fig:1} (the calibration procedure for the PSA retarder is exactly the same as that of the PSG, provided that the angle of the PSG retarder is set to zero \ie, $\alpha^G = \alpha^G_0$). 
%

\subsection{Bundle-adjusted polarimeter calibration}

Given the sequential nature of the previous approach, bias may be accumulated through the procedure (\eg, a wrong calibration of the PSA polarizer will bias the calibration of the PSG retarder, and that of the PSA retarder even more). Moreover, slight displacements of the optical elements between the numerous steps may be another source of bias. Therefore, an integrated calibration method for the joint estimation of all parameters (the three angles $\theta^A_0$, $\alpha^G_0$, and $\alpha^A_0$, and the four parameters $\kappa^{G/A}_{1/2}$ of the two delay functions $\delta^{G/A}_0$ modeled as in \eqref{eq:delay}) would require less manual intervention, and be more accurate. We now introduce such a method, which is inspired by the classic bundle adjustment method widely used in computer vision~\cite{Triggs1999}. 

Let us consider a series of measurements $\prescript{l}{k}{I}_j^i$ obtained under varying wavelength $\left\{\lambda^i\right\}_i$, polarizer angle $\left\{\theta^A_j\right\}_j$, PSG azimuth $\left\{\prescript{}{k}{\alpha}^G\right\}_k$, and PSA azimuth $\left\{\prescript{l}{}{\alpha}^A\right\}_l$ (we used $6$ wavelengths and $8$ different values for the angles taken every $22.5^\circ$ between $0^\circ$ and $157.5^\circ$ for the azimuths $\alpha_0^{G/A}$, and every $45^\circ$ between $0^\circ$ and $315^\circ$ for the polarizer angles $\theta_0^{G/A}$). 

Let us assume that \eqref{eq:Stokes} is satisfied up to additive, zero-mean and homoskedastic Gaussian noise, and denote by $b^i$ the unknown scale parameter for the $i$-th wavelength. 
The maximum likelihood estimate for the set of unknown parameters is thus attained by solving the nonlinear least-squares problem
\begin{equation}
\underset{\substack{\alpha^G_0,\kappa_1^G,\kappa_2^G \\ \alpha^A_0,\kappa_1^A,\kappa_2^A \\ \theta^A_0, \left\{b^i\right\}_{i} }}{\min~}  \sum_{i,j,k,l} \left( b^i \, \prescript{l}{}{A}^i_j(\alpha^A_0,\kappa^A_1,\kappa^A_2,\theta^A_0) \, M \, \prescript{}{k}G^i(\alpha^G_0,\kappa^G_1,\kappa^G_2) - \prescript{l}{k}{I}_j^i \right)^2
\end{equation}
with $\prescript{l}{}{A}^i_j$ the first row of the PSA matrix $A$ in \eqref{eq:Stokes} given, according to \eqref{eq:Muller_retarder}, \eqref{eq:Muller_polarizer} and \eqref{eq:delay}, by
\begin{equation}
\begin{array}{l}
{\tiny
\prescript{l}{}{A}^i_j(\alpha^A_0,\kappa^A_1,\kappa^A_2,\theta^A_0) =  
}
\\
\!\!\!\!\!\!\!\!\!\!{\tiny
\begin{bmatrix*}[l]
1 \\
\cos 2\left(\theta^A_j-\theta^A_0\right)\left(\cos^2 2\left(\prescript{l}{}{\alpha}^A - \alpha^A_0\right) + \cos\left(\dfrac{\kappa^A_1}{\lambda^i} + \dfrac{\kappa^A_2}{\left(\lambda^i\right)^3}\right) \sin^2 2\left(\prescript{l}{}{\alpha}^A - \alpha^A_0\right) \right) \dots \\[-0.5em]
\qquad + \sin 2\left(\theta^A_j-\theta^A_0\right) \left(1 - \cos\left(\dfrac{\kappa^A_1}{\lambda^i} + \dfrac{\kappa^A_2}{\left(\lambda^i\right)^3}\right) \right) \cos 2\left(\prescript{l}{}{\alpha}^A - \alpha^A_0\right)  \sin 2\left(\prescript{l}{}{\alpha}^A - \alpha^A_0\right) \\
\cos 2\left(\theta^A_j-\theta^A_0\right) \left(1 - \cos\left(\dfrac{\kappa^A_1}{\lambda^i} + \dfrac{\kappa^A_2}{\left(\lambda^i\right)^3}\right)\right) \cos 2\left(\prescript{l}{}{\alpha}^A - \alpha^A_0\right)  \sin 2\left(\prescript{l}{}{\alpha}^A - \alpha^A_0\right) \dots \\[-0.5em]
\qquad + \sin 2\left(\theta^A_j-\theta^A_0\right)\left( \cos\left(\dfrac{\kappa^A_1}{\lambda^i} + \dfrac{\kappa^A_2}{\left(\lambda^i\right)^3}\right) \cos^2 2\left(\prescript{l}{}{\alpha}^A - \alpha^A_0\right)  + \sin^2 2\left(\prescript{l}{}{\alpha}^A - \alpha^A_0\right)\right) \\
\cos 2\left(\theta^A_j-\theta^A_0\right) \sin\left(\dfrac{\kappa^A_1}{\lambda^i} + \dfrac{\kappa^A_2}{\left(\lambda^i\right)^3}\right) \sin 2\left(\prescript{l}{}{\alpha}^A - \alpha^A_0\right)   - \sin 2\left(\theta^A_j-\theta^A_0\right) \sin\left(\dfrac{\kappa^A_1}{\lambda^i} + \dfrac{\kappa^A_2}{\left(\lambda^i\right)^3}\right) \cos 2\left(\prescript{l}{}{\alpha}^A - \alpha^A_0\right)
\end{bmatrix*}^\top
}
\end{array}
\label{eq:Mueller_PSA}
\end{equation}
and $\prescript{}{k}G^i$ the Stokes vector exiting the PSG given, according to \eqref{eq:Muller_retarder}, \eqref{eq:Muller_polarizer} and \eqref{eq:delay}, by
\begin{equation}
{\scriptsize
\prescript{}{k}G^i(\alpha^G_0,\kappa^G_1,\kappa^G_2) =  
\begin{bmatrix}
1 \\
\cos\left( \dfrac{\kappa^G_1}{\lambda^i} + \dfrac{\kappa^G_2}{\left(\lambda^i\right)^3} \right) \sin^2 2\left( \prescript{}{k}{\alpha}^G - \alpha^G_0 \right) + \cos^2 2\left( \prescript{}{k}{\alpha}^G - \alpha^G_0 \right)  \\
\left(1 - \cos\left( \dfrac{\kappa^G_1}{\lambda^i} + \dfrac{\kappa^G_2}{\left(\lambda^i\right)^3} \right) \right) \cos 2\left( \prescript{}{k}{\alpha}^G - \alpha^G_0 \right) \sin 2\left( \prescript{}{k}{\alpha}^G - \alpha^G_0 \right) \\
-\sin\left( \dfrac{\kappa^G_1}{\lambda^i} + \dfrac{\kappa^G_2}{\left(\lambda^i\right)^3} \right) \sin 2\left( \prescript{}{k}{\alpha}^G - \alpha^G_0 \right)
\end{bmatrix}.
}
\label{eq:Stokes_in}
\end{equation}

The angles estimated with this integrated approach differ by less than $2^\circ$ from those obtained with the sequential approach, thus we do not reproduce any new calibration result. It is the computation of real-world Mueller matrix measurements which will highlight the significance of this slight difference. 
Still, let us already remark that in the sequential procedure, we used a total of $3 \times 3400 = 10200$ observations per wavelength. In contrast, we used $20$ times less ($512$ per wavelength) observations for the bundle-adjusted method: if the latter is to provide similar results with so much fewer observations then it can be considered as substantially better in terms of simplicity\footnote{Our piezo rotating stages being limited to a speed of $1.5^\circ / sec.$, in our experiments it takes around $1.5\,hrs$ to acquire the $512$ measurements used to calibrate the polarimeter at one particular wavelength, and around $10\,min$ to acquire the $64$ ones used for polarimetric imaging. These numbers might be significantly reduced by using faster rotating stages.}.

\subsection{Polarimetric Imaging}

\begin{table*}[!ht]
\caption{Estimated Mueller matrices and relative error for the air (transmission mode, first row) and a mirror (reflection mode, second row), using the calibration parameters obtained with the sequential (left) and bundle-adjusted methods (right). The bottom figures show the spatial distribution of errors for the mirror, in false colors (blue is zero, yellow is $0.04$).}
\label{tab:air}
\begin{tabular}{cc|cc}
\scriptsize
$M^\mathrm{air}_\mathrm{seq} = \begin{bmatrix}
1.000 & 0.010 & 0.004& 0.001 \\
0.002 & 0.991 & -0.015 & 0.009 \\
-0.007 & 0.010 & 0.992 & -0.004 \\
0.001 & -0.005 & -0.001 & 0.997
\end{bmatrix}$ 
& 
$\frac{\left\| M^\mathrm{air}_\mathrm{seq} - M^\mathrm{air} \right\|_F}{\left\| M^\mathrm{air} \right\|_F} = 0.015$ 
&
\scriptsize
$
M^\mathrm{air}_\mathrm{bun} = \begin{bmatrix}
1.000 & 0.000 & 0.003& -0.003 \\
-0.007 & 1.002 & -0.010 & 0.011 \\
-0.007 & 0.007 & 1.005 & -0.002 \\
0.003 & -0.002 & 0.003 & 0.996
\end{bmatrix}
$
&
$\frac{\left\| M^\mathrm{air}_\mathrm{bun} - M^\mathrm{air} \right\|_F}{\left\| M^\mathrm{air} \right\|_F} = 0.011
$
\\
\hline
\scriptsize
$M^\mathrm{mirr}_\mathrm{seq} = \begin{bmatrix}
1.000 & 0.008 & -0.010& 0.003 \\
0.007 & 0.983 & 0.016 & -0.006 \\
-0.003 & 0.018 & -0.994 & 0.015 \\
-0.002 & -0.005 & -0.016 & -1.002
\end{bmatrix}$
&
$\frac{\left\| M^\mathrm{mirr}_\mathrm{seq} - M^\mathrm{mirr} \right\|_F}{\left\| M^\mathrm{mirr} \right\|_F} = 0.021$
&
\scriptsize
$M^\mathrm{mirr}_\mathrm{bun} = \begin{bmatrix}
1.000 & 0.004 & -0.004& 0.004 \\
-0.001 & 1.000 & 0.000 & -0.005 \\
0.007 & 0.000 & -1.009 & 0.014 \\
-0.003 & -0.000 & -0.017 & -0.998
\end{bmatrix}$
&
$\frac{\left\| M^\mathrm{mirr}_\mathrm{bun} - M^\mathrm{mirr} \right\|_F}{\left\| M^\mathrm{mirr} \right\|_F} = 0.013$ \\
\multicolumn{2}{c|}{
\begin{tabular}{cccc}
\includegraphics[width=0.07\linewidth]{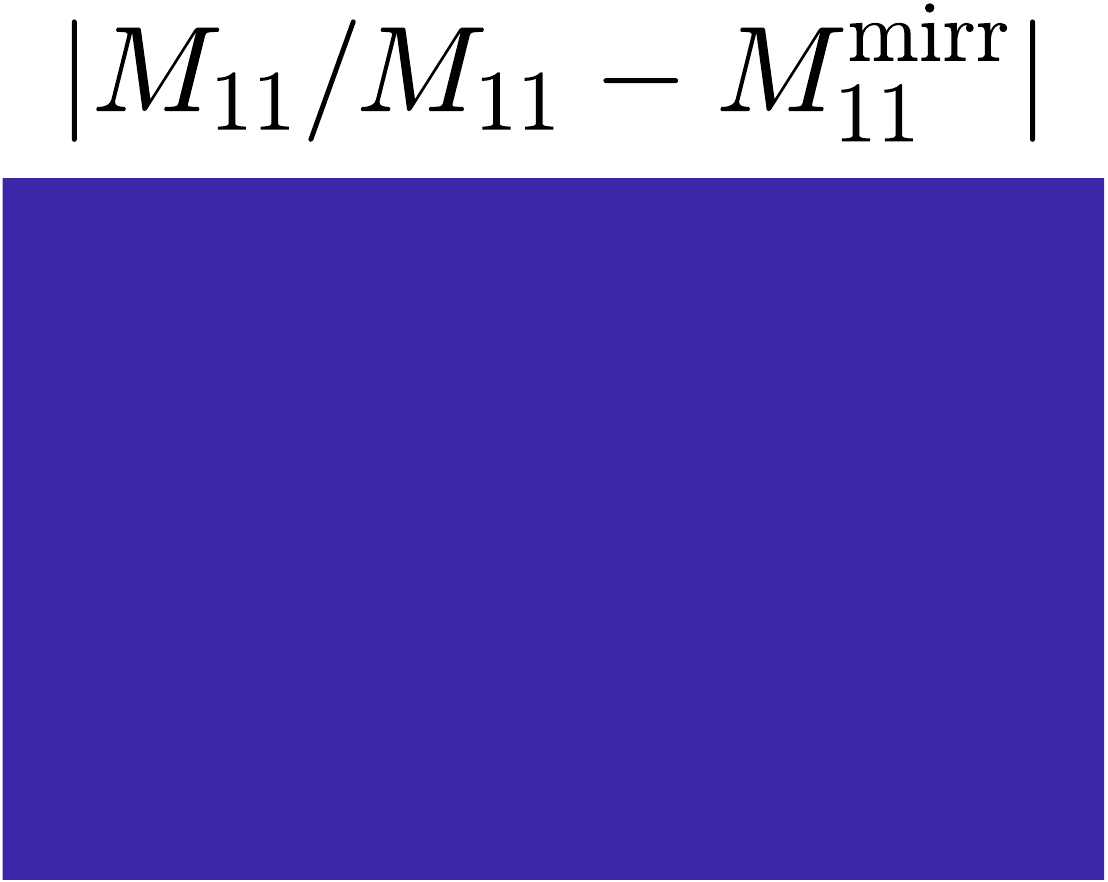} &
\includegraphics[width=0.07\linewidth]{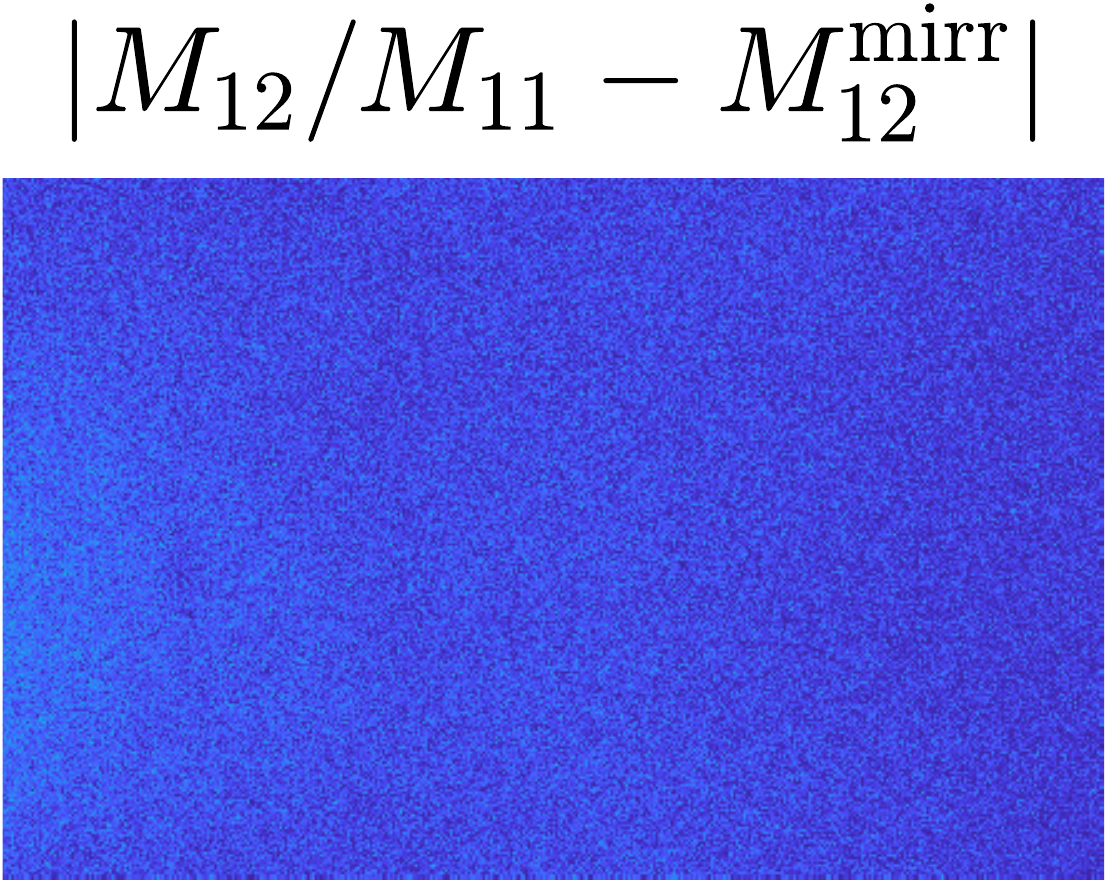} &
\includegraphics[width=0.07\linewidth]{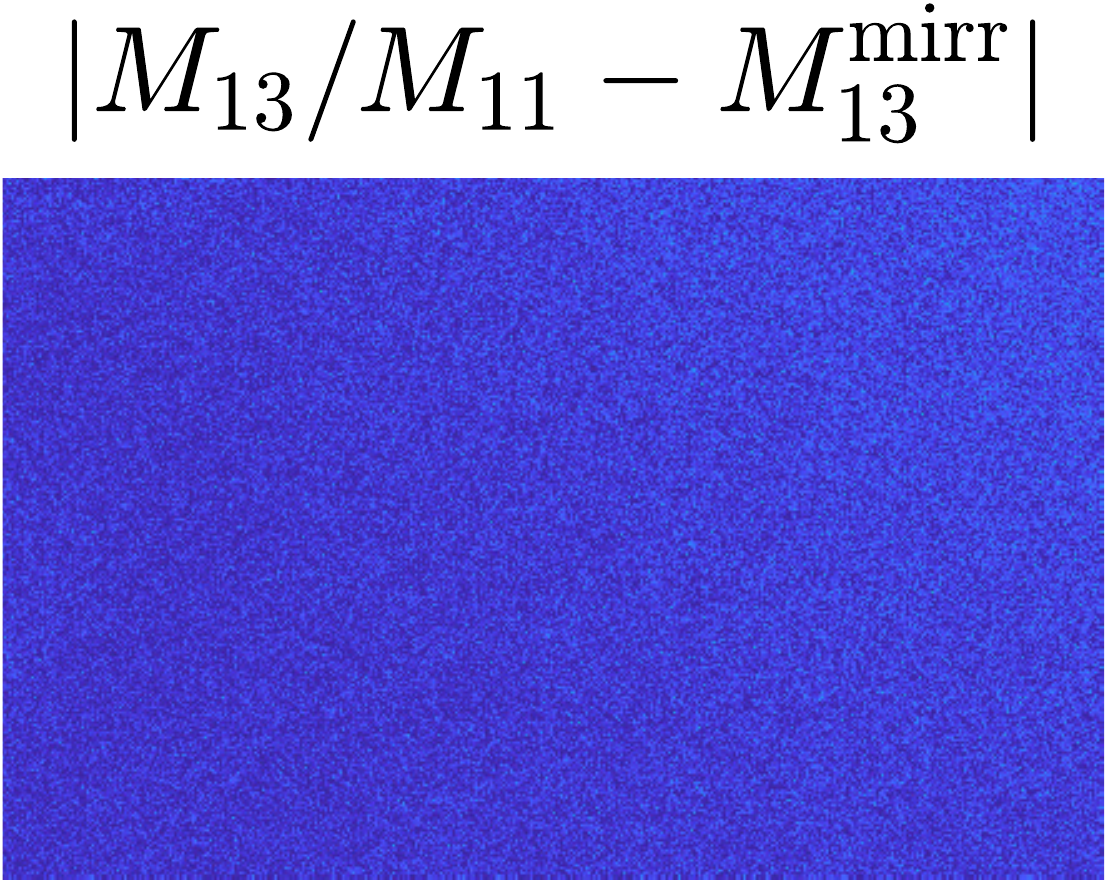} &
\includegraphics[width=0.07\linewidth]{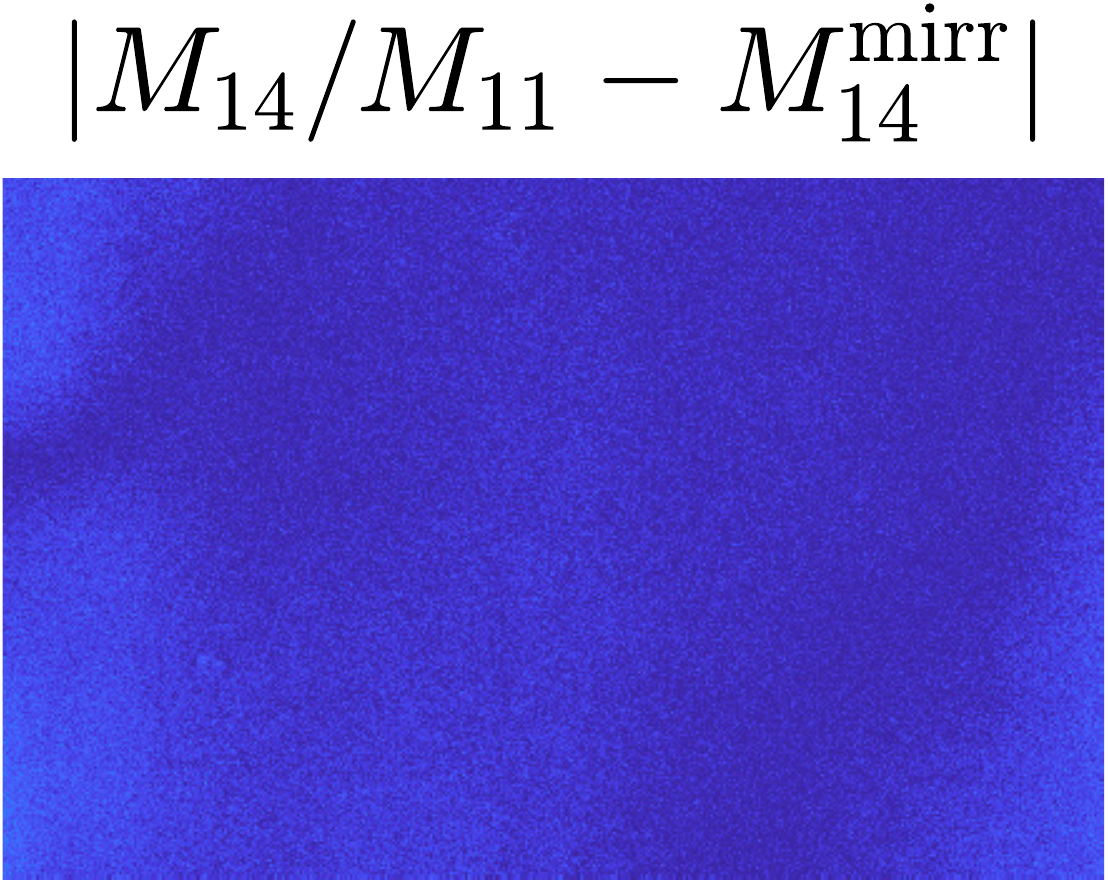} \\
\includegraphics[width=0.07\linewidth]{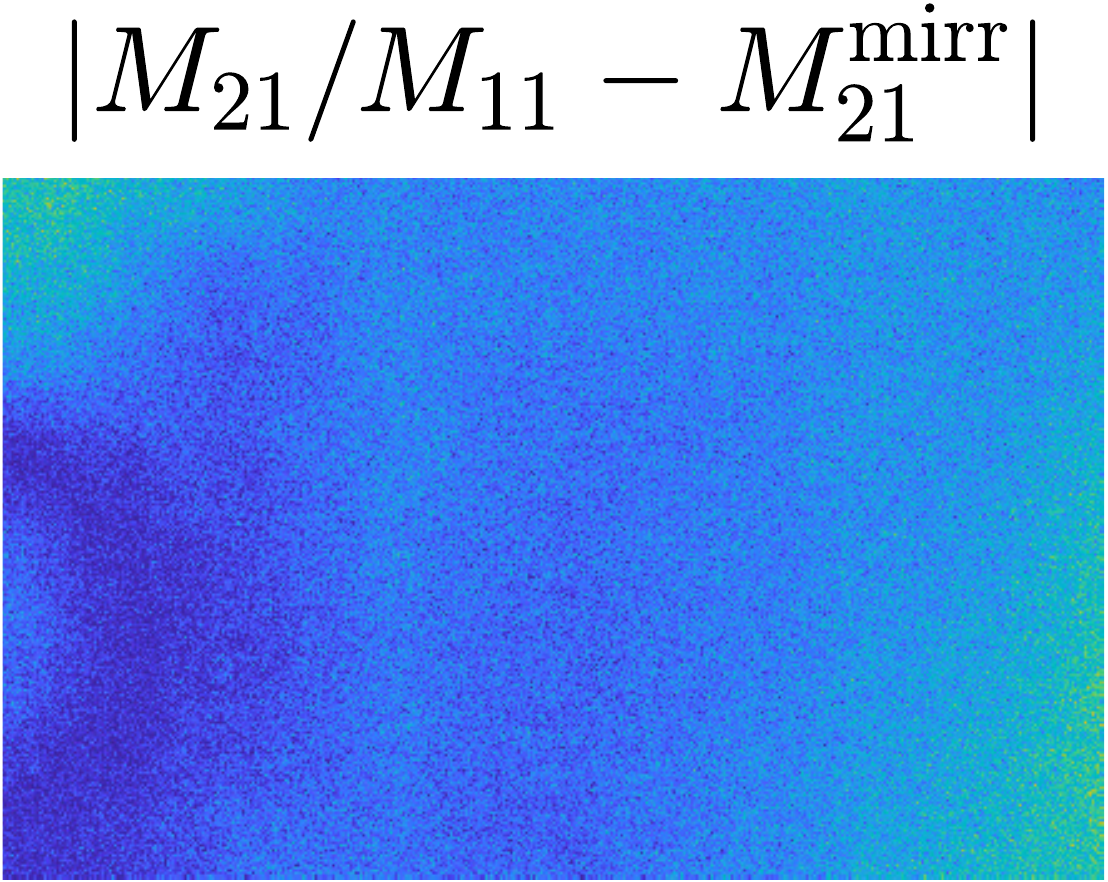} &
\includegraphics[width=0.07\linewidth]{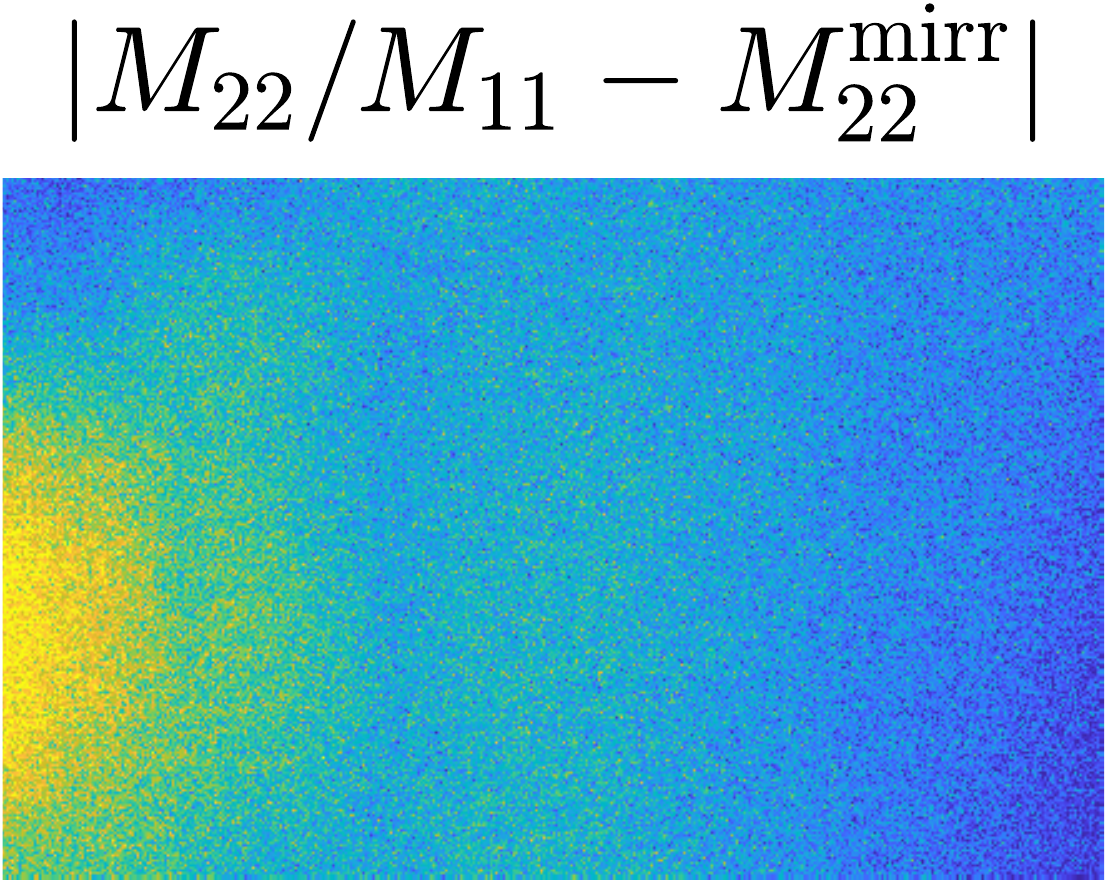} &
\includegraphics[width=0.07\linewidth]{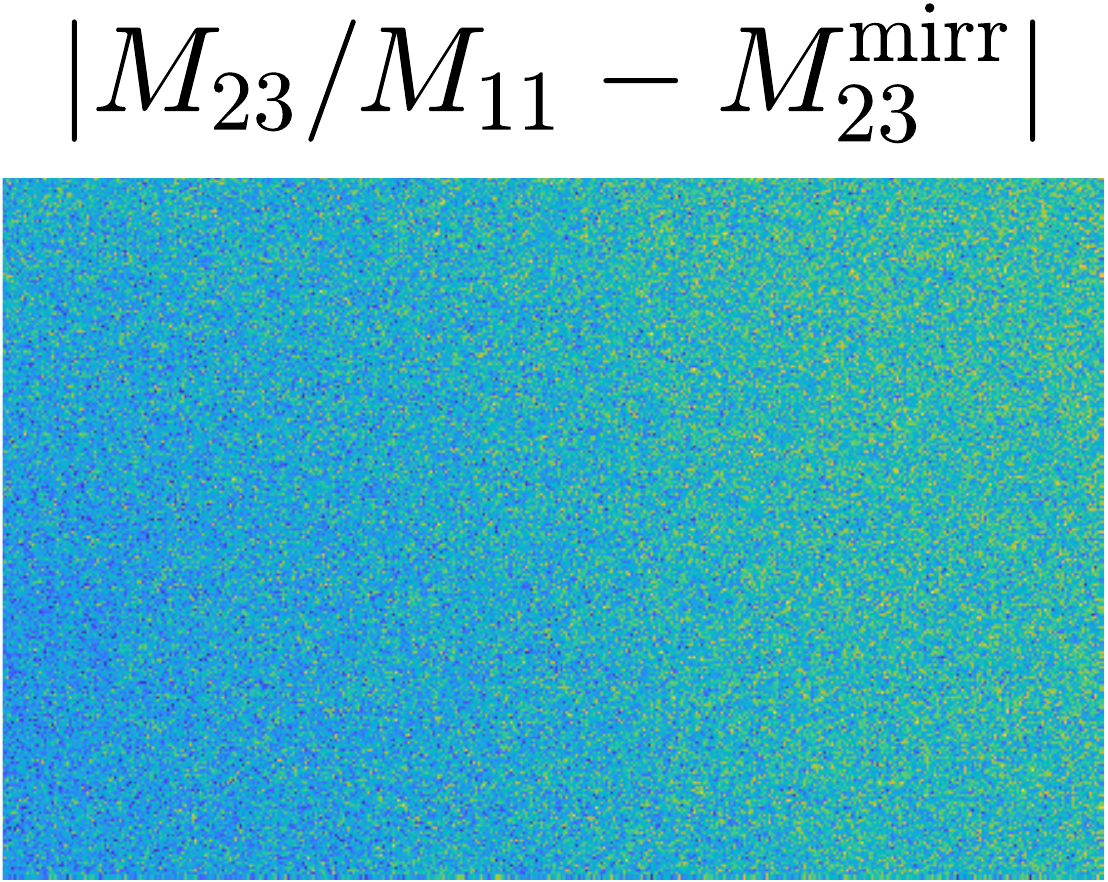} &
\includegraphics[width=0.07\linewidth]{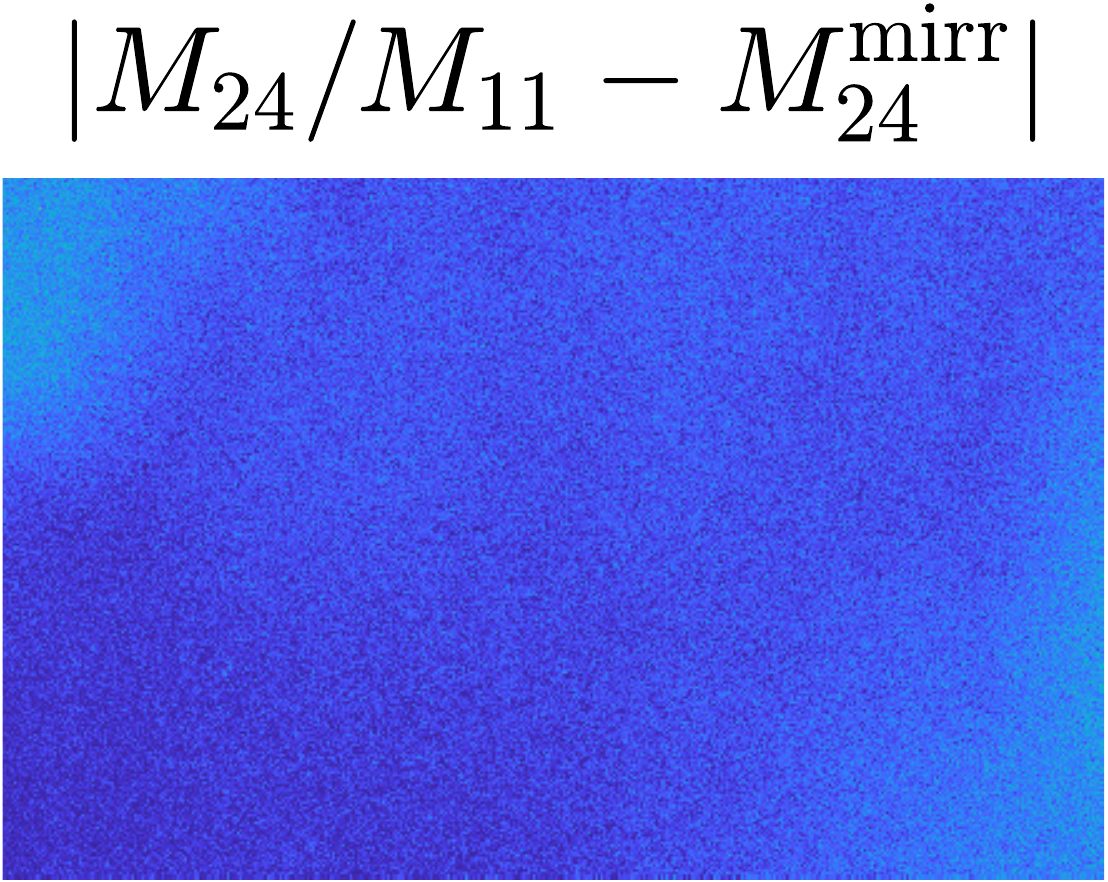} \\
\includegraphics[width=0.07\linewidth]{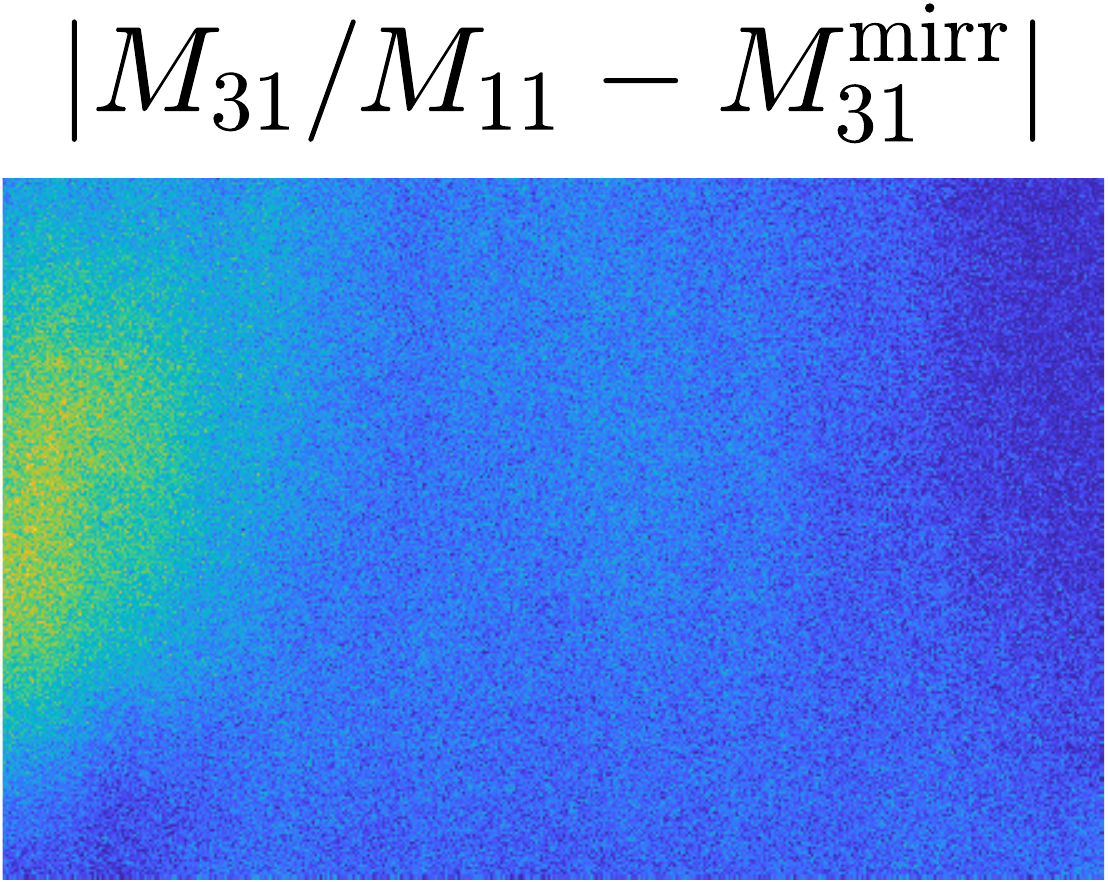} &
\includegraphics[width=0.07\linewidth]{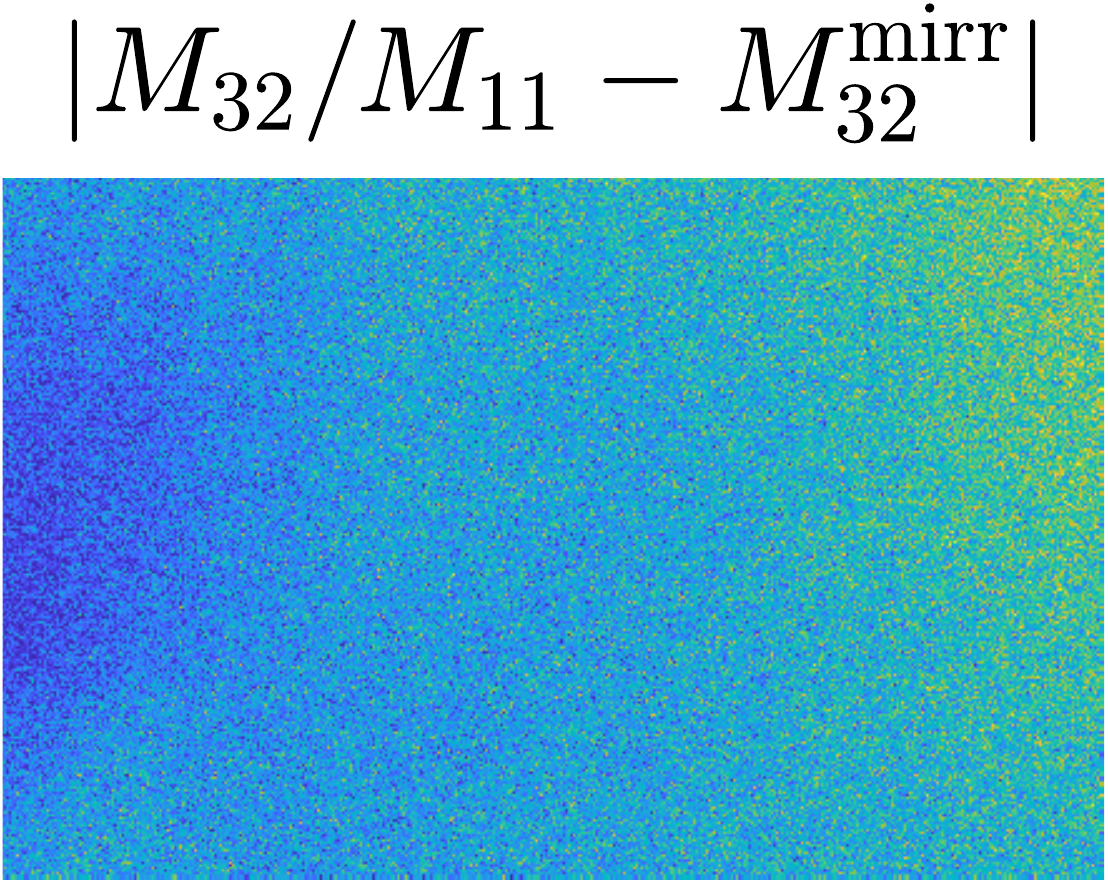} &
\includegraphics[width=0.07\linewidth]{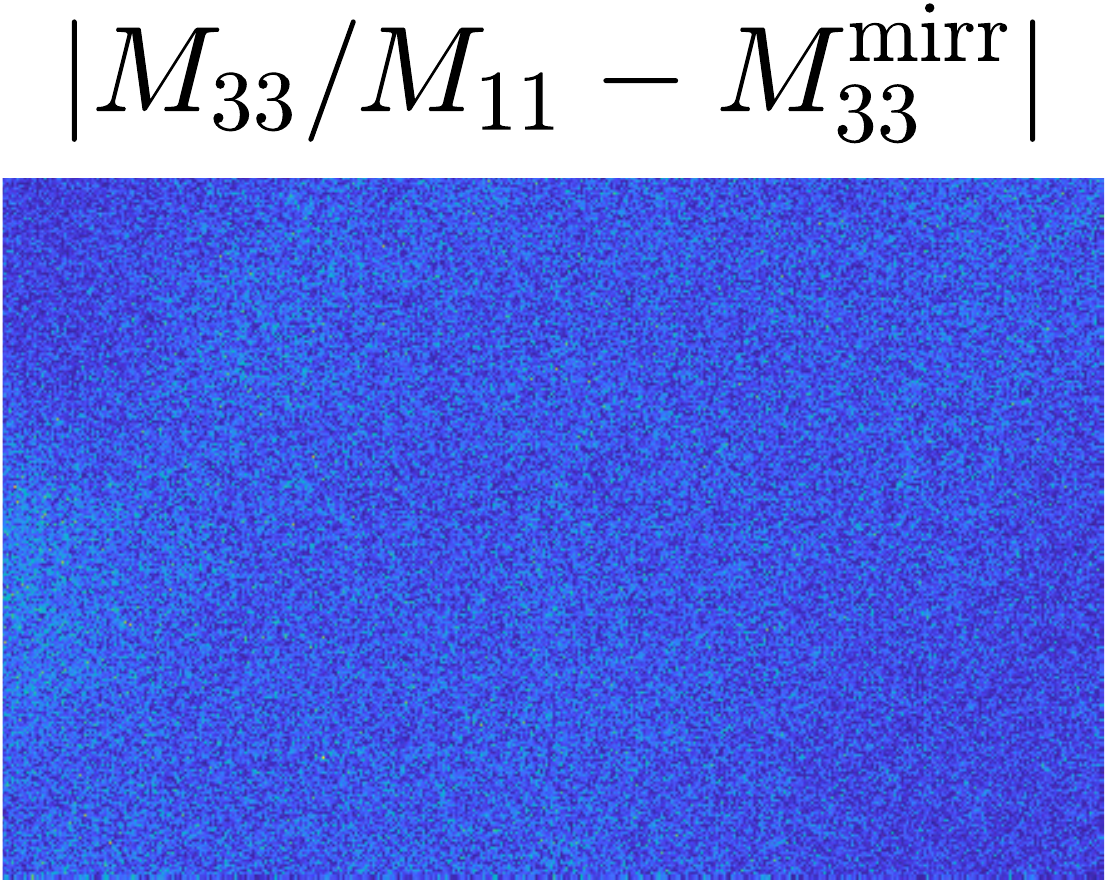} &
\includegraphics[width=0.07\linewidth]{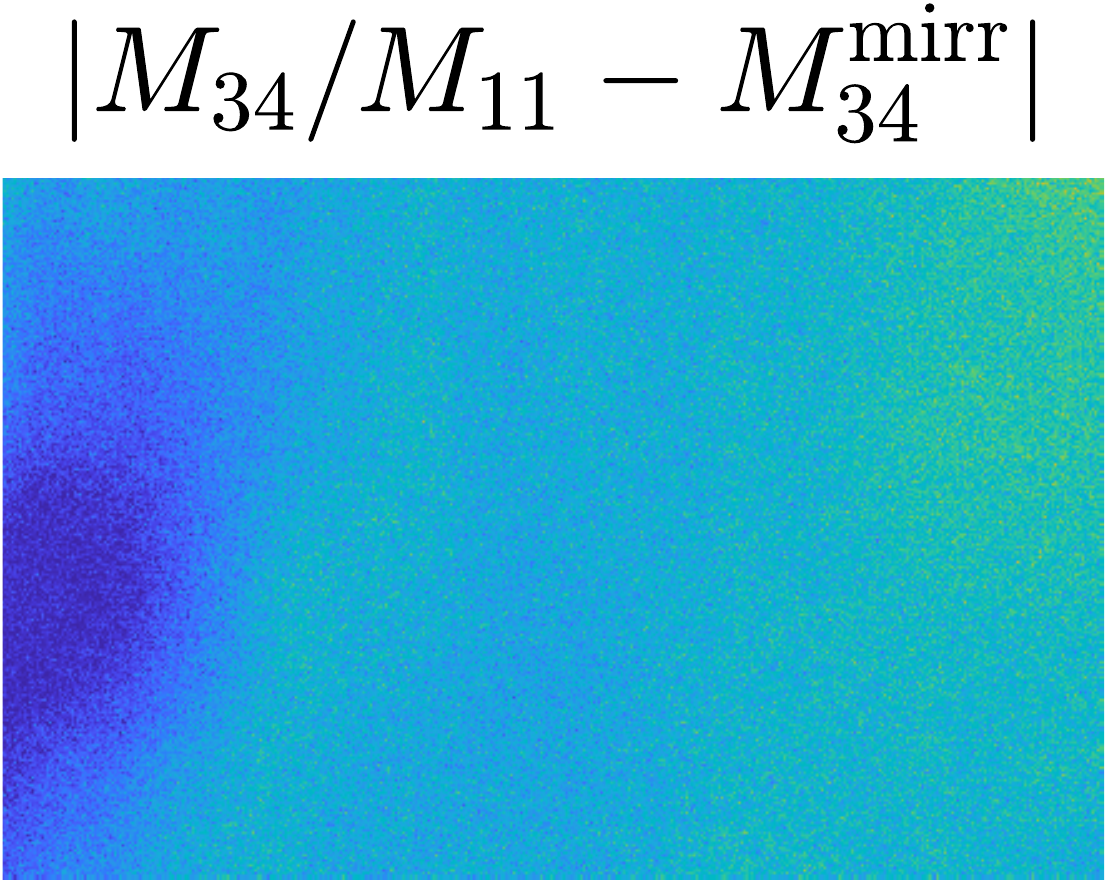} \\
\includegraphics[width=0.07\linewidth]{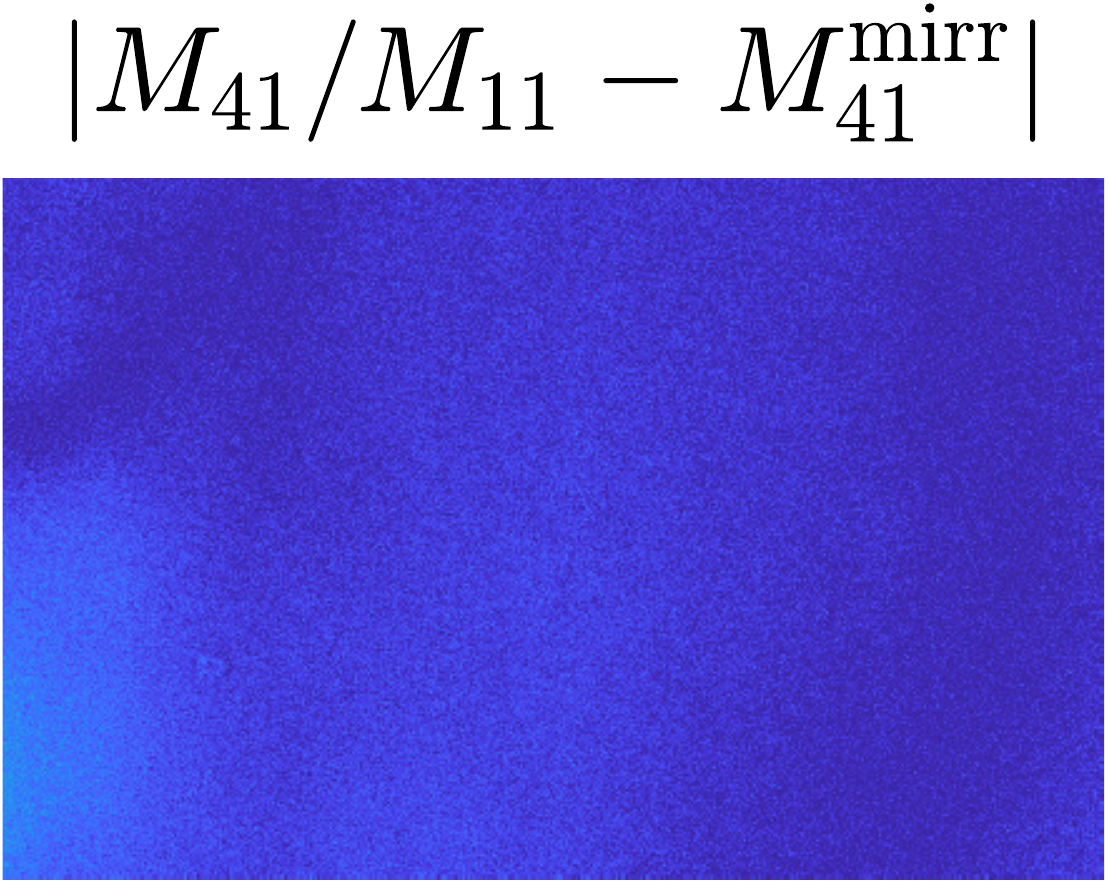} &
\includegraphics[width=0.07\linewidth]{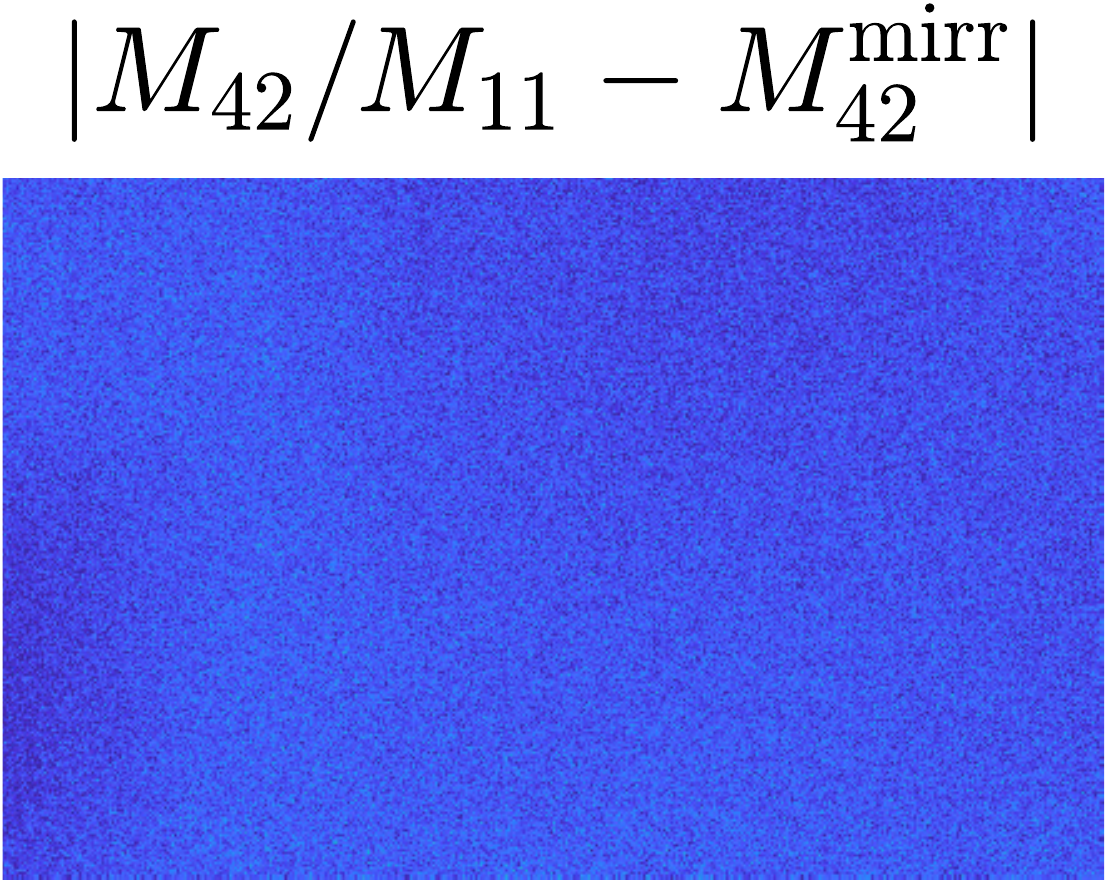} &
\includegraphics[width=0.07\linewidth]{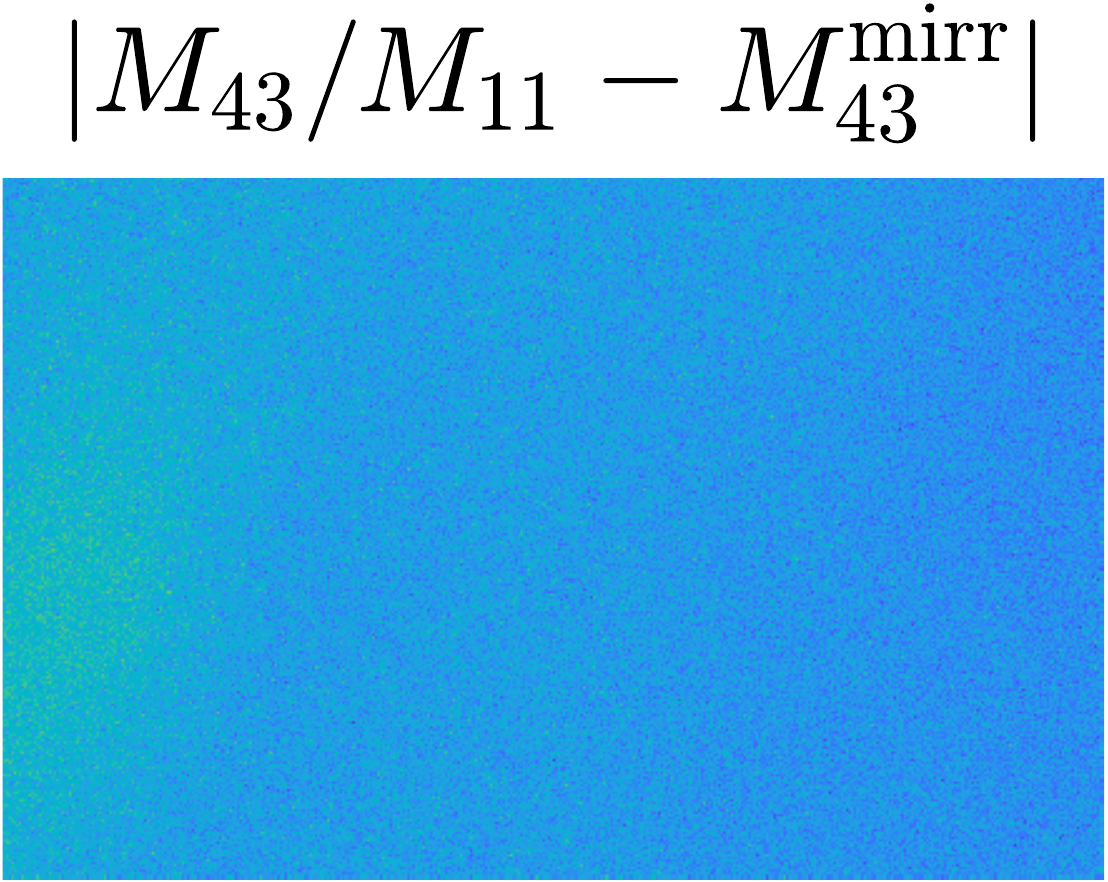} &
\includegraphics[width=0.07\linewidth]{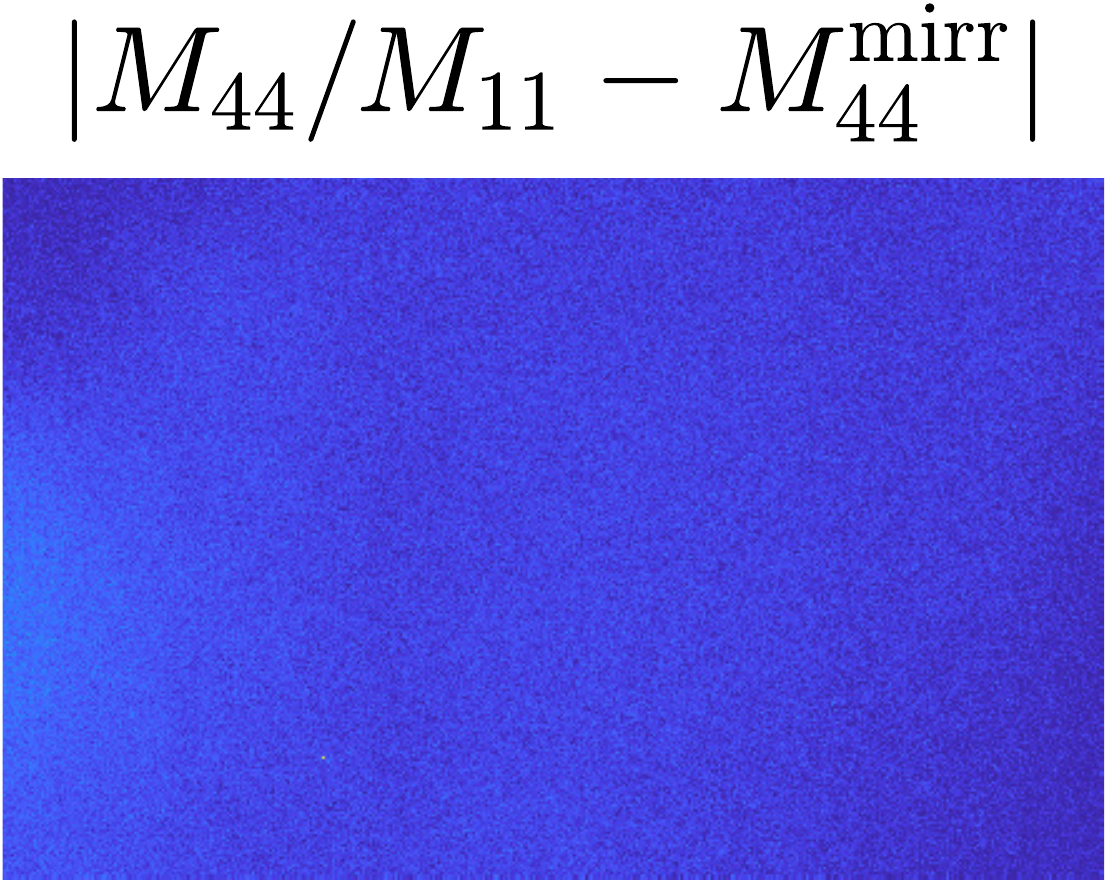} \\
\end{tabular}
}&
\multicolumn{2}{c}{
\begin{tabular}{cccc}
\includegraphics[width=0.07\linewidth]{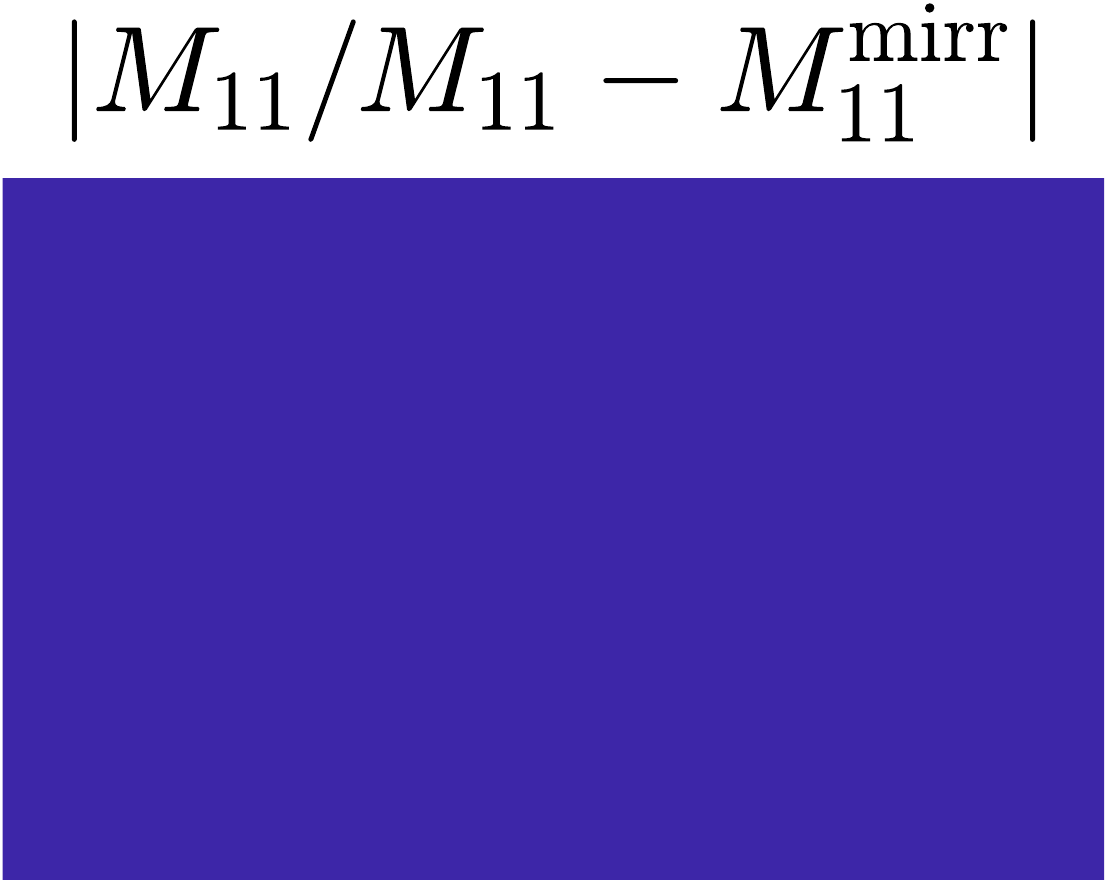} &
\includegraphics[width=0.07\linewidth]{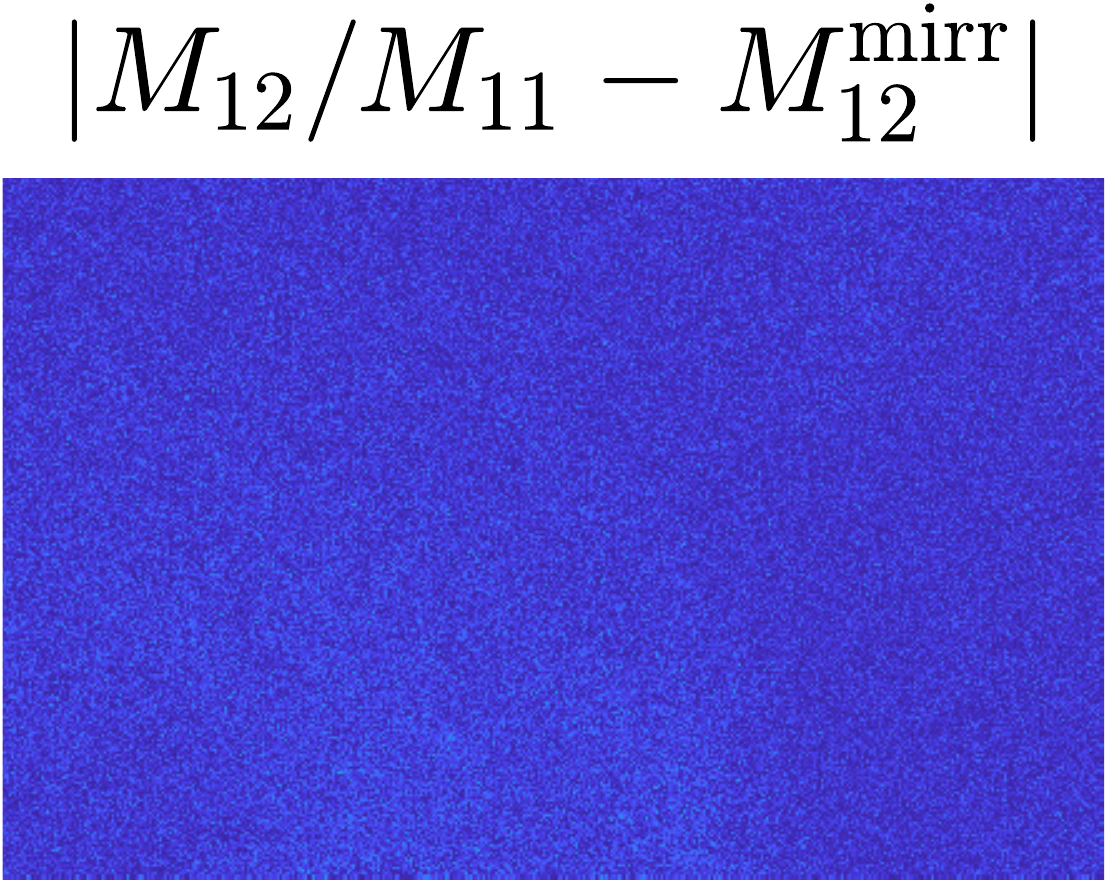} &
\includegraphics[width=0.07\linewidth]{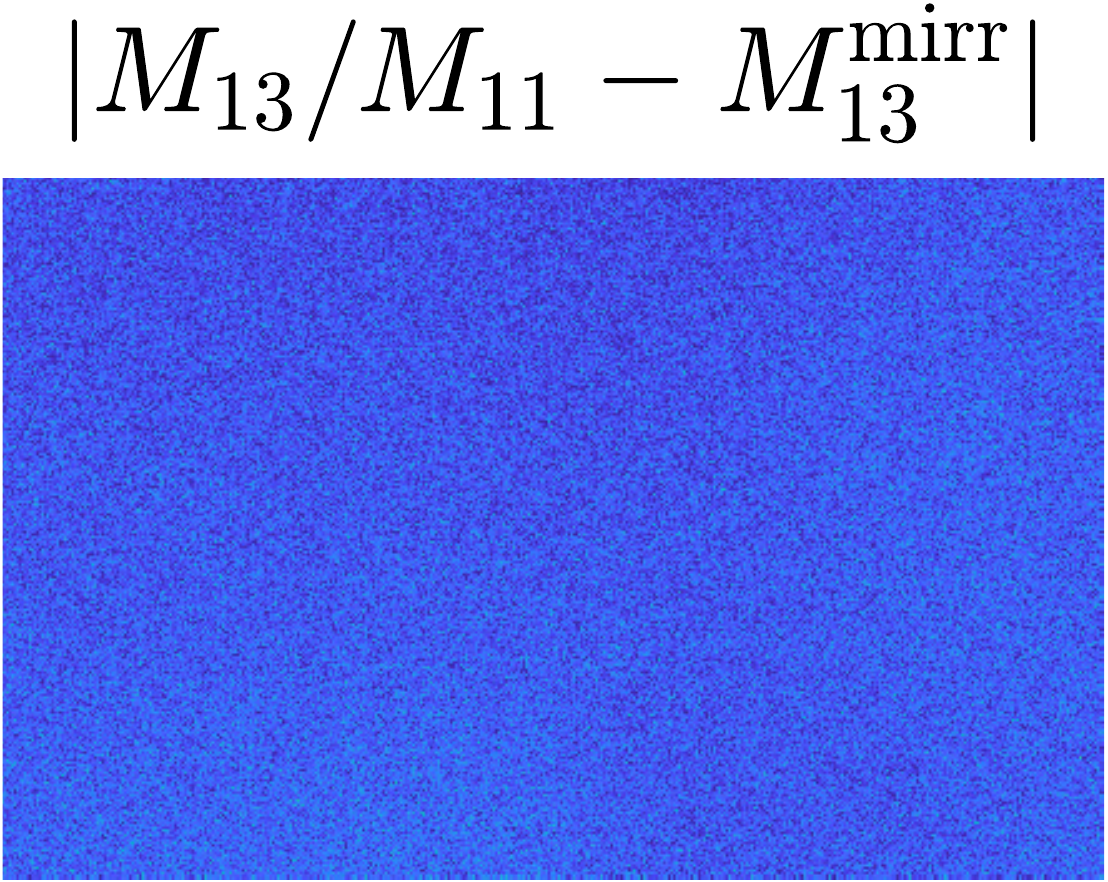} &
\includegraphics[width=0.07\linewidth]{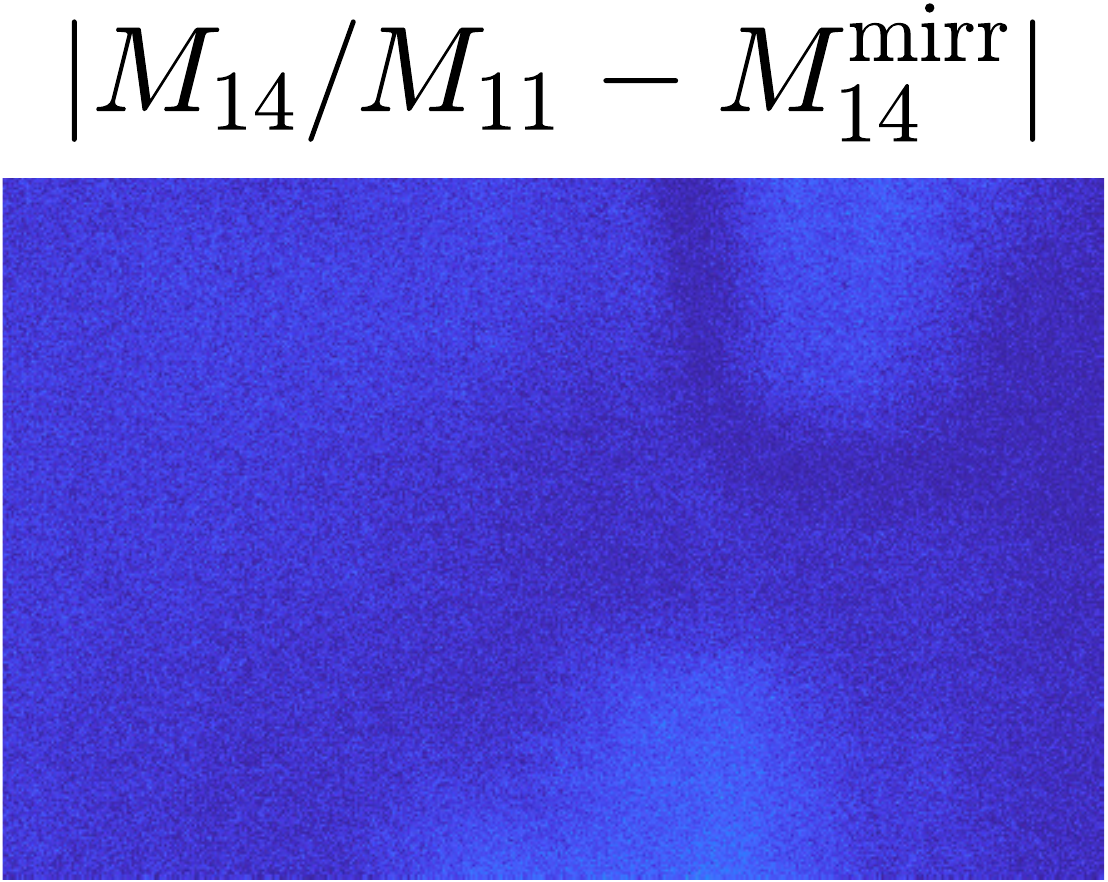} \\
\includegraphics[width=0.07\linewidth]{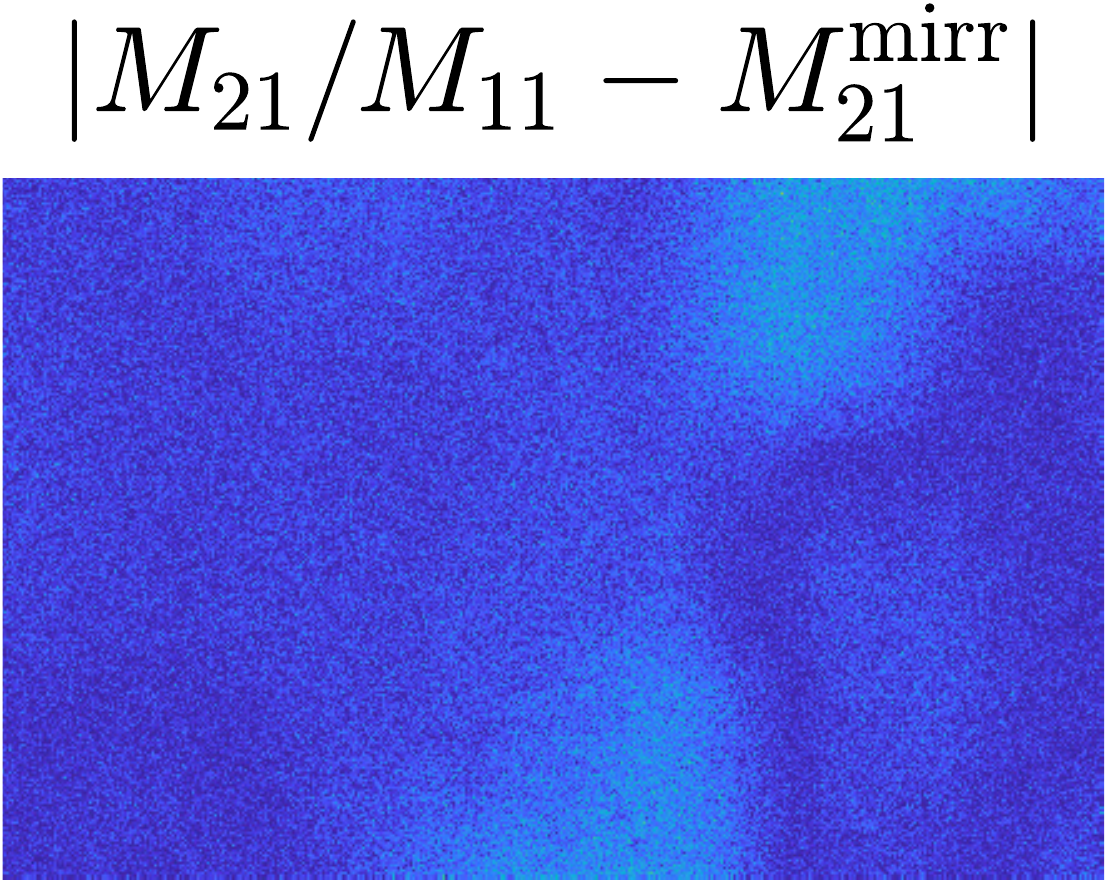} &
\includegraphics[width=0.07\linewidth]{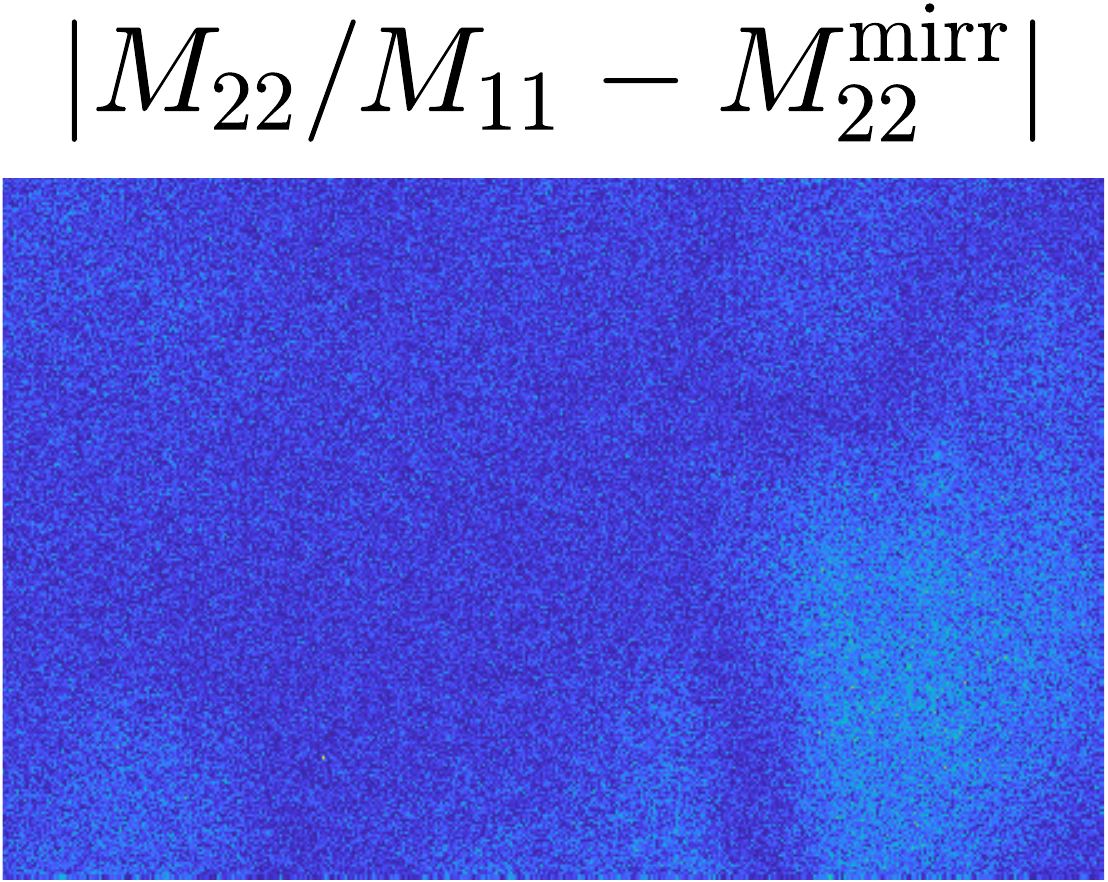} &
\includegraphics[width=0.07\linewidth]{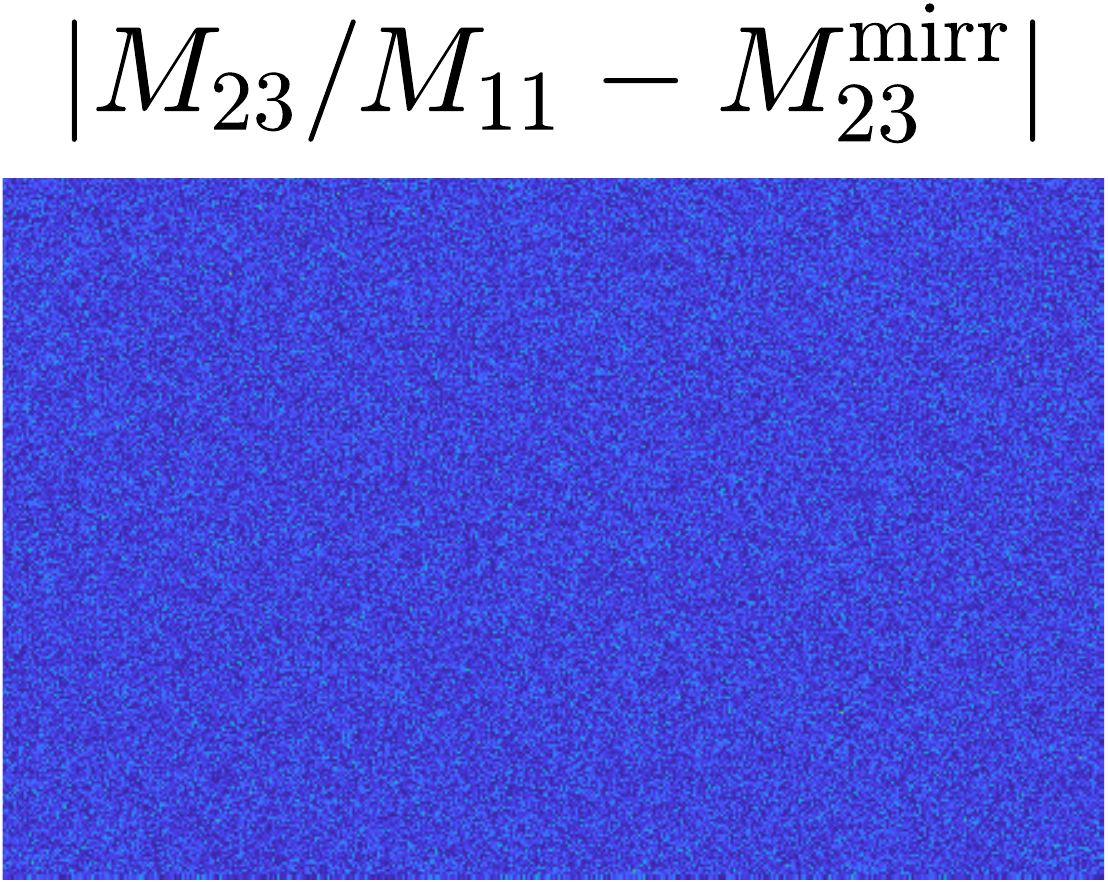} &
\includegraphics[width=0.07\linewidth]{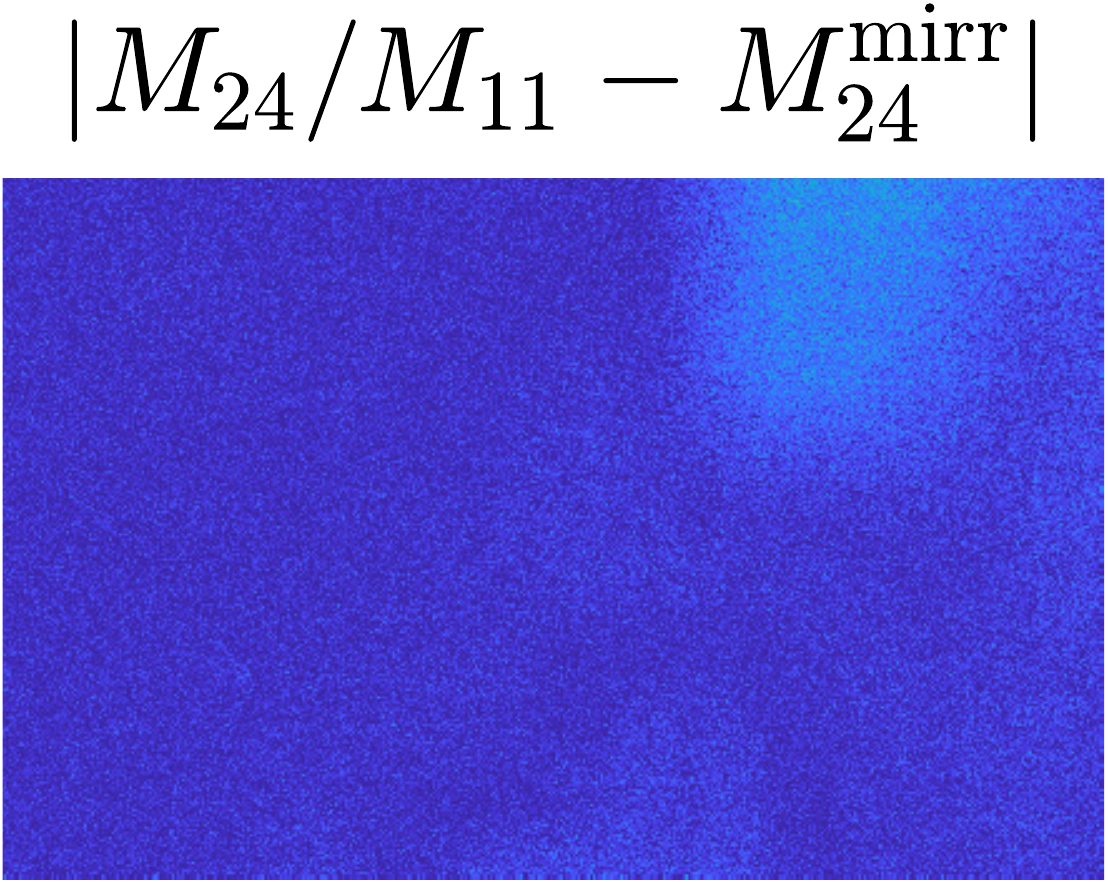} \\
\includegraphics[width=0.07\linewidth]{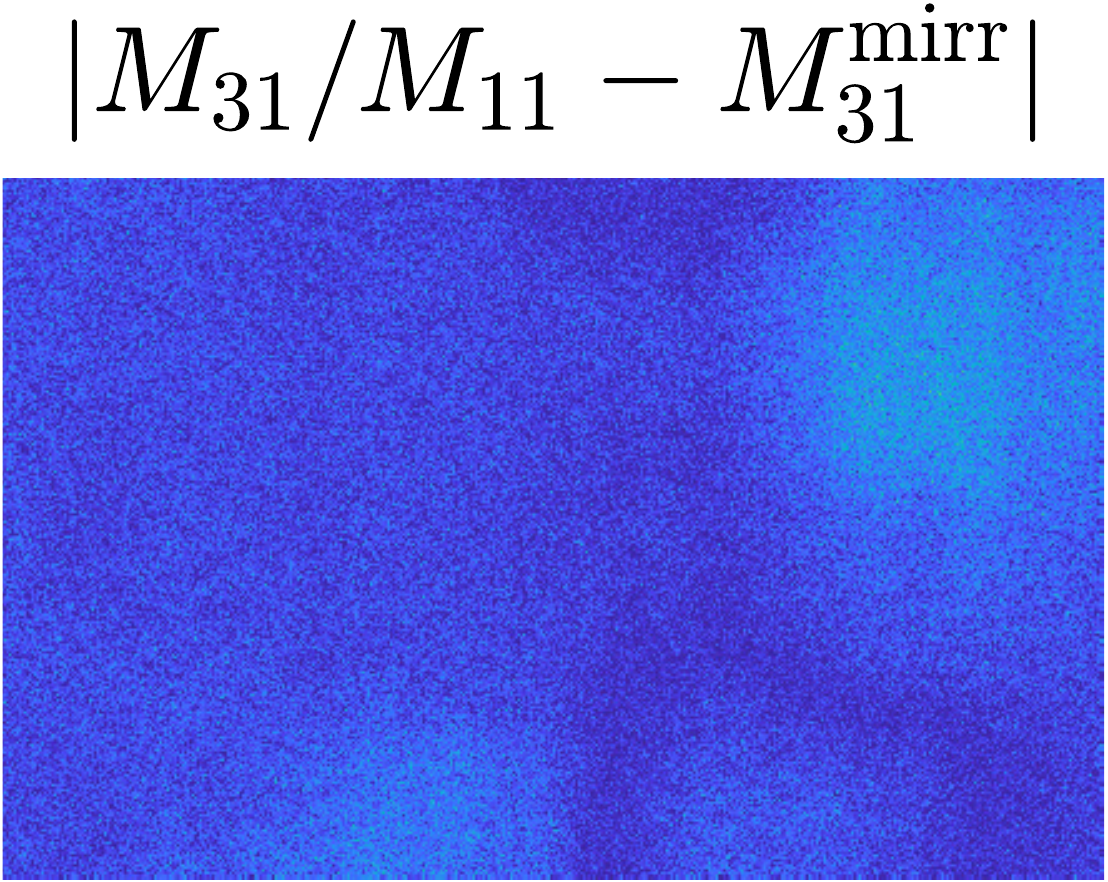} &
\includegraphics[width=0.07\linewidth]{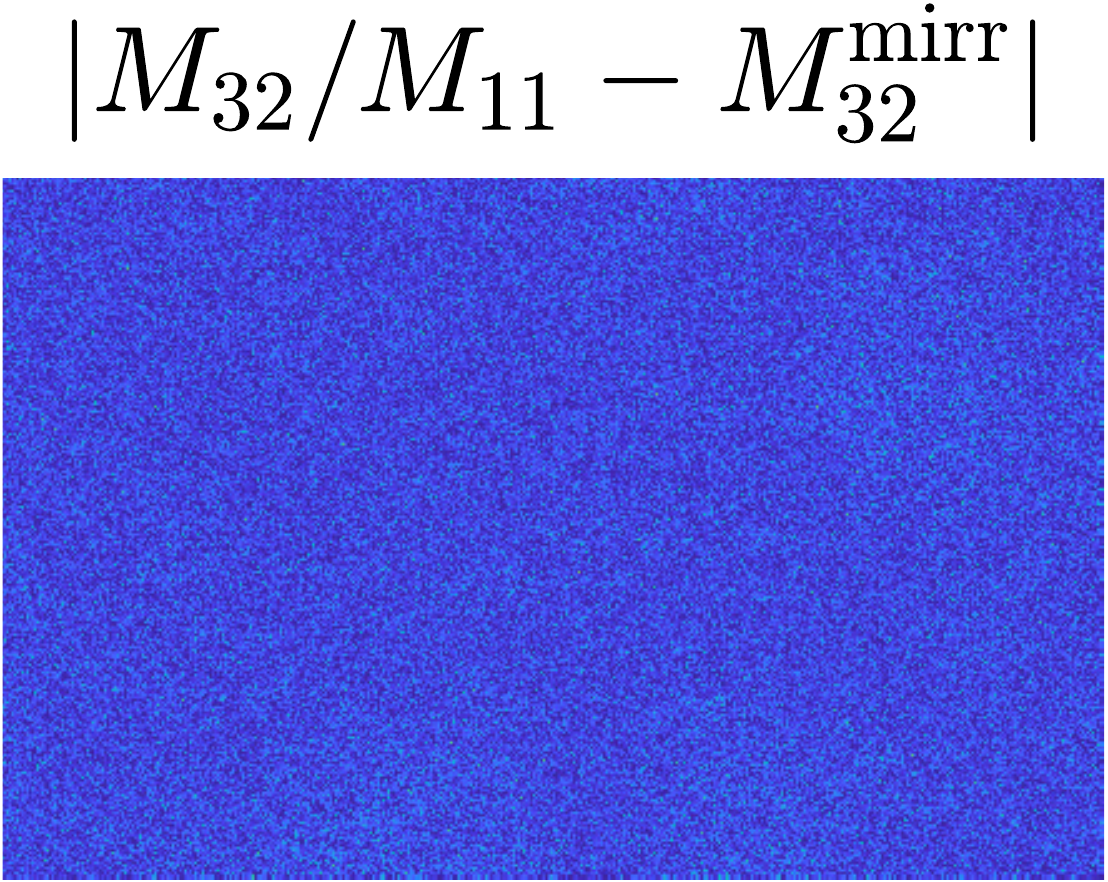} &
\includegraphics[width=0.07\linewidth]{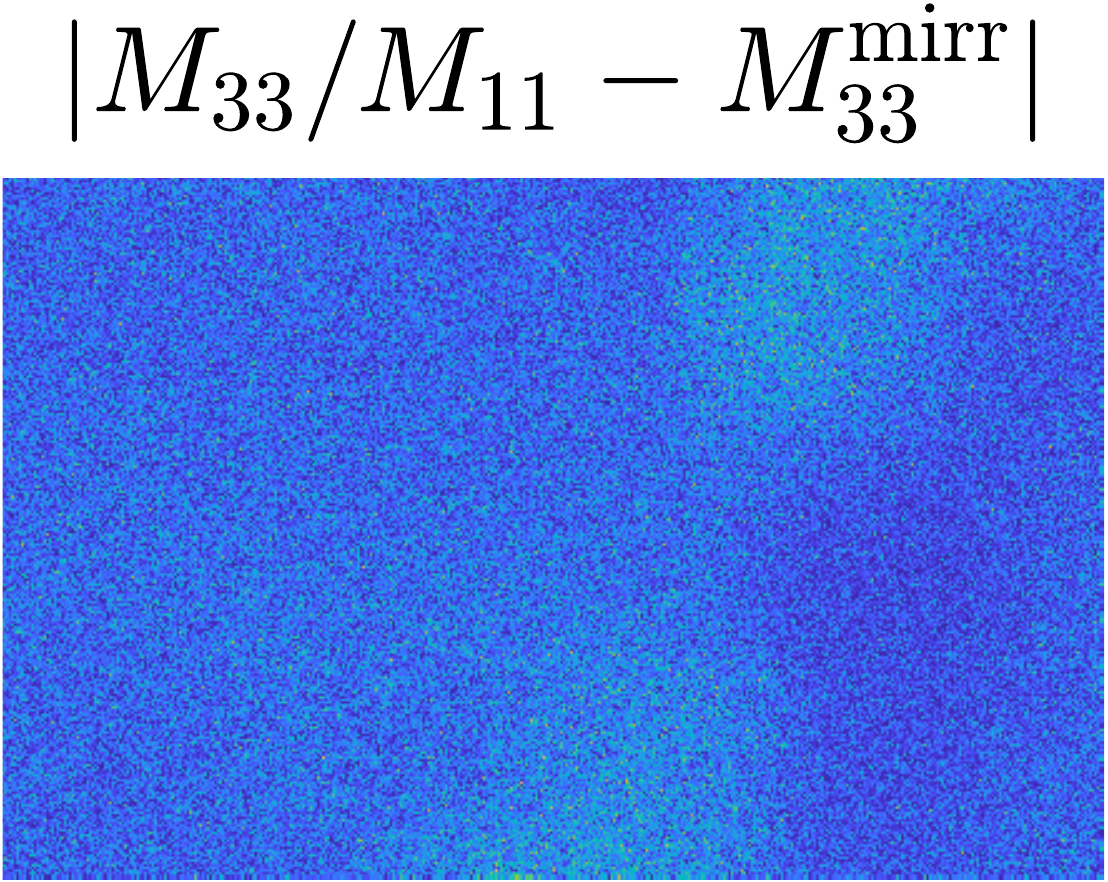} &
\includegraphics[width=0.07\linewidth]{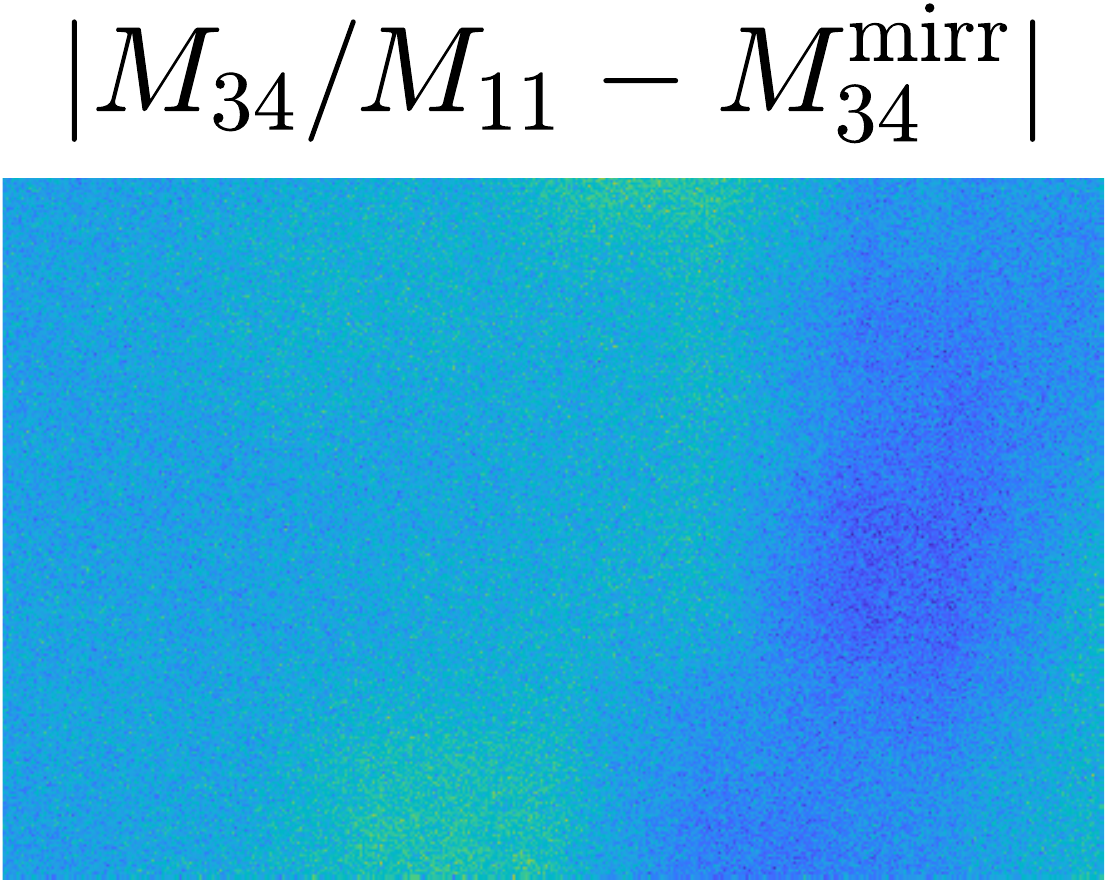} \\
\includegraphics[width=0.07\linewidth]{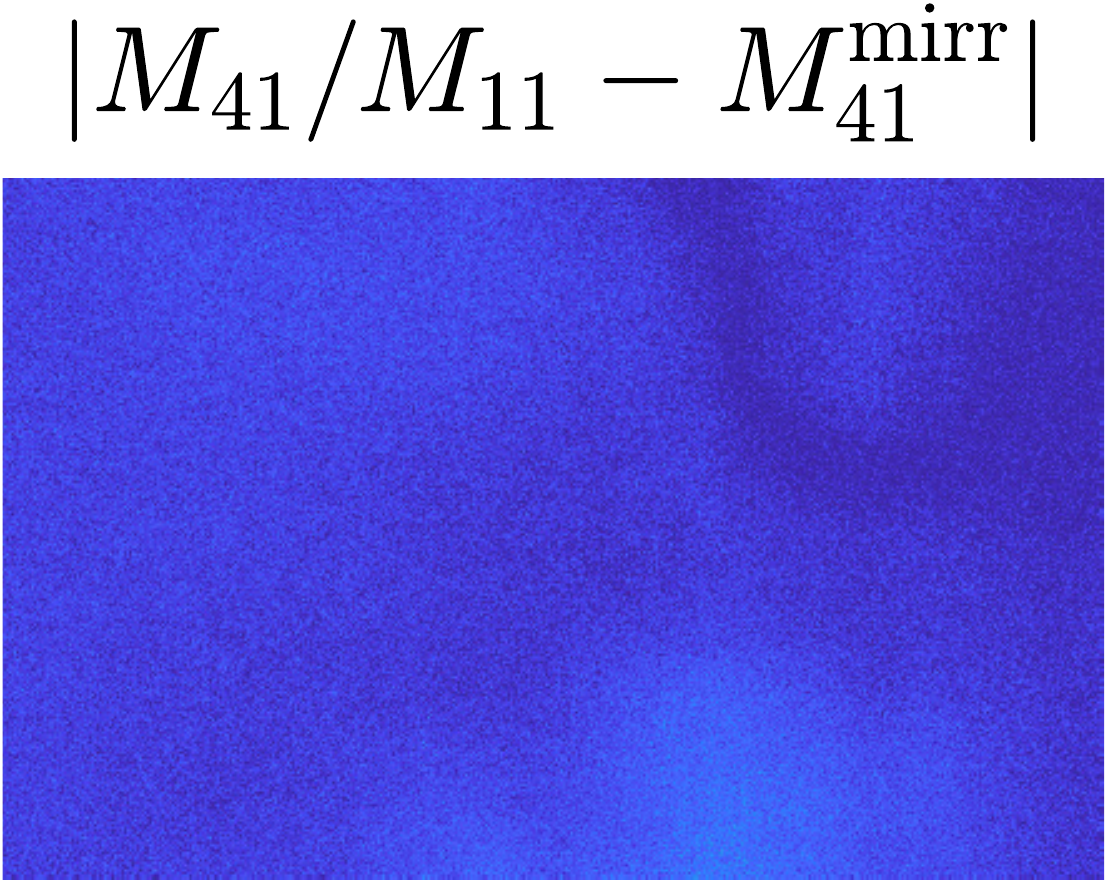} &
\includegraphics[width=0.07\linewidth]{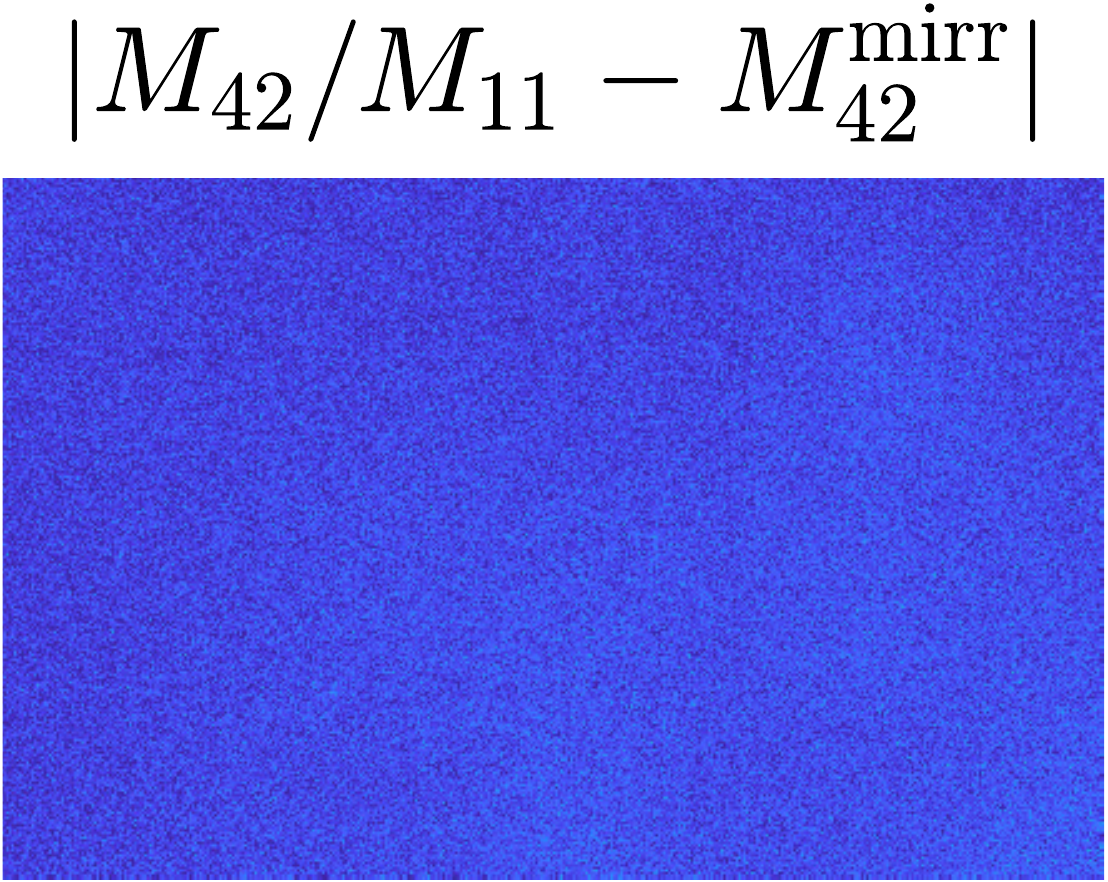} &
\includegraphics[width=0.07\linewidth]{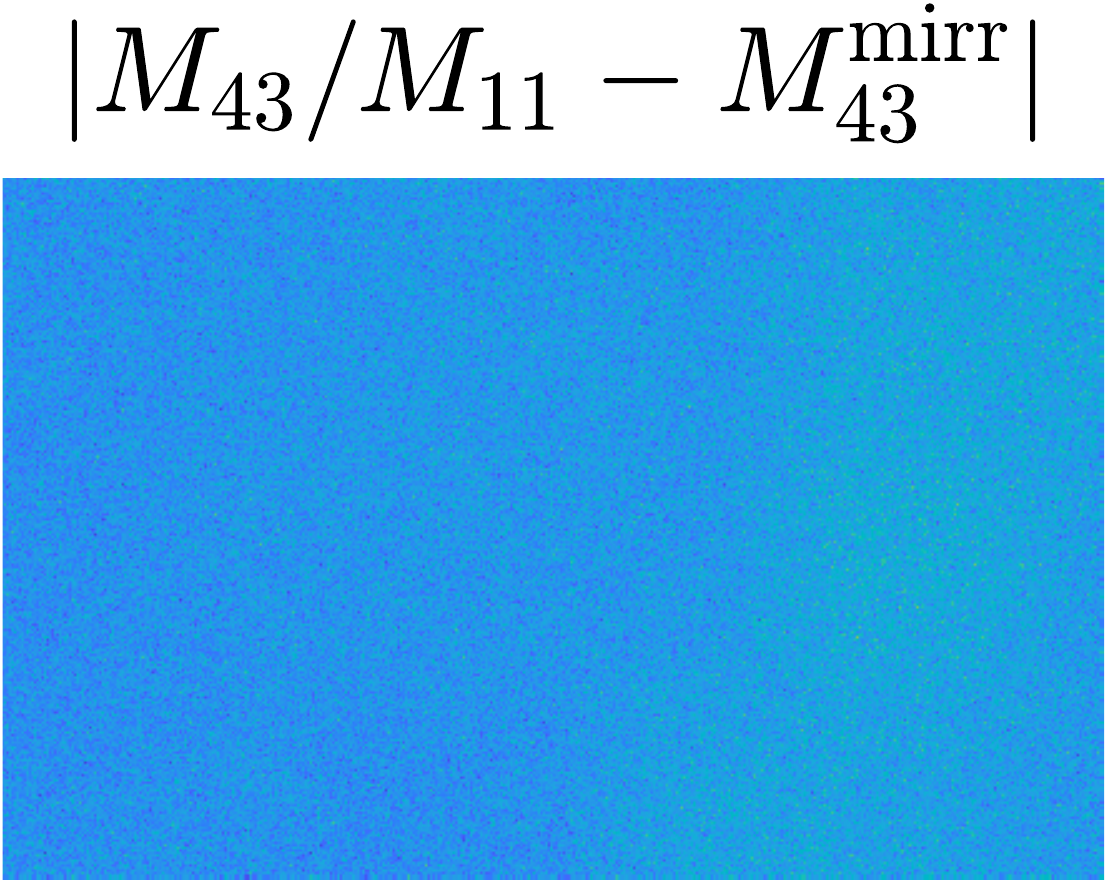} &
\includegraphics[width=0.07\linewidth]{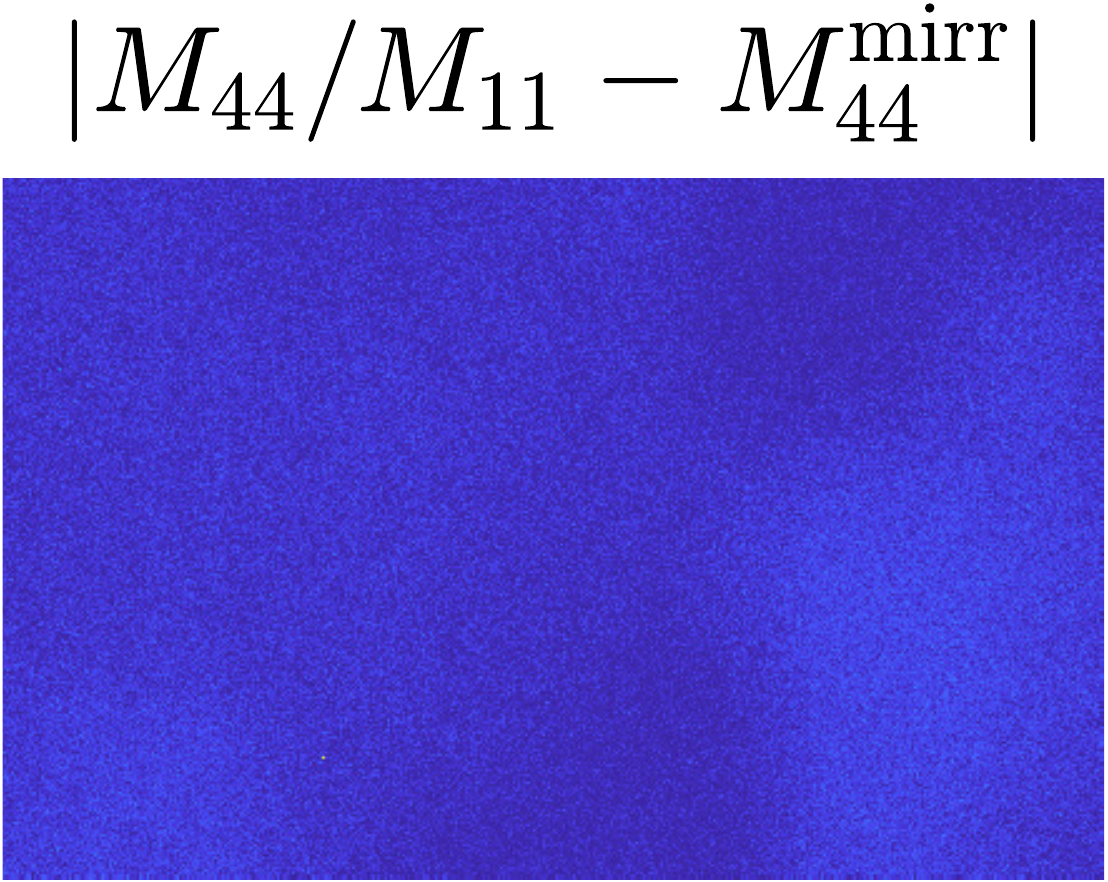} \\
\end{tabular}
}
\end{tabular}
\end{table*}

To measure real-world Mueller matrices, the polarizers are aligned ($\theta^A = \theta^A_0$ and $\theta^G = \theta^G_0$) and the two retarders are controlled through their azimuth $\alpha^G := \alpha^G - \alpha^G_0$ and $\alpha^A := \alpha^A - \alpha^A_0$. The first row in \eqref{eq:Stokes} then turns into the following linear equation in the $16$ unknown coefficients of matrix $M$:
\begin{equation}
{\tiny
I(\alpha^A,\alpha^G) = \underbrace{\begin{bmatrix}
1 \\
1 - \left(1- \cos\delta^A_0 \right) \sin^2 2\alpha^A  \\
\left(1 - \cos\delta^A_0 \right) \cos 2 \alpha^A \sin 2 \alpha^A  \\
\sin\delta^A_0 \sin 2 \alpha^A
\end{bmatrix}^\top}_{:= A(\alpha^A)}
\,
M
\,
\underbrace{
\begin{bmatrix}
1 \\
1 - \left(1- \cos\delta^G_0 \right) \sin^2 2\alpha^G  \\
\left(1 - \cos\delta^G_0 \right) \cos 2 \alpha^G \sin 2 \alpha^G  \\
-\sin\delta^G_0 \sin 2 \alpha^G
\end{bmatrix}}_{:=G(\alpha^G)}.
}
\label{eq:Mueller_polarimetry}
\end{equation}

We acquire $m$ series of intensity measurements under varying PSA retarder angle $\alpha^A$. In each series, we acquire $n$ measurements under varying PSG retarder angle $\alpha^G$. The resulting $mn$ observations\footnote{We used $m=n=8$ angles equally spaced between $0^\circ$ and $157.5^\circ$. This yields conditioning numbers of $3.79$ and $3.99$ for the $64\times 4$ matrices $A$ and $G^\top$ in \eqref{eq:Mueller_polarimetry}. These values are \textit{exactly} the same as those associated with the theoretically optimal~\cite{Compain1999} set of $64$ angles obtained with synchronous variations of both azimuths at a 1:5 speed ratio.} then allow the system of $mn$ equations such as \eqref{eq:Mueller_polarimetry} to be solved in the least-squares sense.


We first calibrated the polarimeter in transmission mode and then estimated the Mueller matrix\footnote{Since estimations are up-to-scale, the estimated Mueller matrices are normalized a posteriori by their first components. Besides, when the medium is spatially homogeneous (\eg, the air or a mirror), we averaged all the intensity measurements over the image domain before calculating the Mueller matrix.} of the air at $540\ nm$ (the expected result is the matrix $M^\mathrm{air}$ equal to identity). Then we calibrated it again in reflection mode and estimated the Mueller matrix of a mirror at $540\  nm$ (the expected result is a diagonal matrix $M^{\mathrm{mirr}}$ with non-zero elements $\left[1,1,-1,-1 \right]^\top$). The results shown in Table~\ref{tab:air} show that the bundle-adjusted calibration method significantly reduces errors. The spatial uniformity of the error distribution further suggests that the remaining errors are mostly due to the accuracy of the detector. 

\begin{figure}[!ht]
\begin{tabular}{cccc}
\includegraphics[height=0.16\linewidth]{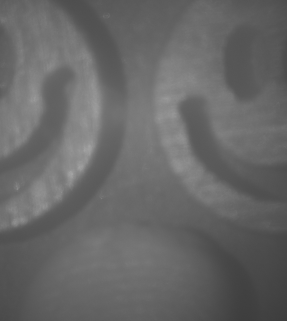} &
\includegraphics[height=0.18\linewidth]{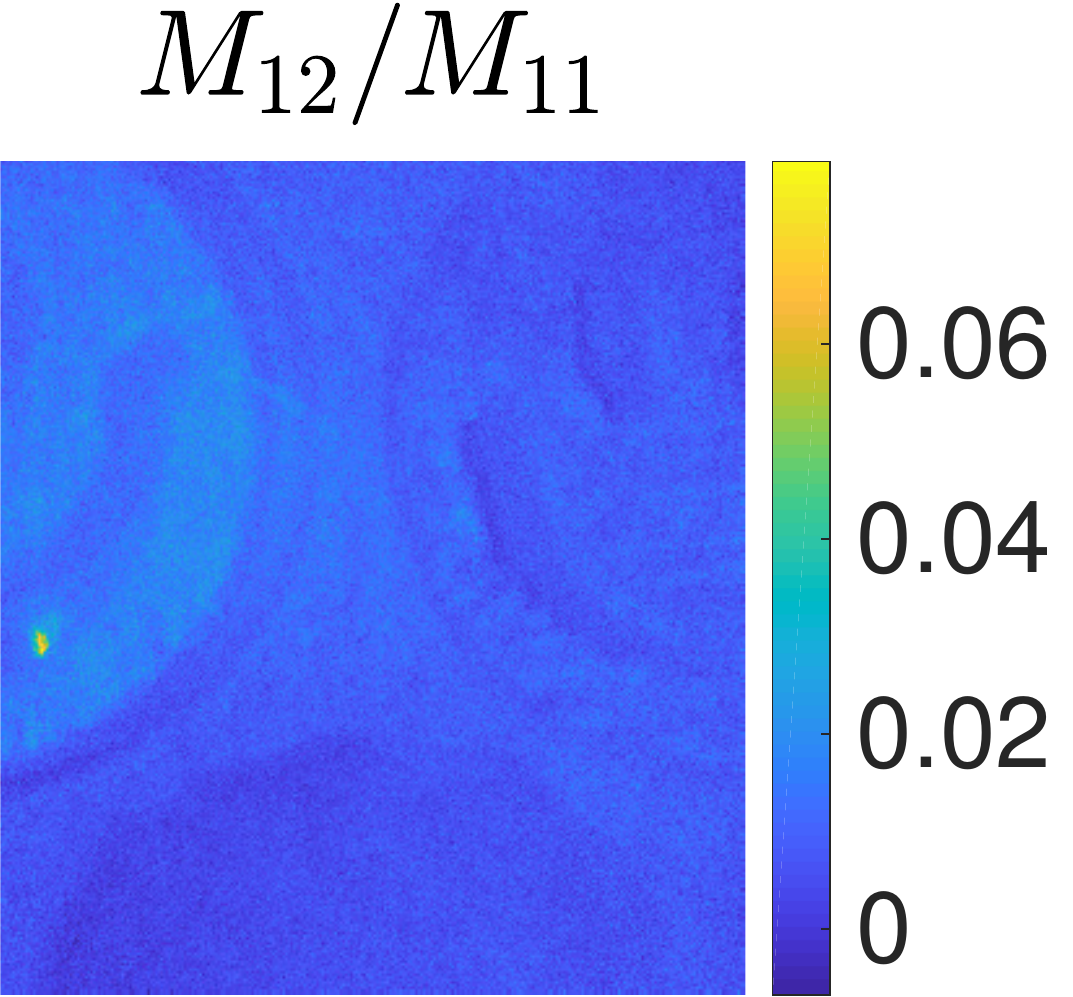} &
\includegraphics[height=0.18\linewidth]{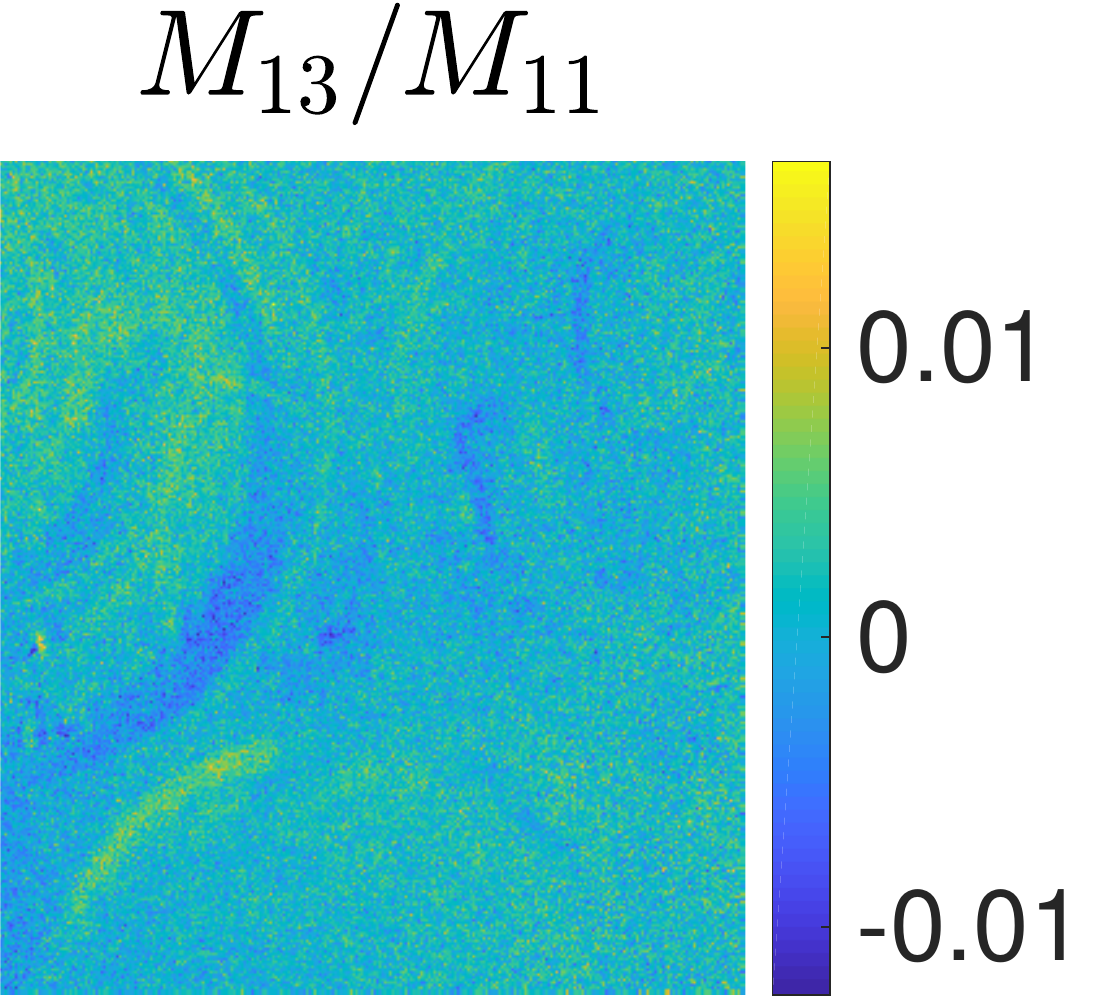} &
\includegraphics[height=0.18\linewidth]{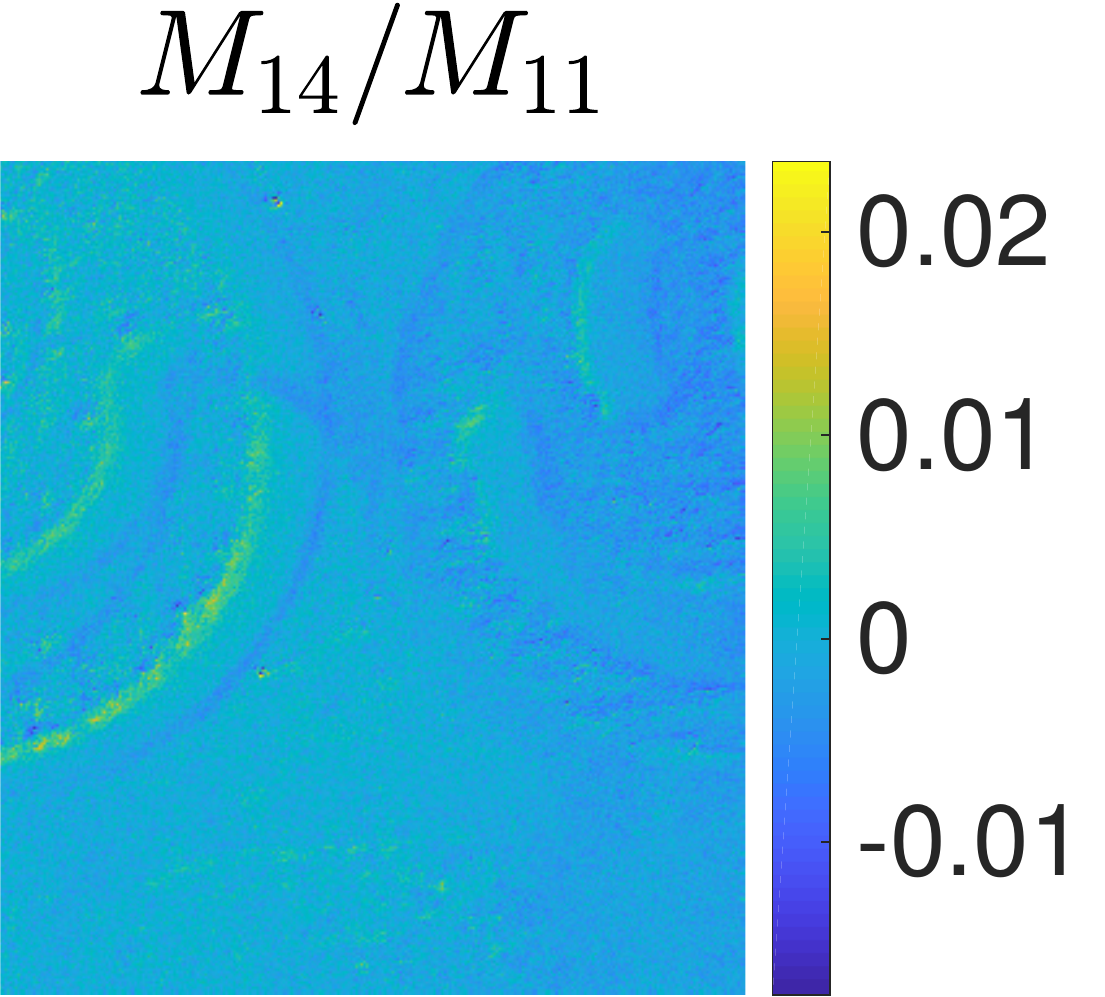} \\
\includegraphics[height=0.18\linewidth]{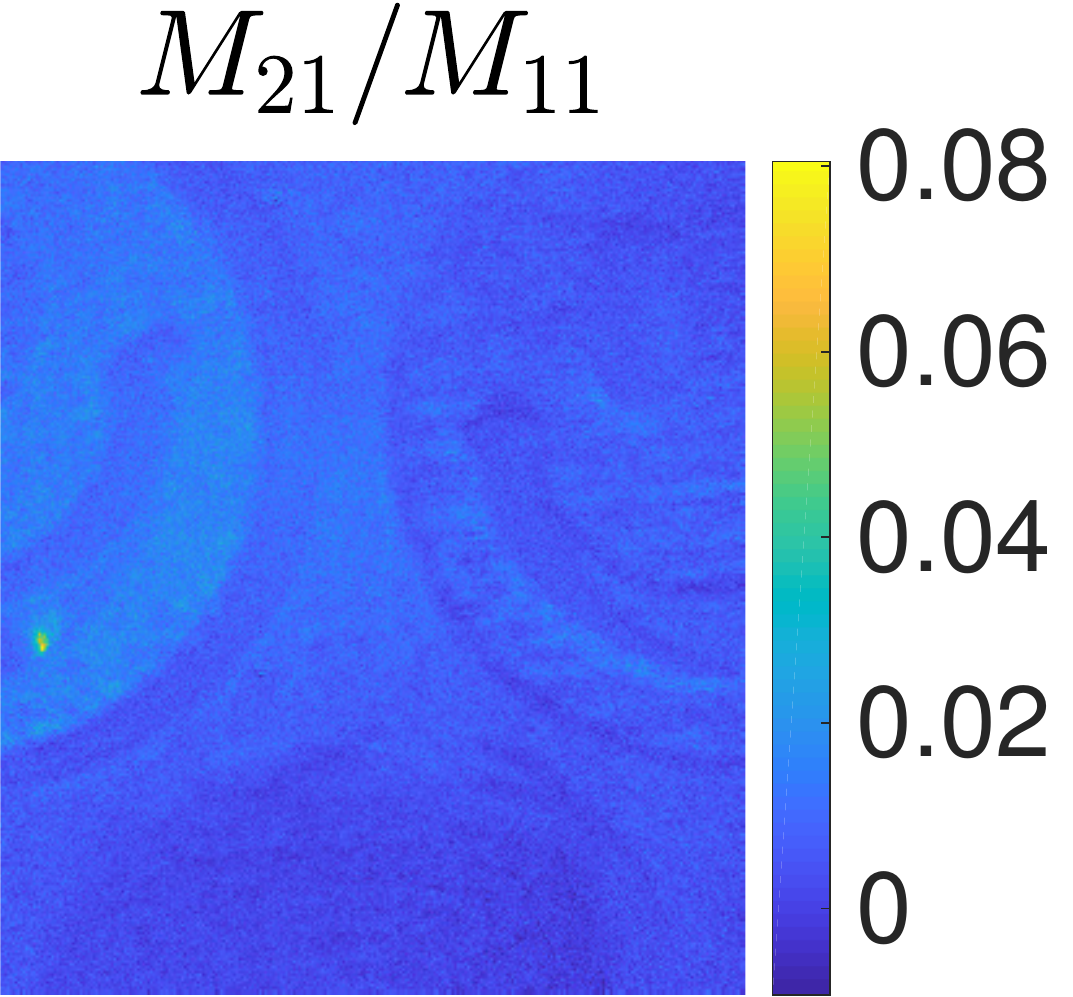} &
\includegraphics[height=0.18\linewidth]{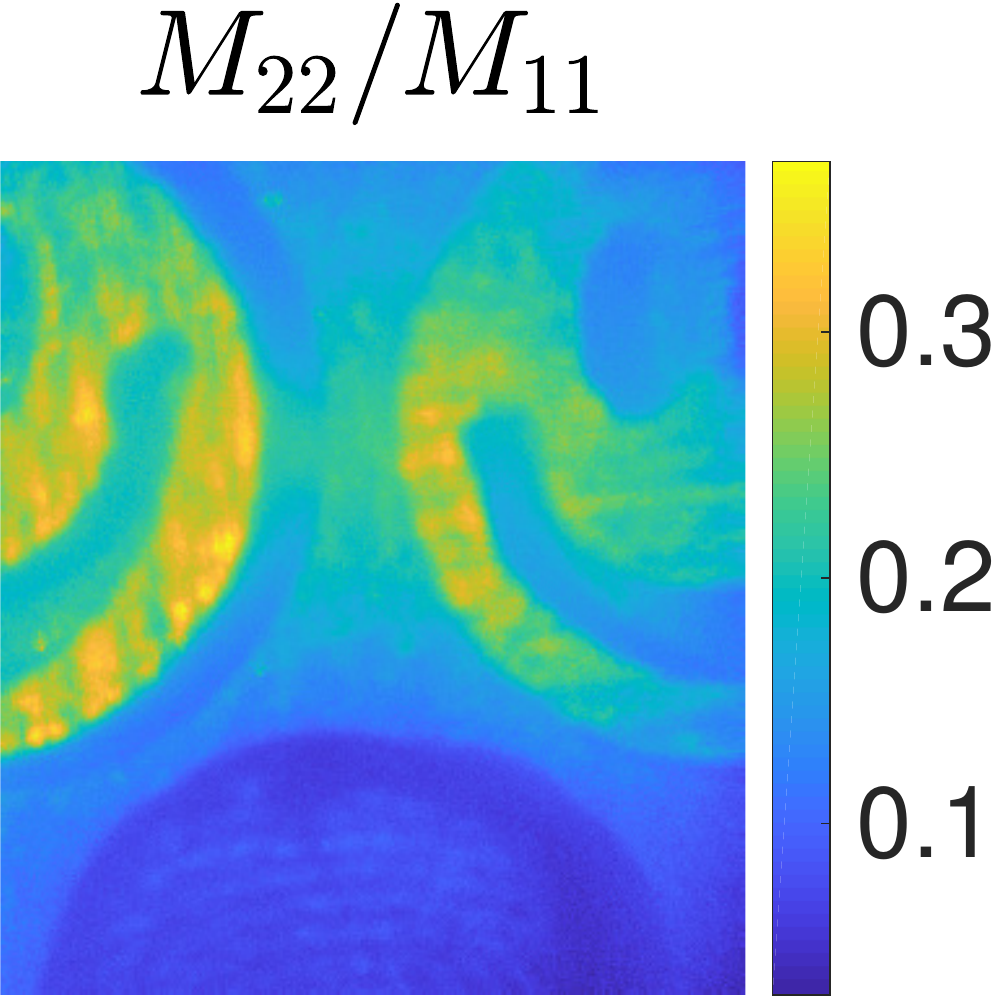} &
\includegraphics[height=0.18\linewidth]{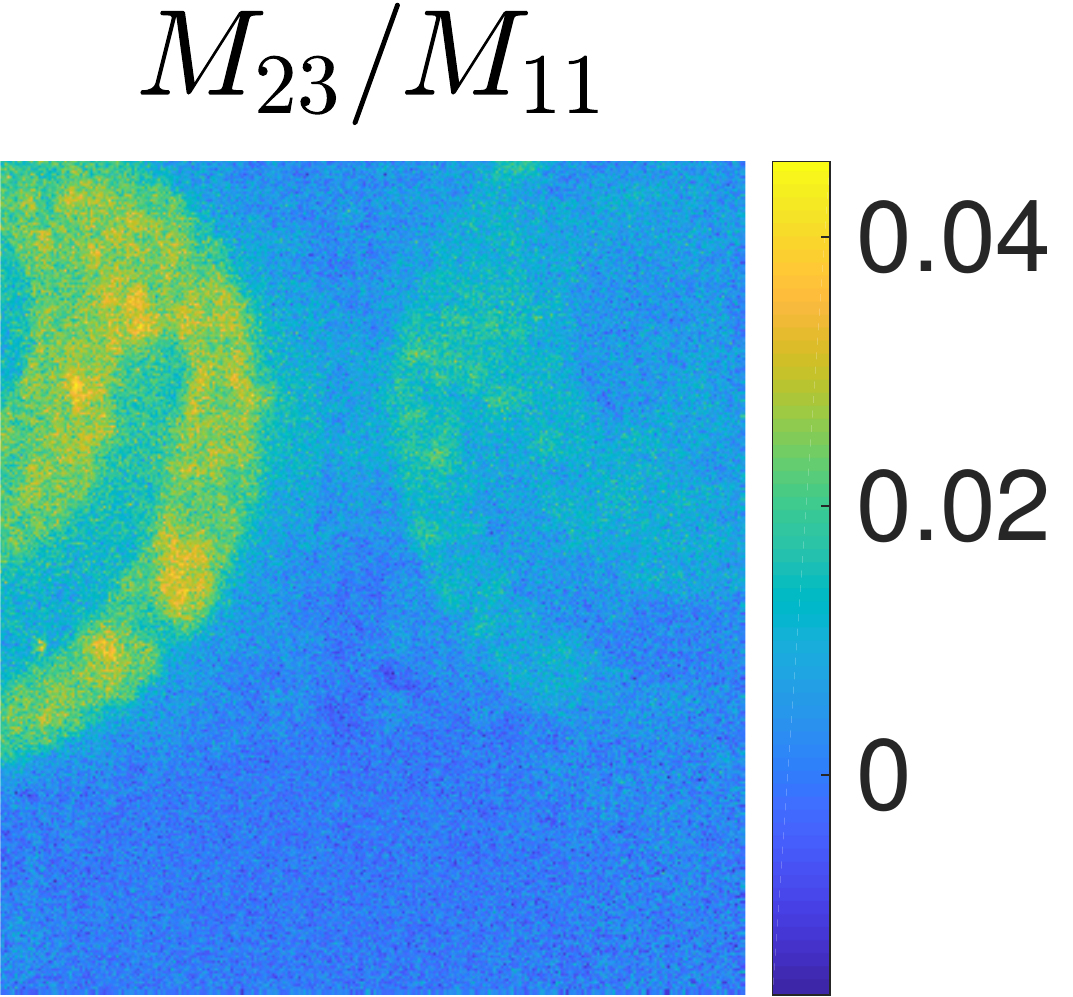} &
\includegraphics[height=0.18\linewidth]{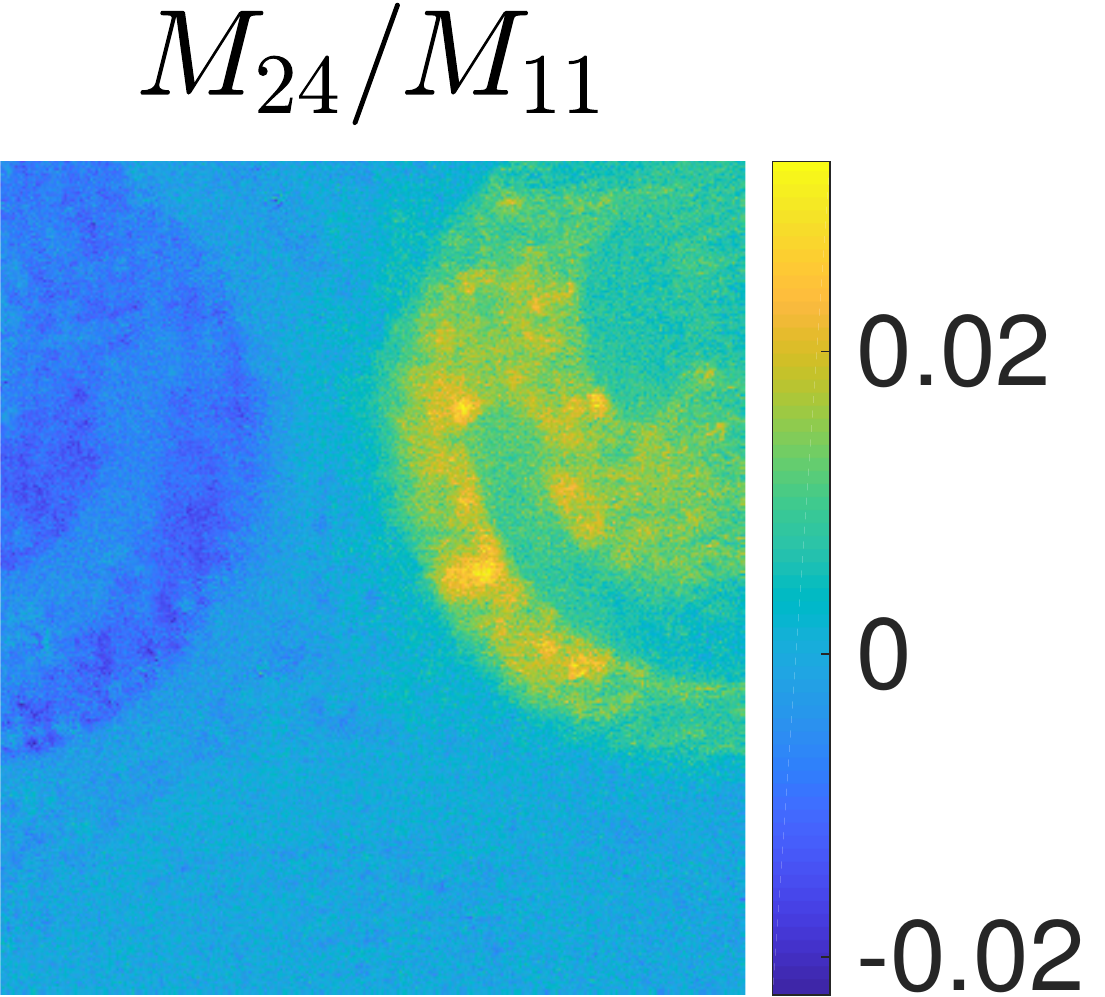} \\
\includegraphics[height=0.18\linewidth]{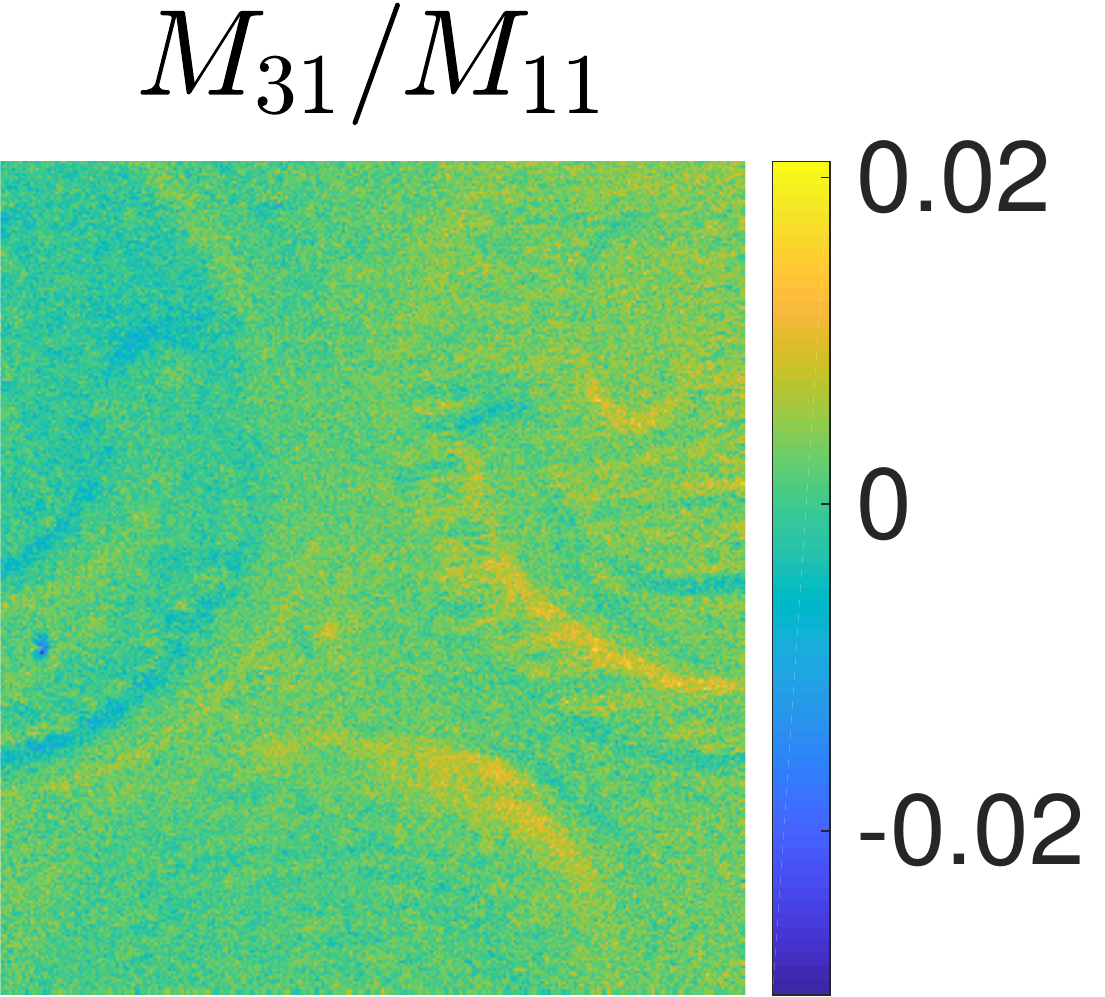} &
\includegraphics[height=0.18\linewidth]{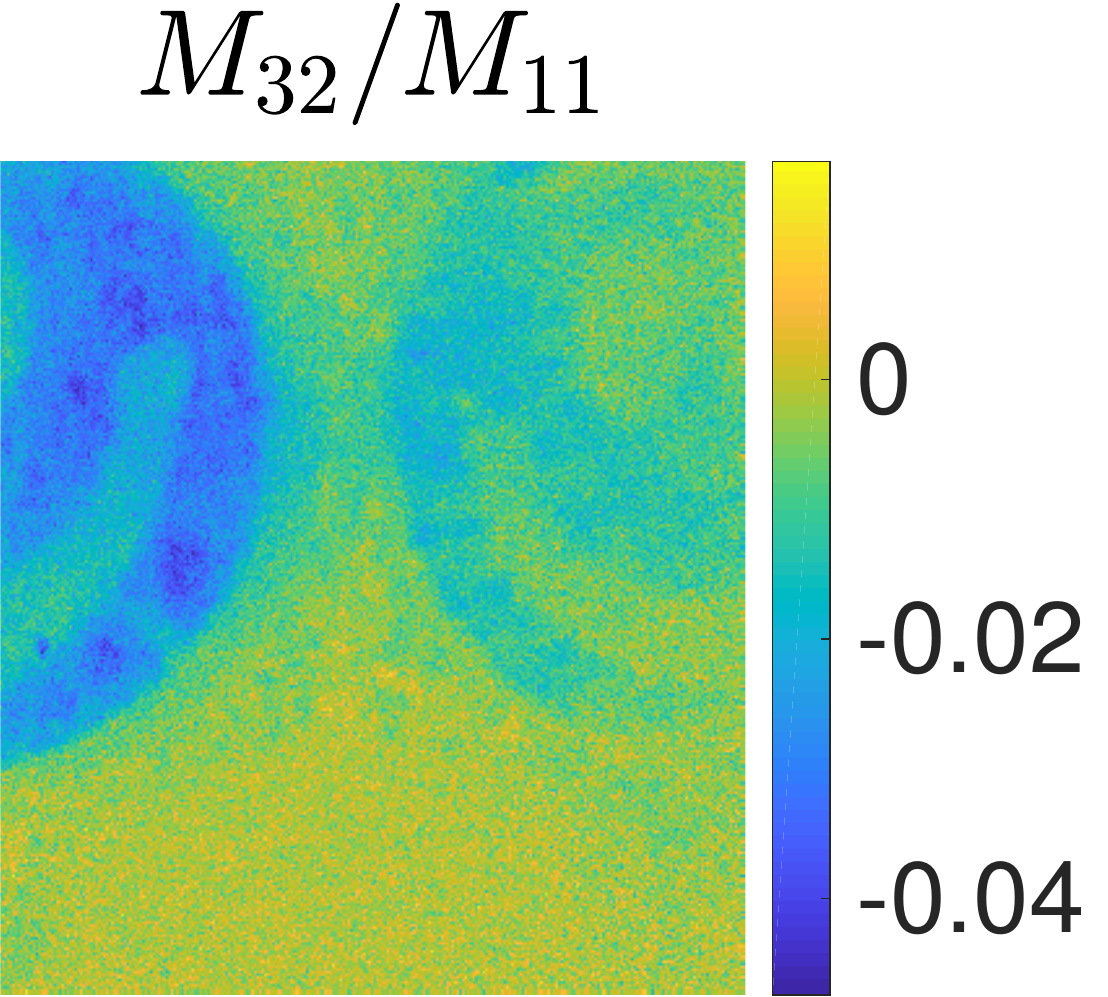} &
\includegraphics[height=0.18\linewidth]{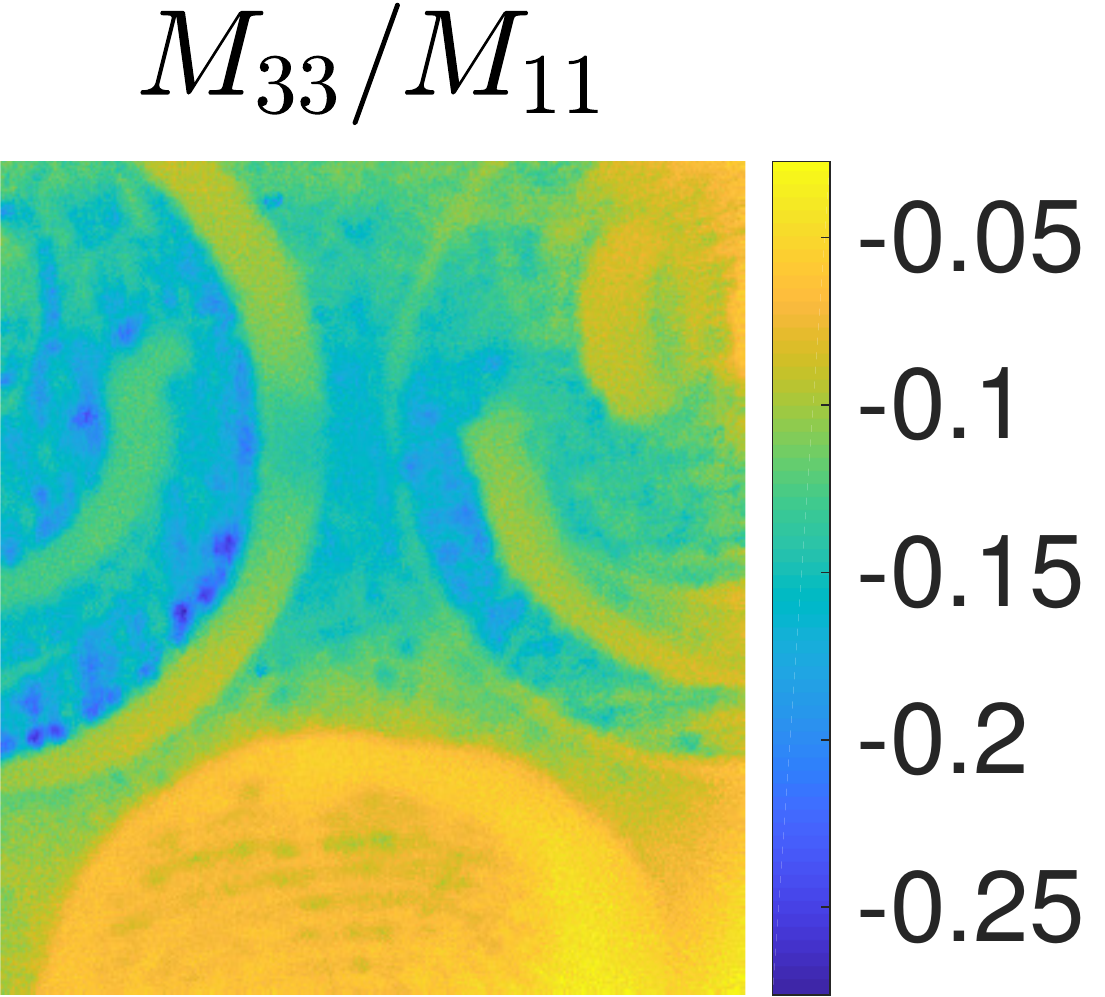} &
\includegraphics[height=0.18\linewidth]{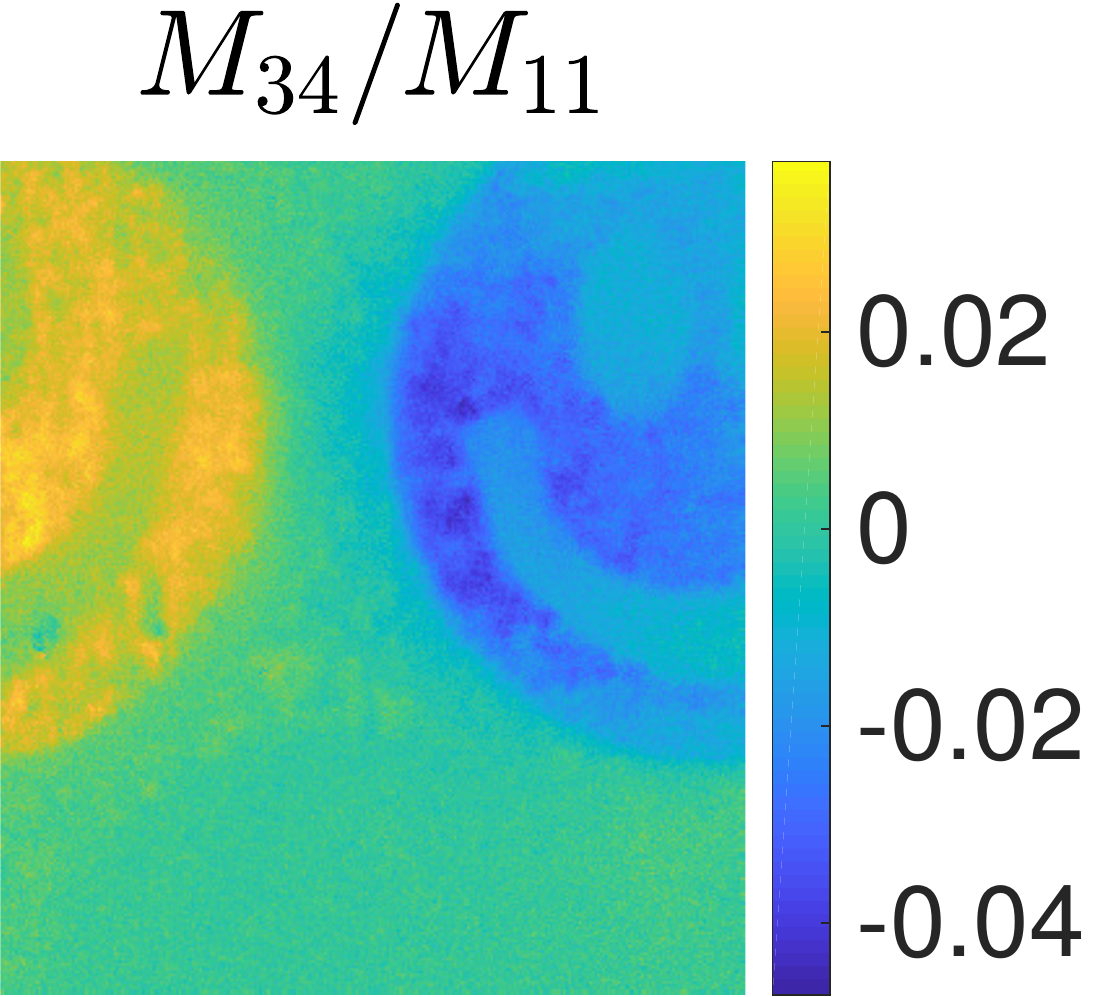} \\
\includegraphics[height=0.18\linewidth]{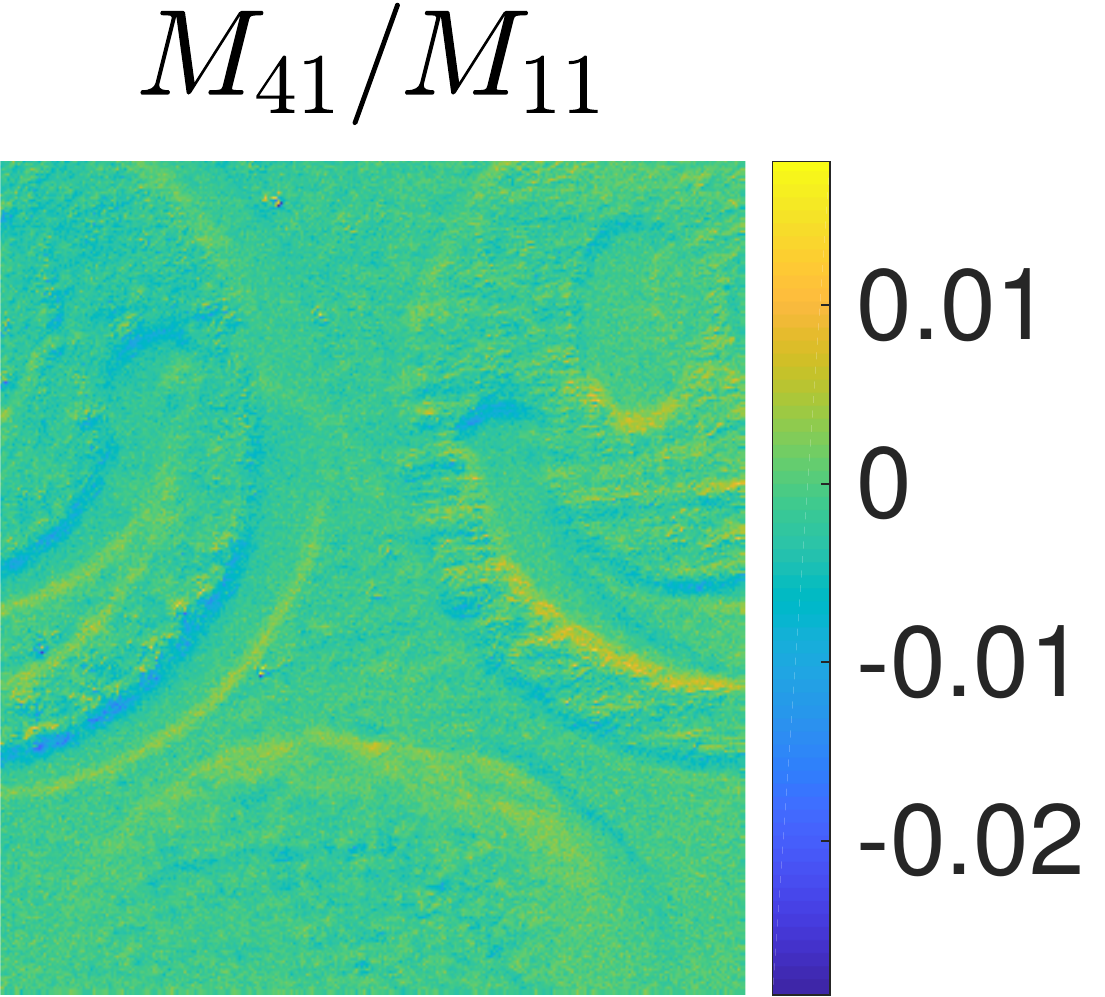} &
\includegraphics[height=0.18\linewidth]{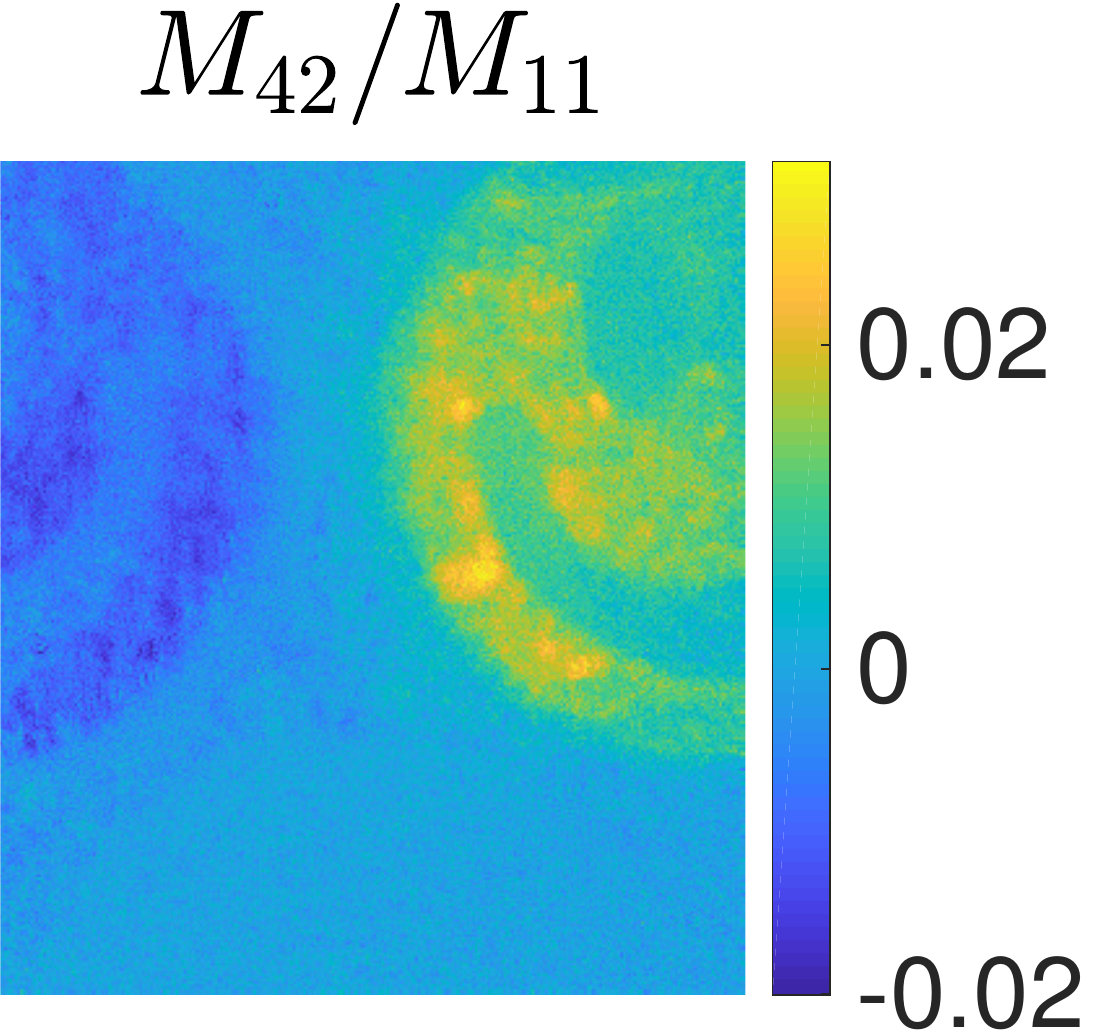} &
\includegraphics[height=0.18\linewidth]{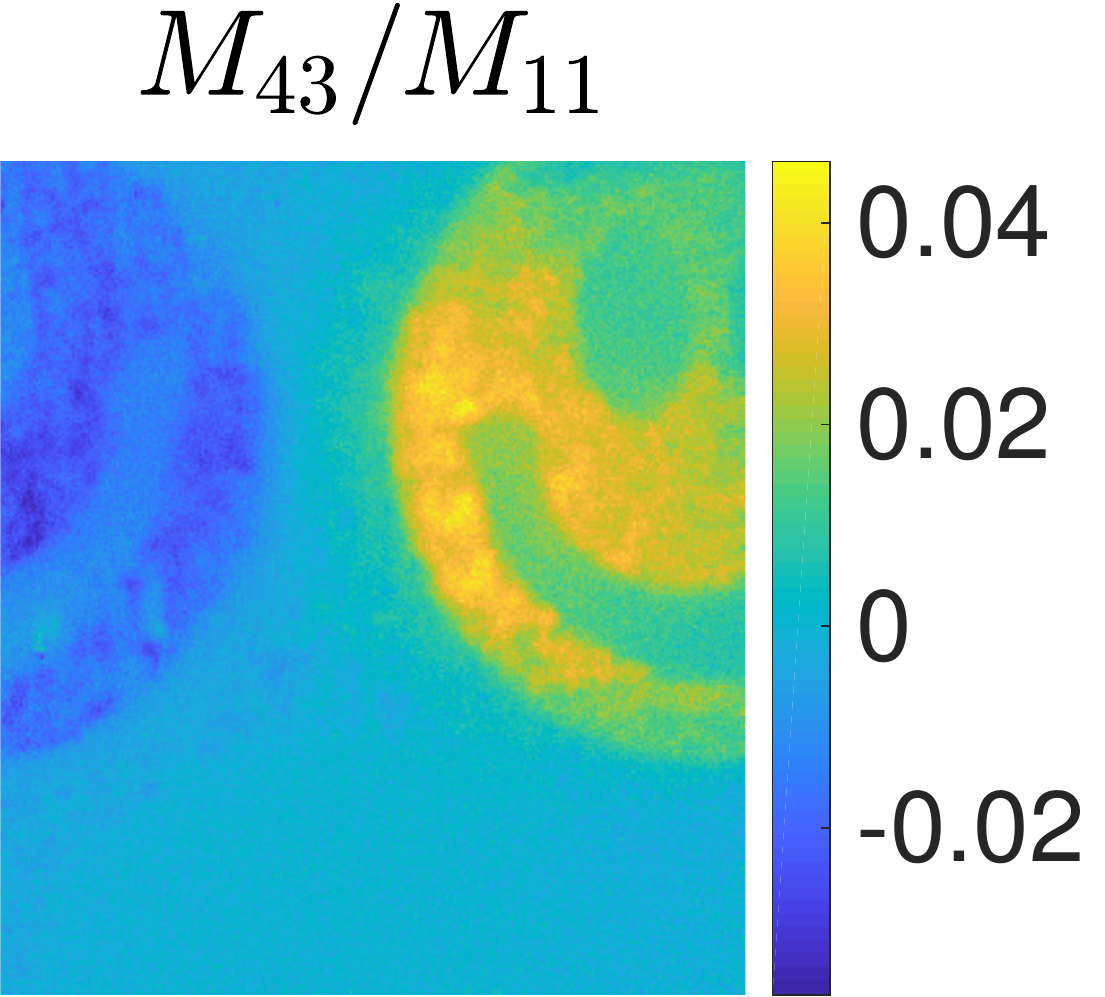} &
\includegraphics[height=0.18\linewidth]{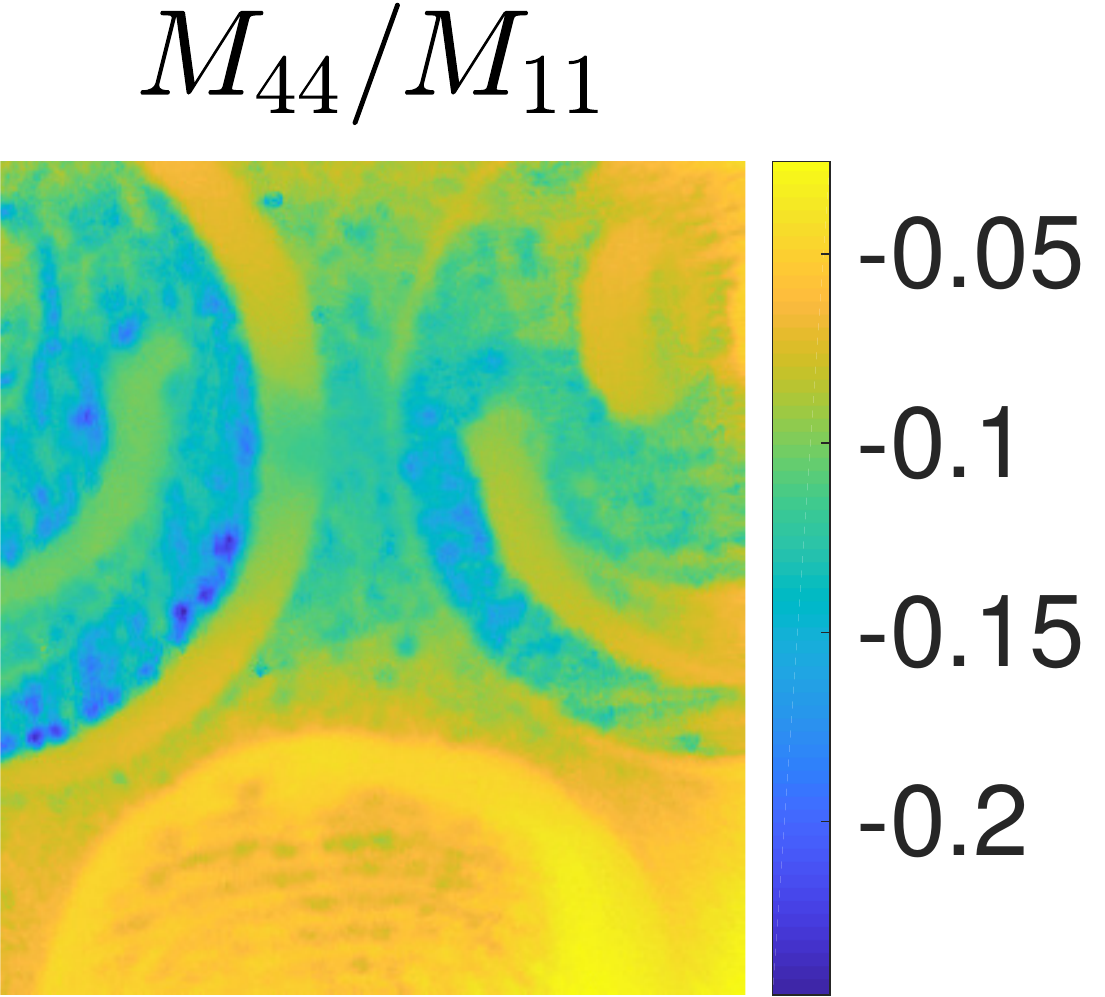} \\
\end{tabular}
\caption{Estimated field of Mueller matrices for a scene containing two wooden objects (top) and a rubber (bottom). The diagonal elements clearly discriminate rubber from wood, while the two types of wood are easily discriminated using $M_{43}$.}
\label{fig:Mueller}
\end{figure}

Figure \ref{fig:Mueller} illustrates the ability of the discussed polarimeter, calibrated in reflection mode, to reconstruct spatially-varying Mueller matrix fields. In this experiment the scene contains three objects whose material can hardly be discriminated from the graylevel image (see the top-left image), but visualization of the Mueller matrix coefficients (for instance, $M_{22}$ and $M_{43}$) makes this task straightforward. This shows the potential of the proposed simplified calibration procedure for Mueller polarimeters in material classification.

\bibliography{biblio}

\begin{thebibliography}{10}
\newcommand{\enquote}[1]{``#1''}

\bibitem{Chipman2009}
R.~A. Chipman, \enquote{Polarimetry,} in \enquote{Handbook of optics, Volume
  II,}  M.~Bass, C.~DeCusatis, J.~Enoch, V.~Lakshminarayanan, G.~Li,
  C.~Macdonald, V.~Mahajan, and E.~Van~Stryland, eds. (McGraw-Hill, Inc.,
  2009), chap.~22.

\bibitem{Qi2017}
J.~Qi and D.~S. Elson, \enquote{Mueller polarimetric imaging for surgical and
  diagnostic applications: a review,} {\protect\JournalTitle{Journal of
  Biophotonics}} \textbf{10}, 950--982 (2017).

\bibitem{Vaughn2012}
I.~J. Vaughn, B.~G. Hoover, and J.~S. Tyo, \enquote{Classification using active
  polarimetry,} {\protect\JournalTitle{Proc. SPIE}} \textbf{8364} (2012).

\bibitem{Goldstein1992}
D.~H. Goldstein, \enquote{Mueller matrix dual-rotating retarder polarimeter,}
  {\protect\JournalTitle{Applied optics}} \textbf{31}, 6676--6683 (1992).

\bibitem{Zallat2012}
J.~Zallat, M.~Torzynski, and A.~Lallement, \enquote{Double-pass
  self-spectral-calibration of a polarization state analyzer,}
  {\protect\JournalTitle{Optics letters}} \textbf{37}, 401--403 (2012).

\bibitem{Compain1999}
E.~Compain, S.~Poirier, and B.~Drevillon, \enquote{General and self-consistent
  method for the calibration of polarization modulators, polarimeters, and
  mueller-matrix ellipsometers,} {\protect\JournalTitle{Applied optics}}
  \textbf{38}, 3490--3502 (1999).

\bibitem{Hauge1978}
P.~S. Hauge, \enquote{Mueller matrix ellipsometry with imperfect compensators,}
  {\protect\JournalTitle{Journal of the Optical Society of America A}}
  \textbf{68}, 1519--1528 (1978).

\bibitem{Goldstein1990}
D.~H. Goldstein and R.~A. Chipman, \enquote{Error analysis of a mueller matrix
  polarimeter,} {\protect\JournalTitle{Journal of the Optical Society of
  America A}} \textbf{7}, 693--700 (1990).

\bibitem{Chenault1992}
D.~B. Chenault, J.~L. Pezzaniti, and R.~A. Chipman, \enquote{Mueller matrix
  algorithms,} {\protect\JournalTitle{Proc. SPIE}} \textbf{1746} (1992).

\bibitem{Bhattacharyya2017}
K.~Bhattacharyya, D.~I. Serrano-Garc{\'\i}a, and Y.~Otani, \enquote{Accuracy
  enhancement of dual rotating mueller matrix imaging polarimeter by
  diattenuation and retardance error calibration approach,}
  {\protect\JournalTitle{Optics Communications}} \textbf{392}, 48--53 (2017).

\bibitem{Carmagnola2014}
F.~Carmagnola, J.~M. Sanz, and J.~M. Saiz, \enquote{{Development of a Mueller
  matrix imaging system for detecting objects embedded in turbid media},}
  {\protect\JournalTitle{Journal of Quantitative Spectroscopy and Radiative
  Transfer}} \textbf{146}, 199--206 (2014).

\bibitem{Collins1999}
R.~W. Collins and J.~Koh, \enquote{{Dual rotating-compensator multichannel
  ellipsometer: instrument design for real-time Mueller matrix spectroscopy of
  surfaces and films},} {\protect\JournalTitle{Journal of the Optical Society
  of America A}} \textbf{16}, 1997--2006 (1999).

\bibitem{Sanz2011}
J.~M. Sanz, J.~M. Saiz, F.~Gonz{\'a}lez, and F.~Moreno, \enquote{Polar
  decomposition of the mueller matrix: a polarimetric rule of thumb for
  square-profile surface structure recognition,} {\protect\JournalTitle{Applied
  optics}} \textbf{50}, 3781--3788 (2011).

\bibitem{Smith2002}
M.~H. Smith, \enquote{{Optimization of a dual-rotating-retarder Mueller matrix
  polarimeter},} {\protect\JournalTitle{Applied Optics}} \textbf{41},
  2488--2493 (2002).

\bibitem{Hu2014}
H.~Hu, E.~Garcia-Caurel, G.~Anna, and F.~Goudail, \enquote{Simplified
  calibration procedure for mueller polarimeter in transmission configuration,}
  {\protect\JournalTitle{Optics letters}} \textbf{39}, 418--421 (2014).

\bibitem{Marquardt1963}
D.~W. Marquardt, \enquote{An algorithm for least-squares estimation of
  nonlinear parameters,} {\protect\JournalTitle{Journal of the society for
  Industrial and Applied Mathematics}} \textbf{11}, 431--441 (1963).

\bibitem{Triggs1999}
B.~Triggs, P.~F. McLauchlan, R.~I. Hartley, and A.~W. Fitzgibbon,
  \enquote{Bundle adjustment -- a modern synthesis,}
  {\protect\JournalTitle{LNCS}} \textbf{1883}, 298--372 (1999).

\end{thebibliography}

\bibliographyfullrefs{biblio}

\end{document}